\documentclass[twocolumn,aps,pra,superscriptaddress,amsmath,amssym,floatfix]{revtex4-1}

\usepackage{graphicx}
\usepackage{bm}
\usepackage{color}
\usepackage{amssymb}
\usepackage{enumerate}
\usepackage{comment}
\usepackage[usenames,dvipsnames]{xcolor}
\usepackage[colorlinks=true,linkcolor=MidnightBlue,citecolor=blue,filecolor=green,urlcolor=Violet]{hyperref}

\def\la{\langle}
\def\ra{\rangle}

\def\be{\begin{equation}}
\def\ee{\end{equation}}

\begin{document}

\newcommand{\bigjprob}{{\mathcal{P}}}
\newcommand{\bigprob}{_{\bm{q}_F}{\mathcal{P}}_{\bm{q}_I}}
\newcommand{\cum}[1]{\llangle #1 \rrangle}       					
\newcommand{\op}[1]{\hat{\bm #1}}                					
\newcommand{\vop}[1]{\vec{\bm #1}}
\newcommand{\opt}[1]{\hat{\tilde{\bm #1}}}
\newcommand{\vopt}[1]{\vec{\tilde{\bm #1}}}
\newcommand{\td}[1]{\tilde{ #1}}
\newcommand{\mean}[1]{\la#1\ra}                  					
\newcommand{\cmean}[2]{ { }_{#1}\mean{#2}}       				
\newcommand{\pssmean}[1]{ { }_{\bm{q}_F}\mean{#1}_{\bm{q}_I}}
\newcommand{\ket}[1]{\vert#1\ra}                 					
\newcommand{\bra}[1]{\la#1\vert}                 					
\newcommand{\ipr}[2]{\left\la#1\mid#2\right\ra}            				
\newcommand{\opr}[2]{\ket{#1}\bra{#2}}           					
\newcommand{\pr}[1]{\opr{#1}{#1}}                					
\newcommand{\Tr}[1]{\text{Tr}(#1)}               					
\newcommand{\Trd}[1]{\text{Tr}_d(#1)}            					
\newcommand{\Trs}[1]{\text{Tr}_s(#1)}            					
\newcommand{\intd}[1]{\int \! \mathrm{d}#1 \,}
\newcommand{\dd}{\mathrm{d}}
\newcommand{\fullint}{\iint \! \mathcal{D}\mathcal{D} \,}
\newcommand{\drv}[1]{\frac{\delta}{\delta #1}}
\newcommand{\partl}[3]{ \frac{\partial^{#3}#1}{ \partial #2^{#3}} }		
\newcommand{\smpartl}[4]{ \left( \frac{\partial^{#3} #1}{ \partial #2^{#3}} \right)_{#4}}
\newcommand{\smpartlmix}[4]{\left( \frac{\partial^2 #1}{\partial #2 \partial #3 } \right)_{#4}}
\newcommand{\limit}[2]{\underset{#1 \rightarrow #2}{\text{lim}} \;}
\newcommand{\funcd}[2]{\frac{\delta #1}{\delta #2}}
\newcommand{\funcdiva}[3]{\frac{\delta #1[#2]}{\delta #2 (#3)}}
\newcommand{\funcdivb}[4]{\frac{\delta #1 (#2(#3))}{\delta #2 (#4)}}
\newcommand{\funcdivc}[3]{\frac{\delta #1}{\delta #2(#3)}}
\definecolor{dgreen}{RGB}{30,130,30}

\title{Chaos in Continuously Monitored Quantum Systems: An Optimal Path Approach}

\author{Philippe Lewalle} 
\email{plewalle@pas.rochester.edu}
\affiliation{Department of Physics and Astronomy, University of Rochester, Rochester, NY 14627, USA}
\affiliation{Center for Coherence and Quantum Optics, University of Rochester, Rochester, NY 14627, USA}
\author{John Steinmetz} 
\affiliation{Department of Physics and Astronomy, University of Rochester, Rochester, NY 14627, USA}
\affiliation{Center for Coherence and Quantum Optics, University of Rochester, Rochester, NY 14627, USA}
\author{Andrew N. Jordan}
\affiliation{Department of Physics and Astronomy, University of Rochester, Rochester, NY 14627, USA}
\affiliation{Center for Coherence and Quantum Optics, University of Rochester, Rochester, NY 14627, USA}
\affiliation{Institute for Quantum Studies, Chapman University, Orange, CA 92866, USA}

\date{\today}

\begin{abstract}
We predict that continuously monitored quantum dynamics can be chaotic. The optimal paths between past and future boundary conditions can diverge exponentially in time when there is time--dependent evolution and continuous weak monitoring. Optimal paths are defined by extremizing the global probability density to move between two boundary conditions, and are then expressed as solutions to a Hamiltonian dynamical system. We investigate the onset of chaos in pure--state qubit systems with optimal paths generated by a periodic Hamiltonian. Specifically, chaotic quantum dynamics are demonstrated in a scheme where two non--commuting observables of a qubit are continuously monitored, and one measurement strength is periodically modulated. The optimal quantum paths in this example bear similarities to the trajectories of the kicked rotor, or standard map, which is a paradigmatic example of classical chaos.  We emphasize connections with the concept of resonance between integrable optimal paths and weak periodic perturbations, as well as our previous work on ``multipaths'', and connect the optimal path chaos to instabilities in the underlying quantum trajectories.
\end{abstract}


\maketitle
\section{Introduction \label{sec-intro}} 

Advances in the fabrication, control, and readout of qubits have propelled rapid progress in the field of quantum information processing \cite{BookNielsen, BookWiseman} over the past two decades. 
This was made possible, in part, through foundational theory work concerning open and continuously--monitored quantum--mechanical and quantum--optical systems \cite{Mensky1979, *BookMensky, *Mensky1998-2, Barchielli1982, Diosi1988, BookCarmichael, Wiseman1993-2, *Wiseman1993, *Wiseman1994, BookPercival, Korotkov1999, *Korotkov2001, *Korotkov2016}, which has led to the contemporary theory of stochastic quantum trajectories (SQTs) \cite{BookWiseman, Brun2001Teach, Jacobs2006, BookBarchielli}. 
Continuous monitoring of quantum systems, where the system is inherently open and experiences measurement--induced backaction and non--unitary dynamics, is an active field of research \cite{Naghiloo2016flor, Ibarcq2016, Vool2016, Ficheux2017}, generating both experimental and theoretical interest related to topics such as feedback control \cite{BookWiseman, Doherty2000, Ahn2002, *Ahn2003, *Ahn2003-2, Sayrin2011, Vijay2012, Ibarcq2013, Rouchon2015, Zhang2017, Gourgy2018, Minev2018}, entanglement generation \cite{Ruskov2003-2, Trauzettel2006, Williams2008, Riste2013deterministic, Roch2014,  Martin2015remote, Chantasri2016, Silveri2016}, and state stabilization \cite{Taylor2017} in qubits. 
These topics have potential applications in larger research efforts toward quantum error correction \cite{BookWiseman, Ahn2002, *Ahn2003, *Ahn2003-2}.

\par Chaos has been researched in a body of literature that is largely independent of that just cited. With the exception of some work concerning chaos in quantum optics \cite{Ackerhalt1985}, the overwhelming majority of the literature on ``quantum chaos'' is concerned with the behavior of quantized versions of systems that have well--defined classically--chaotic analogs \cite{BookGutzwiller,BookHouchesQChaos}. 
In classical mechanics, chaos is defined by exponential sensitivity to changes in initial conditions; therefore, a classically--chaotic system, while mathematically deterministic, is effectively unpredictable in the long term unless it can be initialized with perfect precision. 
Exponential divergence of trajectories is often quantified by computing a Lyapunov exponent (LE), where a positive exponent signals that nearby paths diverge \cite{BookOtt, BookReichl, ClassQuillen, BookTabor, BookStrogatz}. 
Studies of classically--chaotic systems in the quantum regime have resulted in insights about fluctuations in the spectra of many--body quantum systems \cite{BookMehta, LectureBohigas, BookPorter}, semi--classical wavepacket dynamics \cite{LectureSmilansky, LectureGutzwiller, LectureHeller}, and the transition between classical and quantum dynamics \cite{LectureBerry, Zurek1994, Habib1998, Zurek1998, Pattanayak2003, Carvalho2004, Ota2005, Kapulkin2008, Pokharel2016}. 
There is an essential difference in this field, however, in that classical chaos is defined in terms of the divergence of trajectories, which do not exist at all in closed quantum systems. 
Further, the quantum dynamics of wave functions in closed systems are explicitly unitary and applied linearly. 
A handful of works have looked specifically at chaos in open quantum systems~\cite{Carvalho2004,Ott1984}, or the SQTs of continuously--monitored quantum systems with chaotic classical analogs~\cite{Spiller1994, Brun1995, Bhattacharya2000, Carlo2005, Habib2006, Eastman2016}; the latter have primarily focused on theoretical investigations of quantization and measurement of the damped--driven Duffing oscillator, and its classical--to--quantum transition.

\par In the current paper we introduce a fundamentally different kind of quantum chaos, using optimal paths (OPs) to describe the stochastic dynamics induced by quantum measurement \cite{Chantasri2013, Chantasri2015, Areeya_Thesis}.
OPs are defined as the extremal--probability paths which move from an initial state $\mathbf{q}_i$ to a final state $\mathbf{q}_f$ over a traversal time $T$. Although our usage here, to describe chaotic dynamics in a quantum measurement problem, is novel, similar mathematics have been used to study classical stochastic systems (see e.g.~\cite{BookKamenev,Dykman1994, Dykman1994-2, Dykman2001, Dykman2006, Dykman2008}). Both theoretical and experimental investigations of qubit dynamics under various measurement schemes \cite{Weber2014, Jordan2015flor, Lewalle2016, Mahdi2016} have elaborated on the OP concept for quantum measurement, demonstrating good agreement between theory and experiment. The behavior of the open and measured quantum system is expressed in terms of a Hamiltonian dynamical system in the OP approach. This system gives rise to equations of motion for the OPs, which give a well--defined and conceptually clear definition of chaos in continuously monitored quantum systems. 
These OPs for qubits are mathematically analogous to the classical paths in the sense of being extrema of an action \footnote{OPs are effectively a low-noise idealization of the open qubit dynamics, where the noise is directly due to measurement backaction, which is an inherently quantum--mechanical effect. Mathematically, deriving the OPs, or paths in the small--noise limit, is quite similar to deriving classical paths as the limit of a quantum system.}, despite the qubit having no classical analog.
We consider a qubit simultaneously monitored along the non--commuting observables $\sigma_x$ and $\sigma_z$ as an example in which to implement our methods; this is based on the experimental system realized in \cite{Leigh2016}, which is also considered theoretically in Refs.~\cite{Lewalle2016, Chantasri2017, Atalaya2017}. 
We extend our investigation of this system to a case where one of the measurement strengths is varied periodically; in the limit where the periodic measurements become strong, our system bears qualitative similarities to the kicked rotor, or standard map \cite{BookReichl, BookJose, ClassQuillen, Chirikov1969, *Chirikov1979}, a paradigmatic example of classical chaos. 
The kicked rotor has been studied in a quantum context \cite{Carvalho2004, Carlo2005, Lima1991, Bitter2016, *Bitter2017}, but our system, while qualitatively similar at the mathematical level, is physically quite different.

\par Our article is laid out as follows: in section \ref{sec-tm}, we introduce our theoretical methods. 
This includes summaries of the mathematics of OPs generally, and of situations where multiple OP solutions link two boundary conditions (``multipaths'' \cite{Lewalle2016, Mahdi2016}). 
We also propose a definition of the Lyapunov exponent for OP dynamics. 
In section \ref{sec-xzkick} we apply our methods to the two--measurement example mentioned above. 
With that example in mind, we are then able to more formally connect OP chaos with rapid growth in the number of multipath solutions in section~\ref{sec-ch_mani}. 
Some discussion, conclusions, and outlook are included in section \ref{sec-concl}.

\section{Theoretical Model and Methods \label{sec-tm}}
Under continuous measurement, the quantum state is updated through the application of non--unitary operators, which are constructed based on the correspondence between the specific measurement process and the readout signal(s) $\mathbf{r}(t)$. 
In other words, every time a new readout value is acquired, the state $\rho(t)$ is updated by \cite{BookNielsen}
\be \label{stateup}
\rho(t+dt) = \frac{\mathcal{M}_{dt} \rho(t) \mathcal{M}_{dt}^\dag}{\text{tr}\left(\mathcal{M}_{dt} \rho(t) \mathcal{M}_{dt}^\dag\right)},
\ee
where $\mathcal{M}_{dt}(\mathbf{r})$ is the measurement operator. 
The probability density for acquiring a particular readout given a particular quantum state is given by $\wp(\mathbf{r}|\rho) = \text{tr}(\mathcal{M}_{dt} \rho(t) \mathcal{M}_{dt}^\dag)$; this is associated with the function $\mathcal{G}$ we discuss below in the limit $dt \rightarrow 0$. 
The state update can be approximated by expanding \eqref{stateup} to first order in $dt$; this is associated with the function $\mathcal{F}$ used below. 
Such expansions can also lead to the stochastic master equation (SME). The specific operators we use in our subsequent qubit measurement examples are based on a Bayesian update scheme \cite{Korotkov1999, *Korotkov2001, *Korotkov2016}. See appendix \ref{sec-review} for details.

\subsection{Optimal Paths for Pure-State Qubits \label{sec_qubitps}}
\par OPs are defined by extremizing the joint probability of a path $\mathbf{q}(t)$ through quantum state space (e.g. Bloch sphere coordinates for a qubit) and readout(s) $\mathbf{r}(t)$. 
Such a probability, constrained to paths which link a given initial $\mathbf{q}(0)$ and final $\mathbf{q}(T)$ states, may be expressed in terms of a path integral of the form $\int \mathcal{D}[\mathbf{p}] e^S$, where we define the stochastic action $S$, which contains the stochastic Hamiltonian $H$, by
\be \label{sact}
S = \int_0^T dt \left( H(\mathbf{q},\mathbf{p},\mathbf{r},t) - \mathbf{p}\cdot \dot{\mathbf{q}} \right).
\ee
A least action principle $\delta S = 0$ optimizes the path probability, and gives us OPs as solutions to 
\be 
\dot{\mathbf{q}} = \partl{H}{\mathbf{p}}{}, \quad \dot{\mathbf{p}} = - \partl{H}{\mathbf{q}}{}, \quad \partl{H}{\mathbf{r}}{}\bigg|_{\mathbf{r}^\star} = 0,
\ee
which are Hamilton's equations with an additional optimization condition on the measurement readout(s) \cite{Chantasri2013,Chantasri2015}, which defines the optimal readout(s) $\mathbf{r}^\star(\mathbf{q},\mathbf{p})$. The stochastic Hamiltonian can be expressed in the form 
\be \label{stoch}
H = \mathbf{p} \cdot \mathcal{F}[\mathbf{q},\mathbf{r},t] + \mathcal{G}[\mathbf{q},\mathbf{r},t],
\ee
where $\dot{\mathbf{q}} = \mathcal{F}$ can be obtained from a quantum Bayesian scheme \cite{Korotkov1999,*Korotkov2001,*Korotkov2016} by expanding a state update equation to $O(dt)$, or equivalently as the Stratonovich form of a stochastic master equation \cite{BookWiseman, Jacobs2006, BookGardiner2, Gambetta2008} (see appendix \ref{sec-review} or~\cite{Lewalle2016} for a detailed derivation in the context of OPs). 
The ``probability cost-function'' $\mathcal{G}$ \cite{Lewalle2016} is defined by expanding the log--probability $\ln \lbrace P(\mathbf{r}(t)|\mathbf{q}(t)) \rbrace$ for the readout update given state $\mathbf{q}\left(t\right)$, as modeled in the Bayesian formalism. 
A Hamiltonian $H^\star(\mathbf{q},\mathbf{p},t)$ may be obtained by integrating out the readout(s) from the path integral, because the action is Gaussian in $\mathbf{r}(t)$, or equivalently by substituting $\mathbf{r}^\star$ back into $H$. 

\par Generically, when we consider the evolution of a qubit state, $\mathbf{q}$ could include all three of the Bloch sphere coordinates $x$, $y$, and $z$. However, for simplicity, and to focus on new effects, we will limit the space throughout this paper with some simplifying assumptions. 
We suppose that all our states start pure and stay pure (this implicitly includes an assumption that our measurements have ideal quantum efficiency). 
Furthermore, we may constrain the dynamics to the $xz$--plane of the Bloch sphere, so that they may be completely expressed by the polar angle $\theta$ in the $xz$--plane, for $x = \sin\theta$ and $z = \cos\theta$. 
A dispersive qubit readout, as modeled in the Bayesian scheme, results in $\mathcal{F}$ and $\mathcal{G}$ having certain forms and properties, e.g.~$\mathcal{G}$ is quadratic in $\mathbf{r}$; these force the optimal readouts $\mathbf{r}^\star(\theta,p)$ to be linear in $p$. 
Using these relationships, we are able to simplify our stochastic Hamiltonian down to the form \cite{Lewalle2016}
\be \label{hop}
H^\star(\theta,p,t) = a(\theta,t)\:(p^2-1)+b(\theta,t)\:p.
\ee
The angle $\theta$ parameterizes the quantum states and $p$ is the generalized ``momentum'' conjugate to $\theta$. 
The functions $a$ and $b$ are determined by the particulars of any driving and measurements applied to the qubit. 
We will add one more assumption and corresponding notation concerning the time--dependence of $H^\star$, for later use; we suppose that $a$ and $b$ are such that $H^\star$ can be split into
\be 
H^\star(\theta,p,t) = H^{(0)}(\theta,p) + h(\theta,p,t).
\ee
The time--independent term $H^{(0)}$ must be integrable, because its phase space is two--dimensional, and the stochastic energy $E = H^{(0)}$ is conserved; then $h$ can be interpreted as a time--dependent perturbation added to those integrable dynamics.

\subsection{Multipaths and Lagrange Manifolds \label{sec_multi-intro}}
In Refs.~\cite{Lewalle2016,Mahdi2016} we defined ``multipath'' behavior, and identified it in physically--realizable qubit systems. 
We review some definitions and concepts that will be needed in this paper. 
A multipath group of solutions exists when two or more OPs link the same boundary conditions $\theta_0$ and $\theta_T$ (for fixed $T$). A particular Lagrange Manifold (LM) in the OP phase--space, which includes all $p_0$ at one single initial state $\theta_0$, may be used to detect multipaths. 
This LM describes all of the OPs branching out from a particular initial state (the manifold includes all of the different OP possibilities on which we may wish to post--select). 
The manifold will then deform under the Hamiltonian flow of \eqref{hop} over time. 
Multipaths form at final states $\theta_T$ which are represented several times in the final manifold; multipaths form at final boundary conditions where the manifold fails the vertical line test. 
This can happen either due to the formation of a catastrophe \cite{BookArnoldCatastrophe}, which is a fold in the LM, or simply an overlap of the LM with itself mod-2$\pi$, in which case we say the solutions have different winding counts about the Bloch sphere. 
The multipath phenomenon is quite similar to that of optical caustics; just as many rays of light may leave a source with different wave--vectors, and then re--converge on some other location, OPs may leave a given state $\theta_0$ with different $p_0$, and then re--converge on some other state $\theta_T$. 
Regions of $\theta_T$ where the LM overlaps itself are the caustic regions. 
Catastrophes in the manifold specify the boundaries of such regions in the final conditions, where the number of OPs connecting the same $\theta_0$ and $\theta_T$ increases. 
The different $p_0$ are not immediately experimentally accessible for qubit OPs; the generalized momenta merely index different possible optimal readouts $\mathbf{r}^\star$, which may occur according to some probability density.
\par It is useful to define a Jacobian at time $T$ 
\be 
J_T = \partl{\theta_T}{p_0}{},
\ee
for the manifold. 
The LM we use for finding multipaths is defined by $J_0 = 0 \: \forall \: p_0$. 
Catastrophes generating multipaths form where $J_T = 0$, or where its inverse, the ``Van-Vleck determinant'' $V = |J^{-1}| = |\partial^2 S/\partial \theta_0 \partial \theta_T|$, expressed in terms of the stochastic action $S$ \eqref{sact}, diverges \cite{Lewalle2016}. 
The behavior of manifolds we care about here can be described using $J_T$ and a curvature or concavity $C_T = \partial J_T / \partial p_0 = \partial^2 \theta_T / \partial p_0^2$. 

\par A multipath containing two MLPs has been observed in experiment \cite{Mahdi2016}, and in principle a larger number of paths could also be extracted from data given a large enough ensemble of SQTs. 
However, the difficulty of this task increases significantly with the number of approximately--equally--likely paths meeting a given set of boundary conditions. A system with a large number of multipaths becomes significantly less predictable. 

\subsection{Computing Lyapunov exponents for OPs \label{sec-LEdef}}
We now define a measure of OP chaos. 
Based on classical definitions of chaos, we are interested in paths with similar initial conditions which diverge exponentially from each other. 
That is, we consider paths with similar initial states where a distance $D(t) \sim D_0 e^{t\lambda(t)}$ measured between the two paths grows such that $\lambda(t) > 0$ over the time interval of interest. 
The quantity $\lambda(t)$ we have implicitly defined above is the Lyapunov Exponent (LE), which quantifies how quickly paths converge or diverge. 
A number of approaches and conventions for computing LEs can be found throughout the literature, e.g. in \cite{BookOtt, ClassQuillen,  BookReichl}. We will use the simplest definition of the LE,
\be \label{lyap}
\lambda(t) \equiv \frac{1}{t} \ln \left( \frac{D(t)}{D_0} \right),
\ee
obtained directly from above, where we must account for the finite--time nature of this LE, which cannot grow indefinitely due to the bounded nature of the Bloch sphere. 
We define a distance about a path $\theta$, initialized at $\theta_0$ and $p_0$, by using two auxiliary paths $\theta_\pm$ initialized at $\theta_0 \pm \delta\theta_0$, where $\delta \theta_0$ is small (we use $\delta \theta_0 = 0.01$ in subsequent examples) and $p_0$ is fixed ($\delta p_0 = 0$). Using components of distances across the Bloch sphere $\left(\delta x^\pm (t)\right)^2 = \left(\sin \theta(t) - \sin\theta^\pm(t)\right)^2$ and $\left(\delta z^\pm (t)\right)^2 = \left(\cos \theta(t) - \cos\theta^\pm(t)\right)^2$, we define the Euclidean distance as the average of those between the main path and each auxiliary path, i.e.  
\be\begin{split} \label{dist}
D(\theta(t)) =& \tfrac{1}{2}\sqrt{\left(\delta x^+(t)\right)^2 + \left(\delta z^+(t)\right)^2} \\ &+ \tfrac{1}{2}\sqrt{\left(\delta x^-(t)\right)^2 + \left(\delta z^-(t)\right)^2}.
\end{split}\ee 
Using two auxiliary paths offset in opposite directions symmetrizes our distance measure (we have no physical reason to favor the shift being in one direction or the other).
\par Although the distance used in the LE would typically be a distance over all dimensions of phase--space for a classical system (i.e.~would also account for distances between $p$ and some $p_\pm$ over time), we here define our distance in the $\theta$ direction only. 
This is justified for OPs, because the $p$ cannot be measured directly, and we wish to emphasize differences in the quantum state itself. 
Exponential growth in distance, corresponding to chaos, will be sufficiently characterized by $\lambda(t)$ sustaining a positive value over the evolution time of interest because of the finite system size.

\par The way we initialize $D_0$ and define $D(t)$ emphasizes the effect of imperfect state preparation on the OP dynamics. 
Preparing states with the same $p_0$ (although not a physically well--defined task for individual SQTs) amounts to initializing them with similar optimal readout(s) $\mathbf{r}_0^\star$. 
The variation in the initial optimal readout(s) $\delta \mathbf{r}_0^\star \approx \delta\theta_0\cdot \partial_\theta \mathbf{r}_0^\star |_{\theta_0,p_0}$ is(are) on the same order as the small variation $\delta\theta_0$ in the state itself under this scheme. 
Appeals to physical intuition demand that this remain so as long as the states remain similar, because the monitoring of observables leading to readouts is precisely what is used construct the SQT in the quantum state $\theta$ to begin with \footnote{In fact, if we take physically sensible readout signal, but update the state starting from a very wrong initial state, the estimate from the dynamics will still tend back towards the correct state over time, as more readout information is acquired. The states and readouts are necessarily directly connected.}. 
Furthermore, we will see in section~\ref{sec-ch_mani} that this definition lends itself well to connecting the chaos it defines with the multipath behavior given by LMs. 

\section{Two Continuous Measurements with Variable Strengths \label{sec-xzkick}}

We now demonstrate the presence of this kind of quantum chaos in a specific system. 
Consider a qubit simultaneously subjected to weak measurements along $\sigma_x$ and $\sigma_z$ \cite{JordanButt2005, Leigh2016, Chantasri2017, Atalaya2017, Lewalle2016}. 
The measurements are described by characteristic times $\tau_x$ and $\tau_z$, respectively, which determine the time scale on which the bare measurement causes wavefunction collapse. 
If $dt$ is the time to perform a single measurement, then $\tau\gg dt$ denotes a weak measurement and $\tau\lesssim dt$ denotes a stronger measurement, which becomes projective as $\tau/dt \rightarrow 0$. 
In an experiment, $dt$ would reflect the time required to acquire one readout value; the OPs are constructed in the limit of weak and continuous measurements (the limit as $dt \rightarrow 0$). 
Below we will always leave $\tau_x = 1~\mu\mathrm{s}$ fixed, but we will modulate the strength of $\tau_z$, exploring regimes both where $\tau_z \lesssim \tau_x$ (both measurements are still weak; see section~\ref{sec-weakreg} and appendices~\ref{sec-xzperturb} and~\ref{sec-anim_strobo}) and regimes where $\tau_z \ll \tau_x$ (the $z$--measurement is periodically much stronger than the $x$--measurement; see section~\ref{sec-strongreg} and appendix~\ref{sec-anim_strobo}).  
\par Our $z$--measurement will be modulated in strength according to
\be \begin{split} \label{kick}
\tau_z(t) &= \tau_x-A\: g(t) \\
\text{with}\quad
g(t) &= \exp\left[ - \frac{(t-\tfrac{1}{2}\Lambda)^2}{2\tau_m^2} \right] \text{ for } t \in [0, \Lambda],
\end{split} \ee
which is repeated with period $\Lambda$, such that $g(t) = g(t+\Lambda)$ for all $t$. 
It is useful to notate $\gamma_x \equiv 1/\tau_x$, $\epsilon = A/\tau_x$, and set $\gamma_z \equiv 1/\tau_z = \gamma_x(1-\epsilon g(t))^{-1}$, expressing the peak strength of the $z$--measurement in terms of the dimensionless $\epsilon\in [0,1]$. 
We always retain $\tau_m \ll \Lambda$, such that the changes in $\tau_z$ are narrow compared with their period; for the purposes of numerical examples, we use $\tau_m = 25~n\mathrm{s}$.
The form \eqref{kick} amounts to a ``kick'' in the strength of the $z$--measurement, such that the measurement becomes stronger at the peak of the Gaussian (every half--integer microsecond). 
When $\epsilon \ll 1$ ($A \ll \tau_x$), both the $x$-- and $z$--measurements are weak, and when $\epsilon \rightarrow 1$ ($A \rightarrow \tau_x$), the $z$--measurement becomes projective. 
Intuitively, the stronger measurements can cause sudden jumps in the SQTs and OPs, as they induce at least partial state collapse.  
The size of the jumps is obviously related to the kick strength $\epsilon$. 
The relative values of $\tau_x$ and $\Lambda$ also impact the OP jump size, however. 
The OPs do not have to jump when $\tau_x \ll \Lambda$, since diffusion from any state will easily reach the $z$--eigenstates before another kick happens. 
However, jumps are necessary in the opposing regime $\tau_x \gg \Lambda$, since diffusion from an arbitrary state is unlikely to reach the $z$--eigenstates on its own before a kick. 
See appendix~\ref{sec-projlim} for a simplified model of the projective--kick limit, and further details. 
In examples below, we always take $\tau_x = 1~\mu\mathrm{s} = \Lambda$.
Our choice of $\tau_x = 1~\mu\mathrm{s}$ and the kick duration $\tau_m = 0.025~\mu\mathrm{s}$ implies that at $\epsilon = 0.975$ we pass the point where $2 \cdot \text{min}(\tau_z)$ fits within one standard deviation of its kick peak (meaning that this is approximately the value of $\epsilon$ where a single kick lasts long enough compared to $\tau_z$ to collapse the state to the eigenstates of $\sigma_z$).

\subsection{Stochastic Hamiltonian for Two Non-Commuting Qubit Measurements}
The stochastic Hamiltonian which generates OPs for the two--measurement system is defined by \eqref{stoch} and
\be \begin{split} \label{fxz}
\mathcal{F} &=  \frac{r_x}{\tau_x} \cos\theta - \frac{r_z}{\tau_z}\sin\theta \\
\mathcal{G} &= -\frac{r_x^2 - 2 r_x \sin\theta + 1}{2\tau_x}-\frac{r_z^2 - 2 r_z \cos\theta + 1}{2\tau_z}
\end{split} \ee
(see ~\cite{Lewalle2016} and/or ~\cite{Chantasri2013,Chantasri2017}). With the optimal readouts $r_x^\star =\sin\theta + p \cos\theta$ and $r_z^\star =\cos\theta- p \sin \theta$ substituted in (or integrated out), we obtain $H^\star = (p^2-1)\: a(\theta,t) + p \:b(\theta,t)$ with
\be \begin{split} \label{hxz}
a(\theta,t) &\equiv \frac{\sin^2\theta}{2\tau_z(t)} + \frac{\cos^2\theta}{2\tau_x}, \\
b(\theta,t) &\equiv \sin\theta \cos\theta \left( \frac{1}{\tau_x} - \frac{1}{\tau_z(t)} \right).
\end{split} \ee
When the measurement strengths are equal (i.e.~$\tau_x = \tau = \tau_z$), as is approximately true for all time when $\epsilon \ll 1$, and at the times between kicks regardless of the value of $\epsilon$, the stochastic Hamiltonian reduces to that of a simple rotor,
\be \label{hrotor}
H_{rot}^\star = \frac{p^2-1}{2\tau}.
\ee
As discussed in appendix \ref{sec-diffuse}, and in other works \cite{Leigh2016,Chantasri2017,Atalaya2017,Lewalle2016}, this rotor Hamiltonian corresponds to simple diffusion of the state on a circle. 
The rotor Hamiltonian is integrable because the stochastic energy $E = H$ is conserved, as is $p$. Since $H^\star_{rot}$ is a function of $p$ only, we may furthermore regard the pair $\lbrace \theta, p \rbrace = 1$ as its action--angle coordinates.

\par We now introduce the kick \eqref{kick} into $H^\star$. Notice that $\gamma_z$ can be expanded in powers of $\epsilon$ such that
\be \label{xz_kick}
\gamma_z = \frac{\gamma_x}{1 - \epsilon g(t)} = \gamma_x\sum_{n=0}^\infty \left[ \epsilon g(t) \right]^n.
\ee
When $\epsilon \ll 1$, the kick barely changes the measurement strength, and gives OP dynamics described by a small perturbation to those of $H_{rot}^\star$.
The full stochastic Hamiltonian $H^\star$ can be expanded in powers of $\epsilon$ as
\be 
H^\star(\theta,p,t) = H^{(0)}(p) + \sum_{n=1}^{\infty} \epsilon^n H^{(n)}(\theta,p,t),
\ee
where $H^{(0)} = (p^2-1)/2\tau_x = H^\star_{rot}$ and 
\be  \label{hexpansion}
H^{(n\geq1)}= g^{n}(t)\left(\frac{p^2-1}{2\tau_x} \sin^2\theta - \frac{p}{\tau_x}\sin\theta\cos\theta\right).
\ee
In the notation of previous sections, we have $H^\star = H^{(0)}(p) + h(\theta,p,t)$, where the time--dependent part $h$ containing the kicks has been decomposed in powers of $\epsilon$. Our aim below will be to explore the dynamics first for small $\epsilon \ll 1$, and then for stronger ($\epsilon \rightarrow 1$) measurements. 
Links to, and explanations of, a series of supplemental animations illustrating the dynamics across the full range of $\epsilon$ can be found in appendix~\ref{sec-anim_strobo}. 

\par We make a few more general remarks before analyzing the dynamics in detail. 
First, at the level of SQTs, the integrable part of the Hamiltonian $H^\star_{rot}$ describes isotropic diffusion about the Bloch sphere, consistent with experimental findings \cite{Leigh2016} (see also Refs.~\cite{Lewalle2016,Chantasri2017}, and appendix \ref{sec-diffuse}). 
However, when $\tau_x \neq \tau_z$, diffusion towards one set of measurement eigenstates is favored over the other, and the SQTs diffuse anisotropically. The diffusion constants generically grow larger as $\tau$ shrinks. 
See appendix~\ref{sec-diffuse} for further details. Second, the system we have constructed greatly resembles the kicked--rotor or standard map, a system which is often used as a pedagogical example of classical chaos. 
The classical kicked rotor \cite{BookJose, BookReichl, ClassQuillen, Chirikov1969, *Chirikov1979} is derived by adding periodic $\delta$--kicks to the Hamiltonian for a simple rotor; such a perturbation, which destroys conservation of $E$ and $p$, is known to make the rotor's dynamics become chaotic, especially for stronger kick strengths. 
The system we have described above is also a rotor disturbed by a periodic force, and it most resembles the classical kicked rotor in the limit where $\epsilon\rightarrow 1$ and $\tau_m \rightarrow 0$ \footnote{The correspondence between our two--measurement kicking and the classical kicked rotor is still not exact. For instance the phase space for the quantum simulation is identical over $\theta \in [0,\pi]$ and $\theta \in[\pi,2\pi]$, because the eigenstates of the kicking measurement are important in the physical situation, and the dynamics are symmetric about them (at least as long as qubit relaxation may be neglected)} (see appendix~\ref{sec-projlim} for details). 
The onset of chaos in the kicked rotor and similar mappings is well understood \cite{BookOtt, BookReichl, BookJose, ClassQuillen}, and below we will show that our $H^\star$ generates qualitatively similar dynamics, with $\epsilon$ playing a role similar to the kick--strength parameter of the standard map.

\subsection{Resonances in the weak measurement regime \label{sec-weakreg}}

\begin{figure}
\begin{picture}(200,138)
\put(-22,0){\includegraphics[width = .75\columnwidth,trim = {13pt 0pt 30pt 20pt},clip]{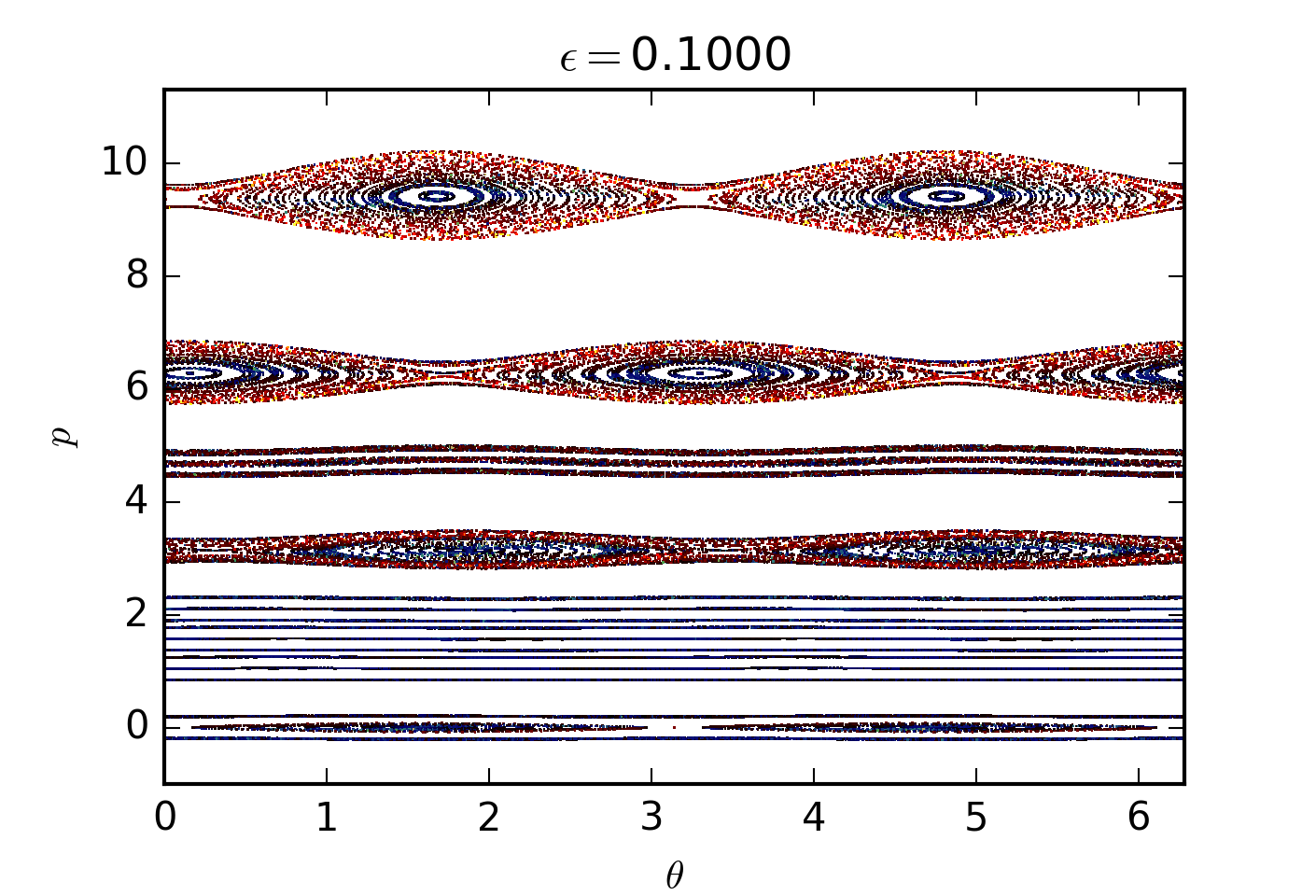}}
\put(167,10){\includegraphics[height = 120pt]{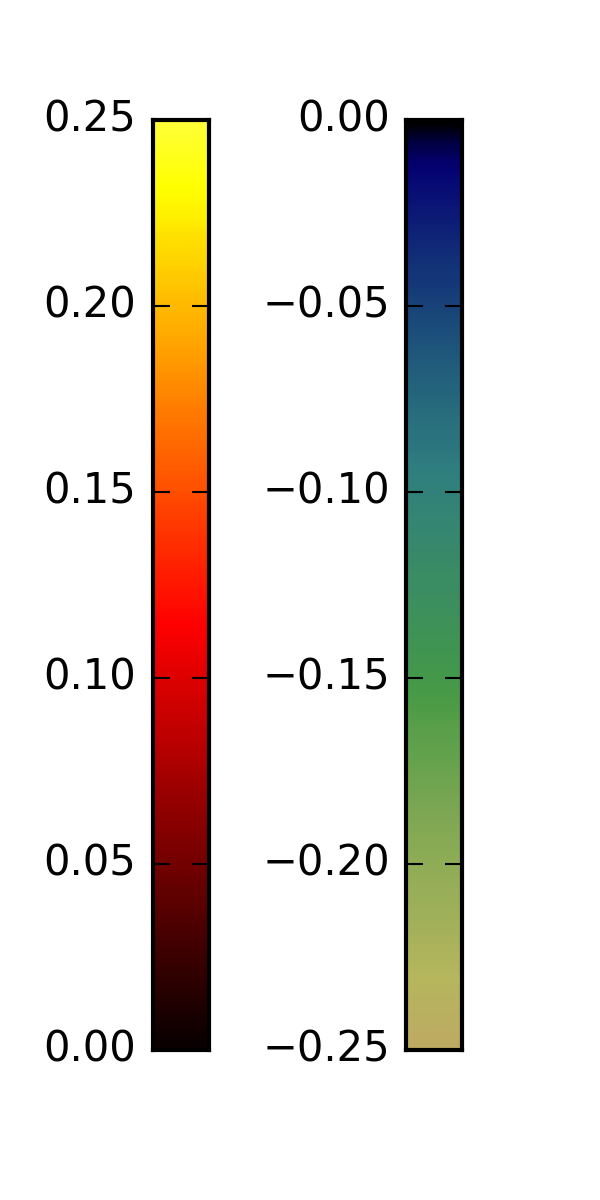}}
\put(179,5){\small $\lambda$ (MHz)}
\end{picture}
\caption{ (Color online) We show a stroboscopic phase portrait of $H^\star$ for $\epsilon = 0.1$. Paths are initialized across $\theta$ for $p = 0, \: \pi/3, \: \pi/2, \: 2\pi/3, \: \pi, \: 3\pi/2, \: 2\pi,$ and $3\pi$, along with small offsets ($\pm0.2$) about each $p_0$ to improve the plotted resolution of features in the phase--space. Points are plotted in between kicks every $\Lambda = 1~\mu\mathrm{s}$ from $T = 0 \rightarrow 100~\mu\mathrm{s}$ to construct the image. Colors are assigned based on the LE $\lambda(t)$ \eqref{lyap}. We see that integrable rotor tori are destroyed at the $p = k\pi$ resonances, where $k$ is an integer, giving way to alternating hyperbolic and elliptic fixed points with new periodic islands; the other initial conditions shown are not impacted substantially. A formal derivation in support of this result, and further remarks, appear in appendices \ref{sec-resdetail} and \ref{sec-xzperturb}. This image is drawn from a larger animation which can be found in the supplements described in appendix \ref{sec-anim_strobo}. Compare the pattern of islands and tori above with the shape of the LM displayed in Fig.~\ref{fig-LMreson}(c).}\label{fig-earlyreson}
\end{figure}

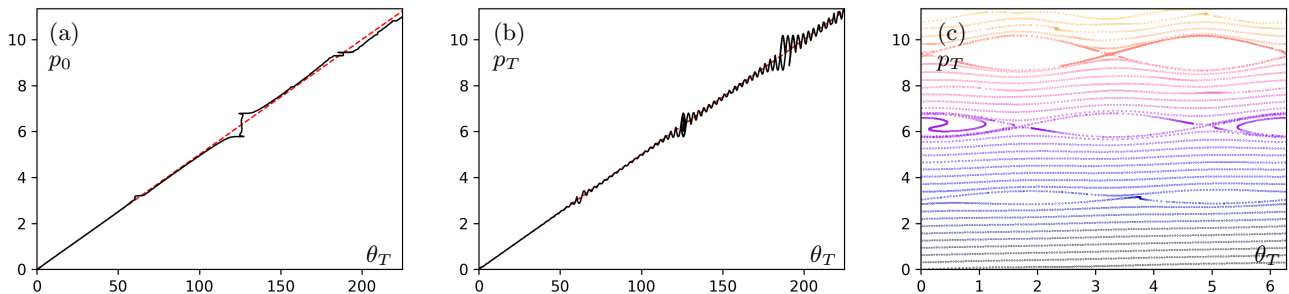
\begin{figure*}
\begin{tabular}{ccc}
\includegraphics[width = .32\textwidth]{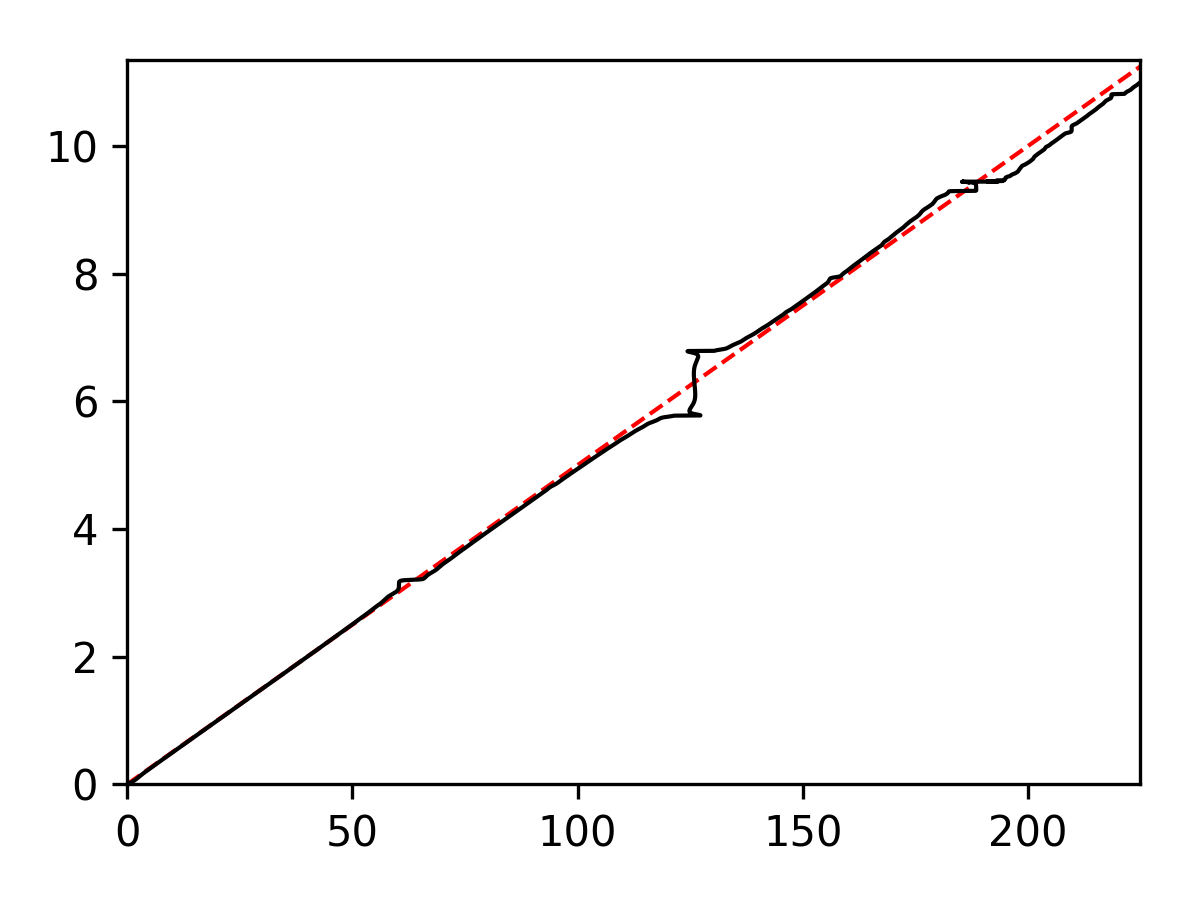} &
\includegraphics[width = .32\textwidth]{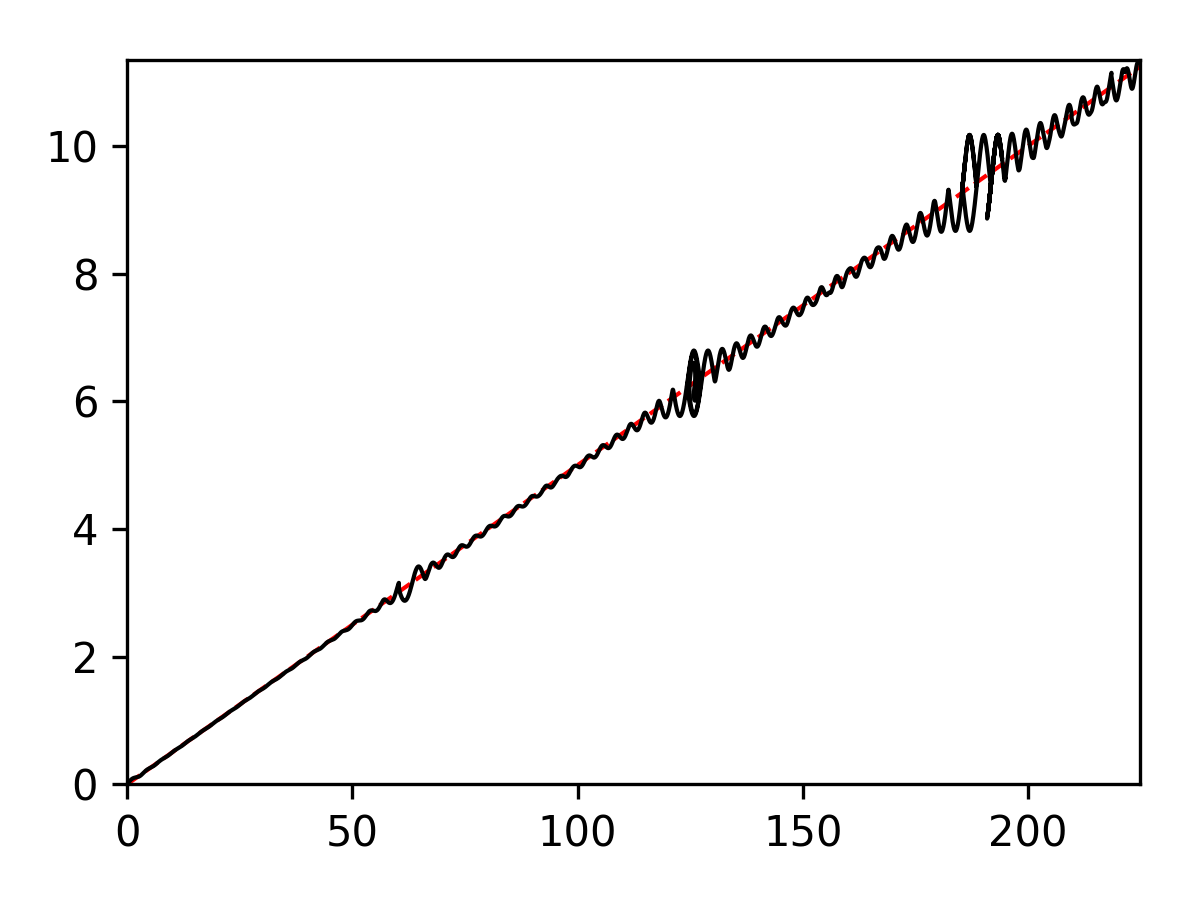} &
\includegraphics[width = .32\textwidth]{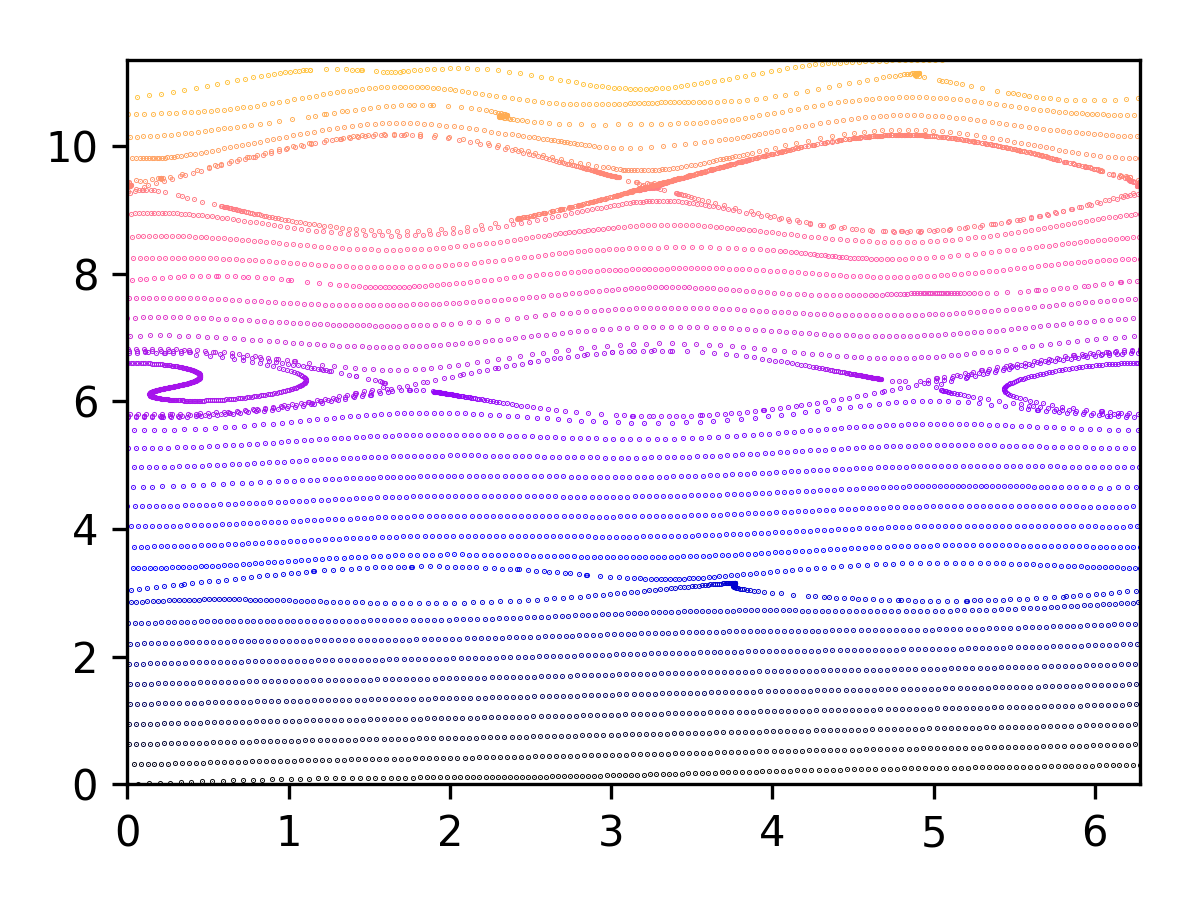}
\end{tabular}
\begin{picture}(0,0)
\put(-481,45){(a)}
\put(-313,45){(b)}
\put(-145,45){(c)}
\put(-481,35){$p_0$}
\put(-313,35){$p_T$}
\put(-145,35){$p_T$}
\put(-361,-39){$\theta_T$}
\put(-193,-39){$\theta_T$}
\put(-25,-39){$\theta_T$}
\end{picture}
\caption{(Color online) We show plots of the LM initialized at $\theta_0= 0$, after $T = 20~\mu\mathrm{s}$, for $\epsilon = 0.1$, $\tau_x = 1~\mu\mathrm{s} = \Lambda$, and $\tau_m = 0.025~\mu\mathrm{s}$. In (a) we plot $\theta_T$ against $p_0$, in (b) we plot $\theta_T$ against $p_T$, and in (c) we repeat plot (b), but with $\theta_T$ plotted mod-$2\pi$ since states separated by an angle of $2\pi$ on the Bloch sphere are the identical. Lighter colors denote higher winding numbers. In (a) and (b), the black line shows the LM for $H^\star = H^{(0)} + h$, and the dashed red line shows the LM for $H^{(0)}$ only. Deviations in the LM due to the perturbations, relative to the integrable case, are primarily restricted to values $p \approx k \pi$ for integer $k$, that is, in the neighborhood of resonances between $H^{(0)}$ and the weak applied kicks. We further note that the representation in (c) highlights the enormous similarity between the LM and stroboscopic phase portrait shown in Fig.~\ref{fig-earlyreson}; the same patterns highlighting flat paths, except at islands forming around resonance zones, are clearly visible in both images.}\label{fig-LMreson}
\end{figure*}

\begin{figure*}
\begin{picture}(500,150)
\put(0,5){\includegraphics[height=.22\textheight]
{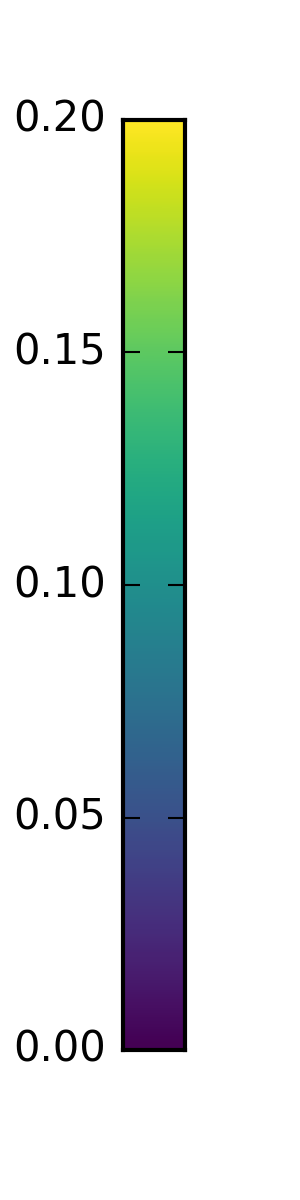}}
\put(40,0){\includegraphics[height=.22\textheight]{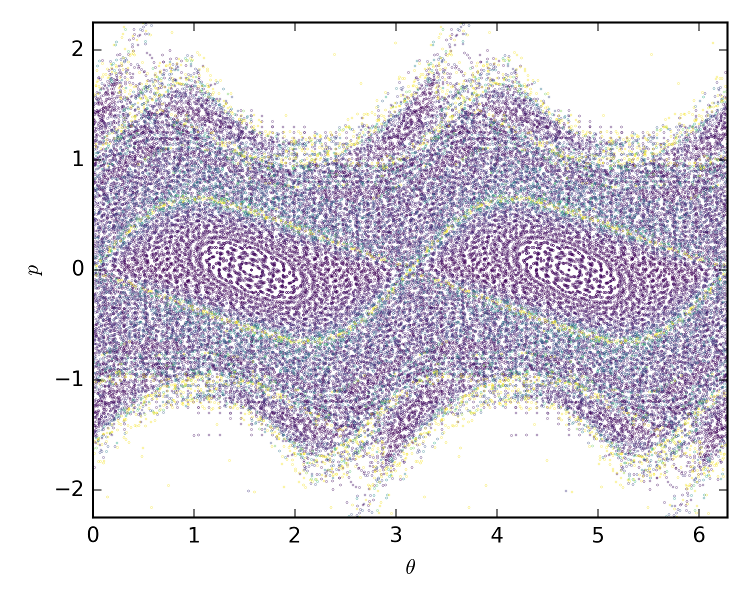}}
\put(235,0){\includegraphics[height=.22\textheight]{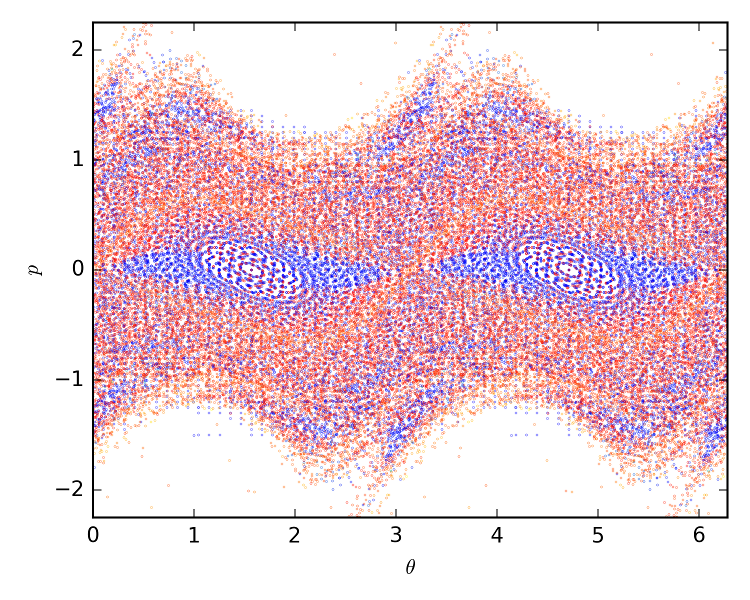}}
\put(430,5){\includegraphics[height=.22\textheight]{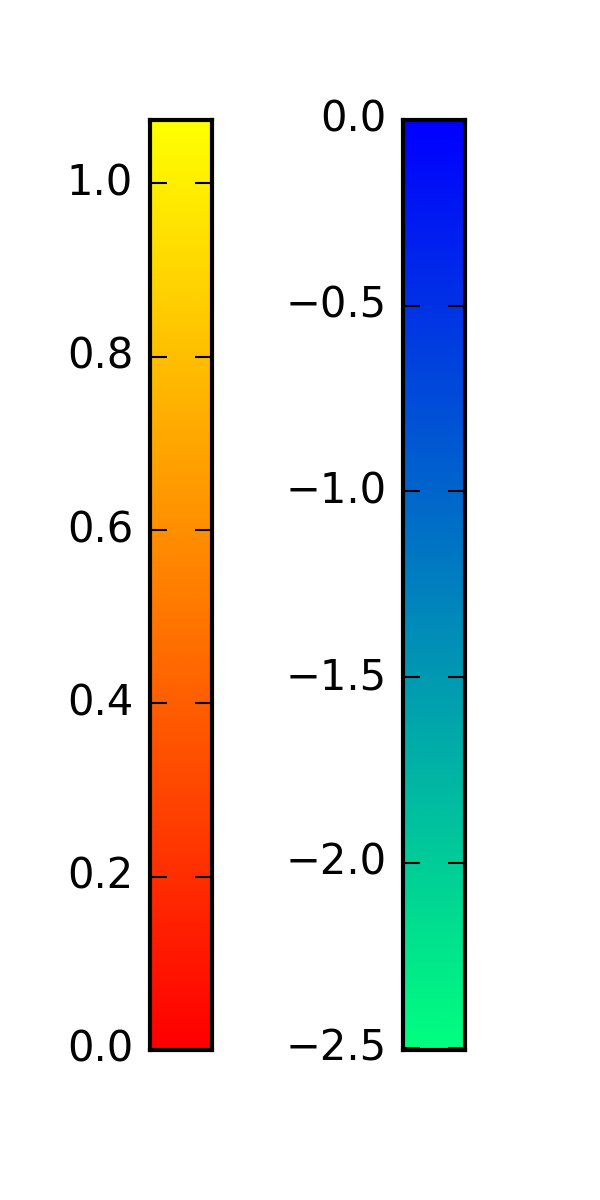}}
\put(70,28){(a)}
\put(265,28){(b)}
\put(15,144){$D$}
\put(450,144){$\lambda$ (MHz)}
\put(180,131){\footnotesize $\epsilon = 0.95$}
\put(375,131){\footnotesize $\epsilon = 0.95$}
\end{picture} \\
\begin{picture}(500,150)
\put(0,5){\includegraphics[height=.22\textheight]
{vartau_LYtest6_2017-11-02_15-50-03.png}}
\put(40,0){\includegraphics[height=.22\textheight]{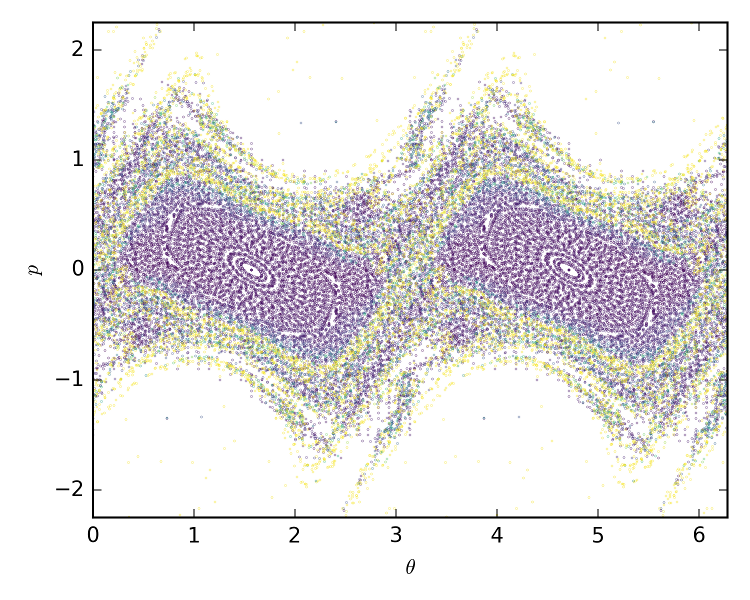}}
\put(235,0){\includegraphics[height=.22\textheight]{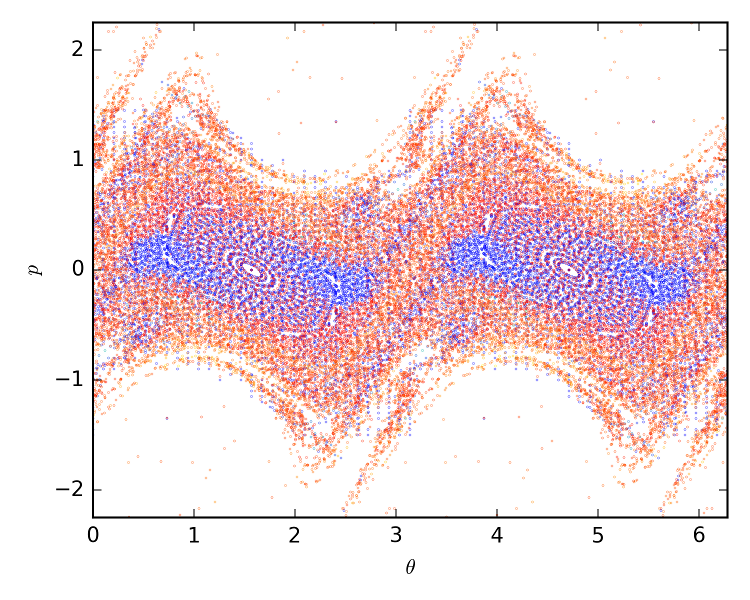}}
\put(430,5){\includegraphics[height=.22\textheight]{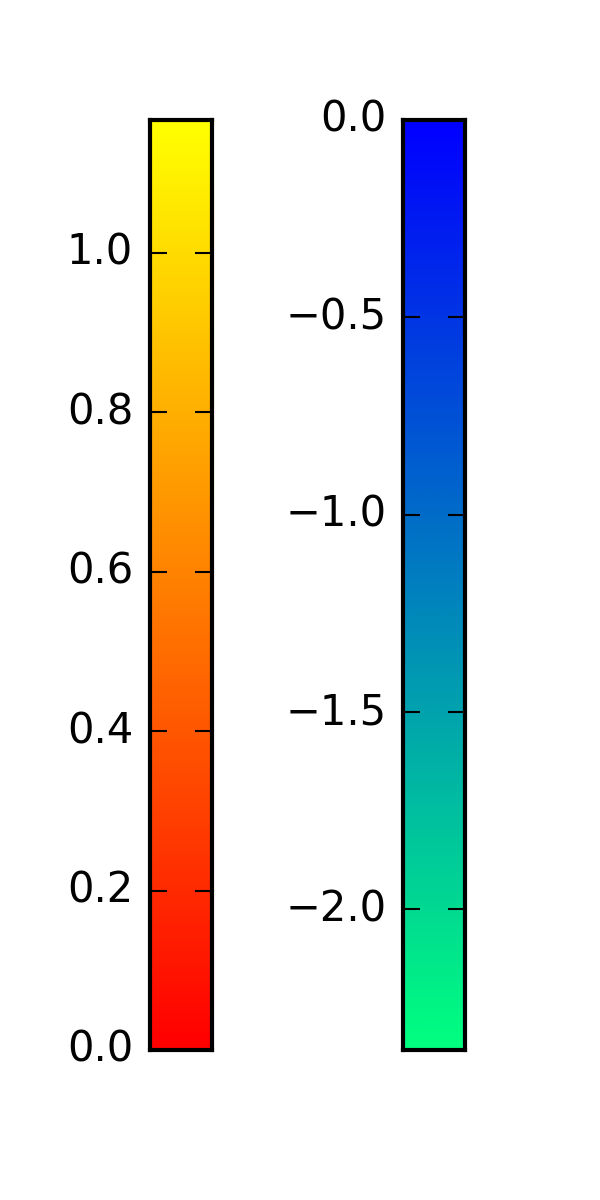}}
\put(70,28){(c)}
\put(265,28){(d)}
\put(15,144){$D$}
\put(450,144){$\lambda$ (MHz)}
\put(180,131){\footnotesize $\epsilon = 0.98$}
\put(375,131){\footnotesize $\epsilon = 0.98$}
\end{picture} \\
\begin{picture}(500,150)
\put(0,5){\includegraphics[height=.22\textheight]
{vartau_LYtest6_2017-11-02_15-50-03.png}}
\put(40,0){\includegraphics[height=.22\textheight]{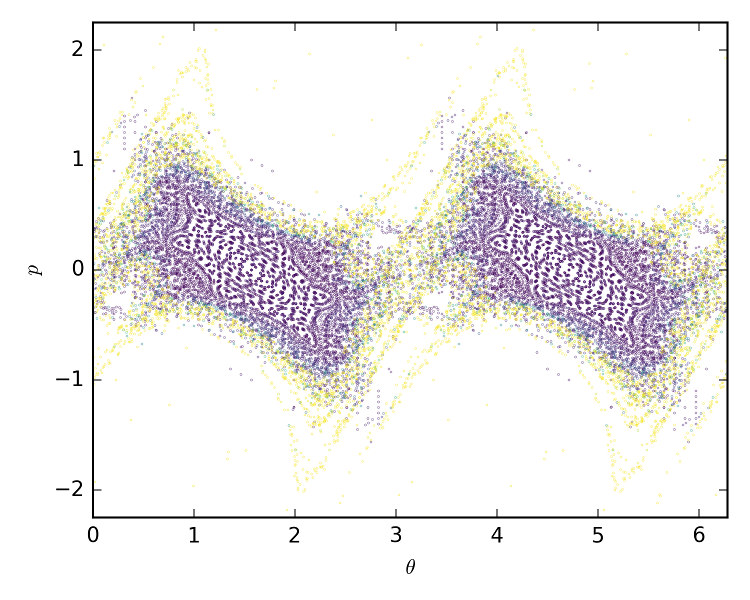}}
\put(235,0){\includegraphics[height=.22\textheight]{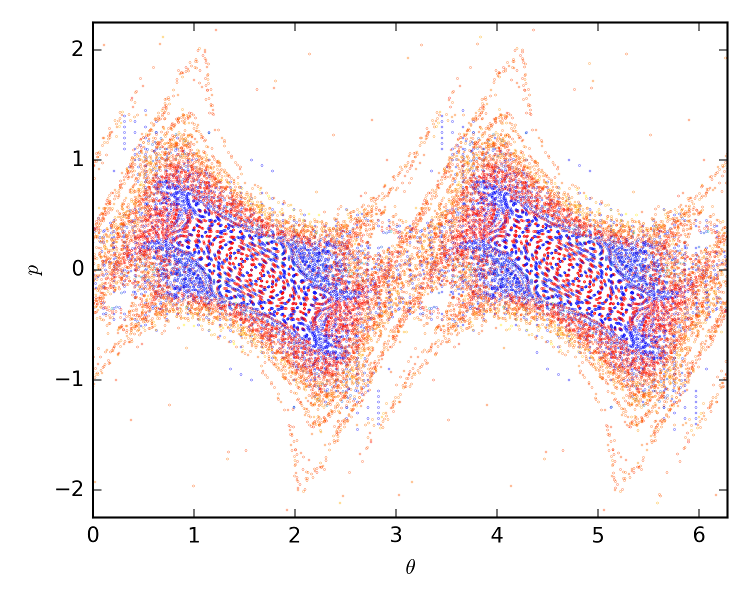}}
\put(430,5){\includegraphics[height=.22\textheight]{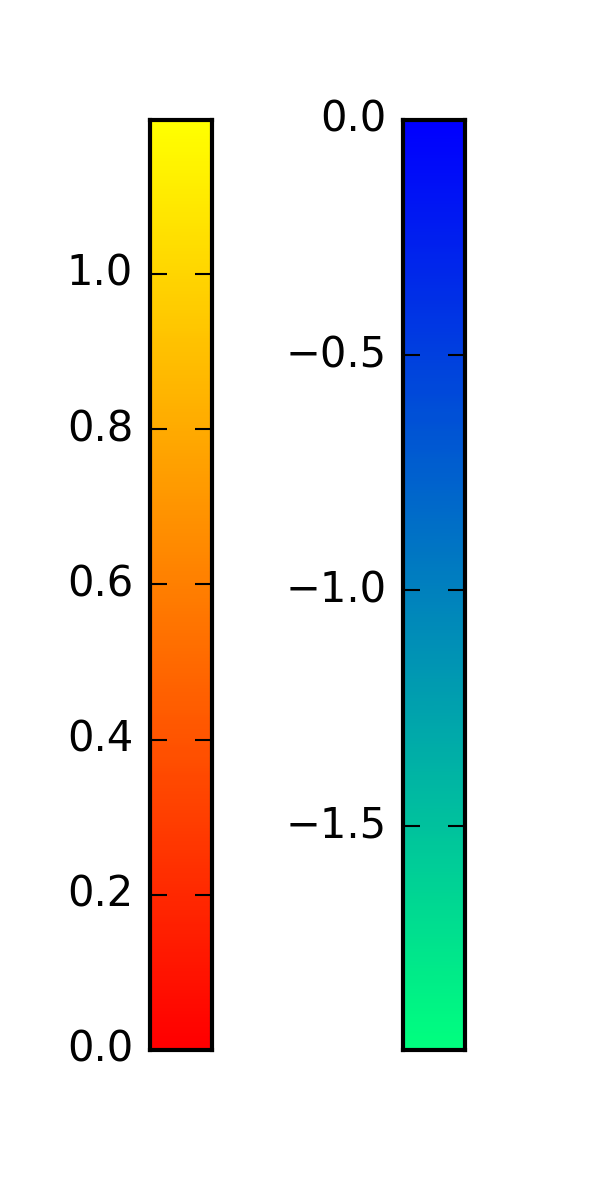}}
\put(70,28){(e)}
\put(265,28){(f)}
\put(15,144){$D$}
\put(450,144){$\lambda$ (MHz)}
\put(180,131){\footnotesize $\epsilon = 0.99$}
\put(375,131){\footnotesize $\epsilon = 0.99$}
\end{picture}
\caption{(Color online) We show stroboscopic phase portraits for the Hamiltonian \eqref{hxz} subject to periodic strong measurements according to \eqref{xz_kick}. Points are plotted at the moment when $H^\star$ most closely resembles \eqref{hrotor}, halfway between kicks. We use $\tau_m = 25~\mathrm{ns}$ and $\tau_x =1~\mu\mathrm{s} = \Lambda$. The simulation was run from $T = 0 \rightarrow 15~\mu\mathrm{s}$, including 15 ``kicks'', to generate the figures. We show $\epsilon = 0.95$ (a,b), $\epsilon = 0.98$ (c,d), and $\epsilon = 0.99$ (e,f). Color denotes the LEs \eqref{lyap} for the paths in (b,d,f), computed according to the distance \eqref{dist} shown in (a,c,e). We are particularly interested in examples where the LE is large as well as positive, implying that $D(t)$ grows to be much larger than $D_0 \approx 0.01$ within a modest duration $T$. Note that the yellow color for $D \geq 0.2$ corresponds to differences in angle on the Bloch sphere greater than approximately $8^{o}$. The original integrable rotor orbits are nowhere to be found, but islands formed by the $p = 0$ resonance are still clearly visible in the phase portrait. As $\epsilon$ grows, those periodic islands are gradually destroyed, turning into a chaotic sea as more internal resonances propagate out, destroying the remaining stable tori. Examples of individual paths with large $D$ and $\lambda$ from the $\epsilon = 0.99$ case are shown in Fig.~\ref{fig-xzexamples}. For supplemental animations showing the evolution between these images, see appendix~\ref{sec-anim_strobo}.} \label{fig-xzps}
\end{figure*}

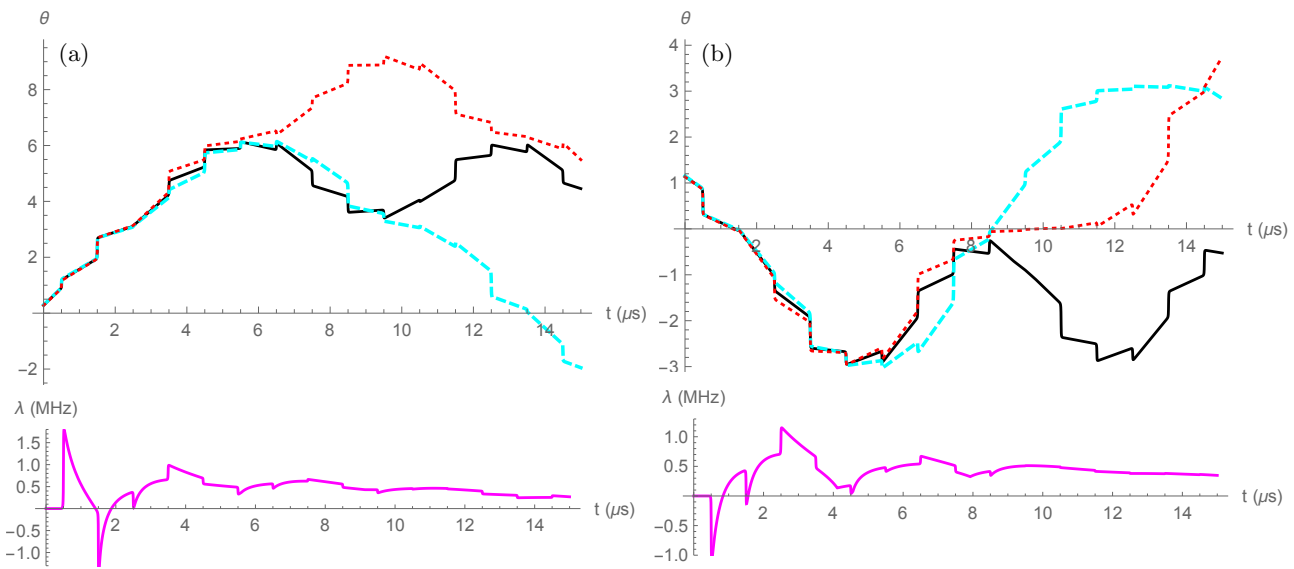
\begin{figure*}
\begin{tabular}{cc}
\includegraphics[width=.97\columnwidth]{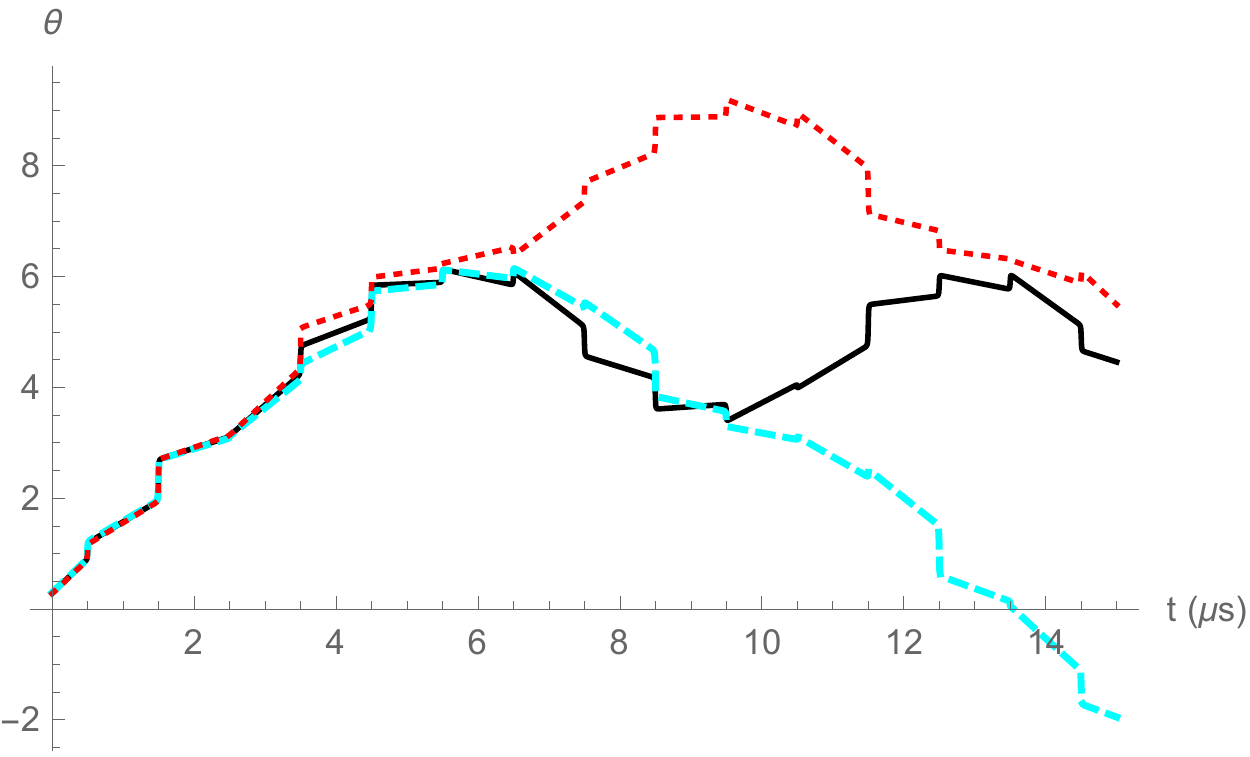} &
\includegraphics[width=.97\columnwidth]{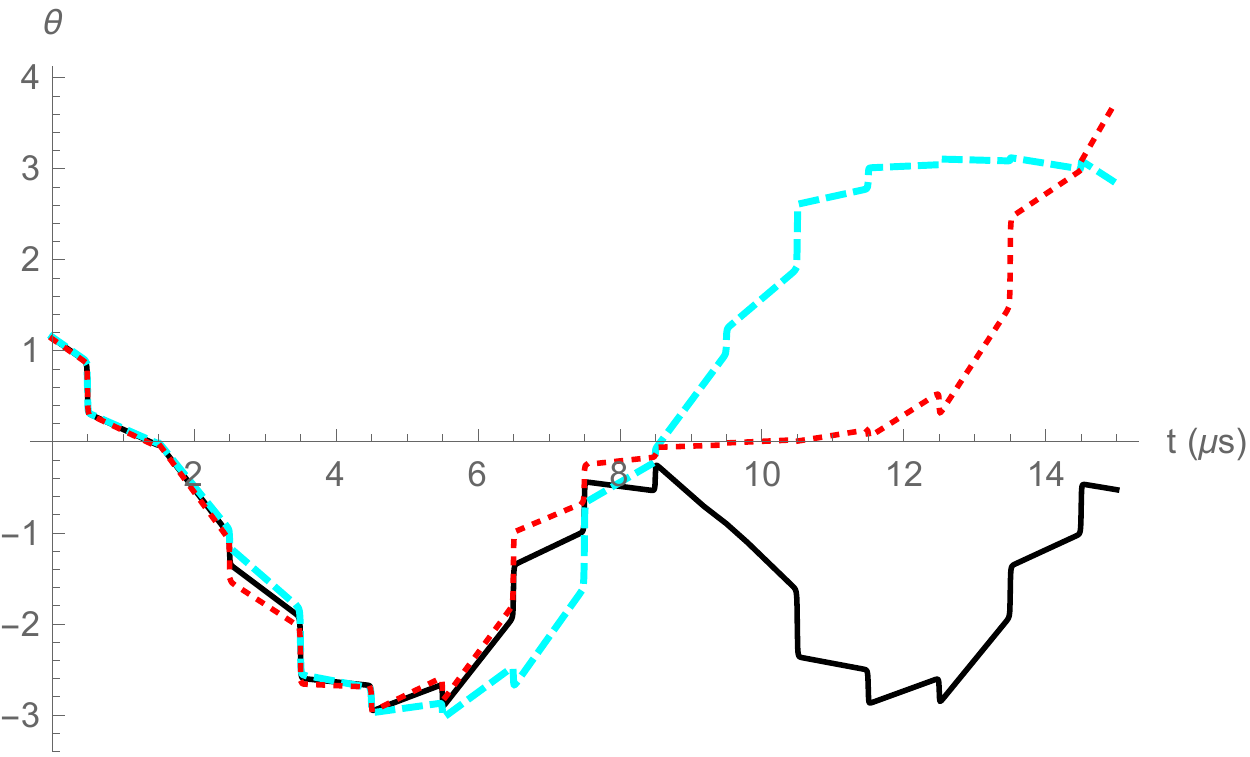} 
\end{tabular}
\\ \begin{picture}(490,64)
\put(245,62){\color{white} \rule{1cm}{12pt}}
\put(0,0){\includegraphics[width = .97\columnwidth]{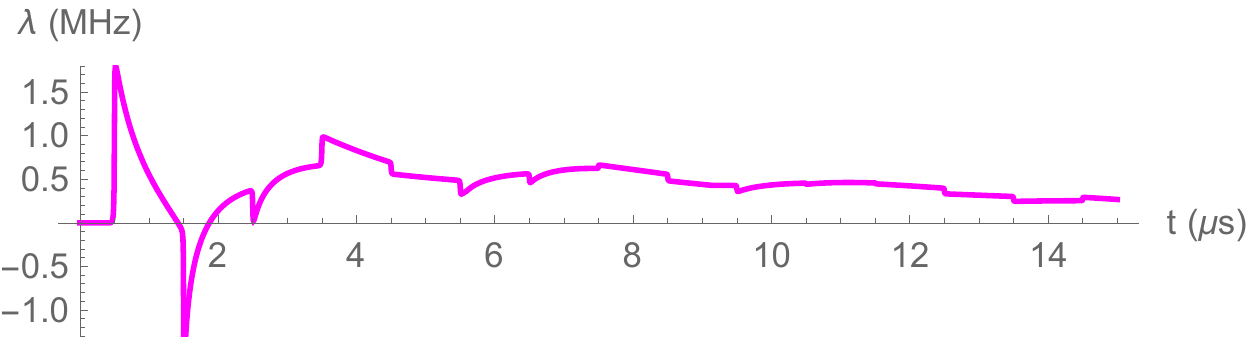}}
\put(245,0){\includegraphics[width = .97\columnwidth]{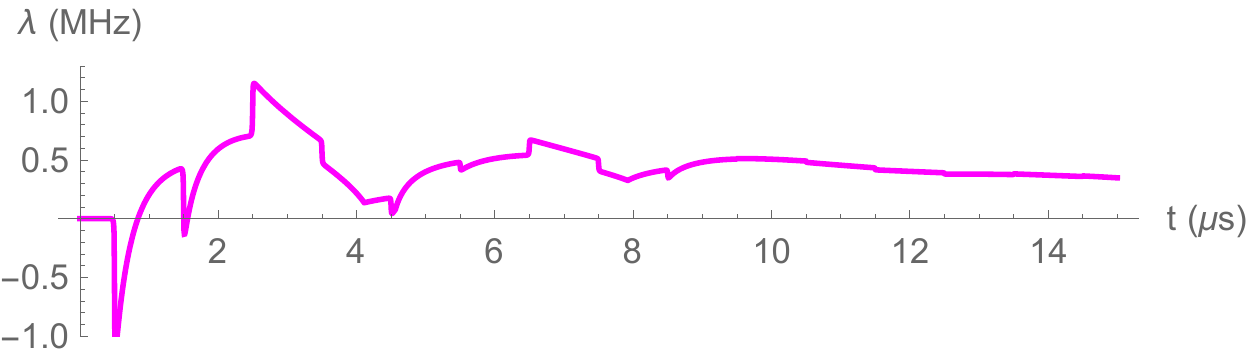}}
\put(20,192){(a)}
\put(263,192){(b)}
\end{picture}
\caption{(Color online) We show chaotic paths generated by (a) $\theta_0 = 0.286$ and $p_0 = 1.227$, and (b) $\theta_0=1.142$ and $p_0 = -0.545$, in the system defined by \eqref{hxz} and \eqref{xz_kick}. We use $\tau_x = 1~\mu\mathrm{s} = \Lambda$, $\tau_m = 0.025~\mu\mathrm{s}$, and $\epsilon = 0.99$ in both examples. In each plot, the paths generated by those initial conditions themselves are shown in solid black, whereas those from $\theta_0+0.01$ are shown in dashed cyan, and those generated by $\theta_0-0.01$ are shown in dotted red. The dashed and dotted paths are those used to compute the distance \eqref{dist} about the solid black curves, which in turn defines the LE \eqref{lyap}. The finite time LEs are plotted on the axes below their respective path groups. In each of these examples, small variations in the state appear by $T \approx 5~\mu\mathrm{s}$, the paths grow far apart in the quantum state space by $T \approx10~\mu\mathrm{s}$, and the trend continues to $T\approx15~\mu\mathrm{s}$ and beyond. This growth in $D$ corresponds to LEs with sustained positive values large enough to generate sizeable values of $t\cdot\lambda(t)$ over the evolution time of interest.}\label{fig-xzexamples}
\end{figure*}

\begin{figure*}
\begin{picture}(400,375)
\put(-50,250){\includegraphics[width=.32\textwidth,trim={10pt 10pt 0 0},clip]{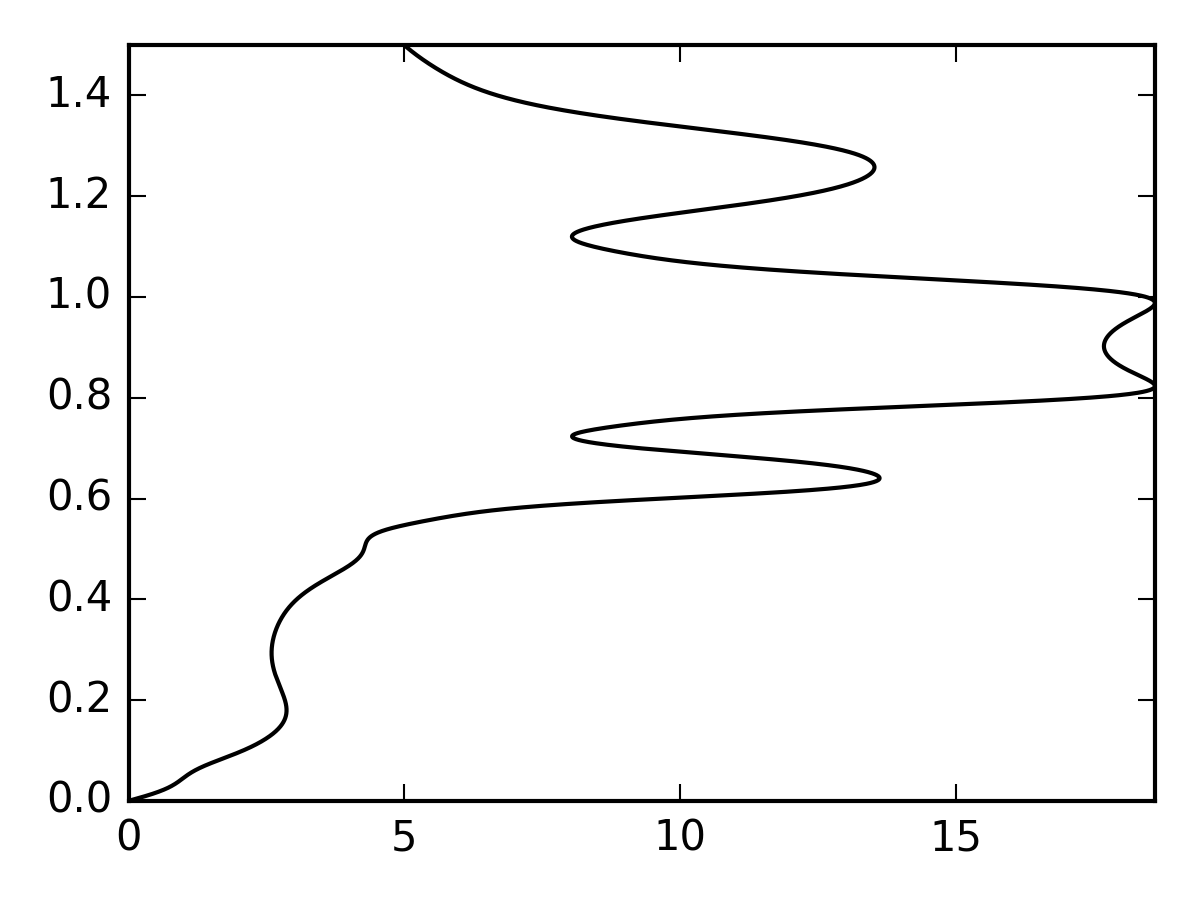}}
\put(-35,353){(a)}
\put(-33,343){$p_0$}
\put(95,262){$\theta_T$}
\put(70,353){\footnotesize $T = 3~\mu\mathrm{s}$}
\put(120,250){\includegraphics[width=.32\textwidth,trim={10pt 10pt 0 0},clip]{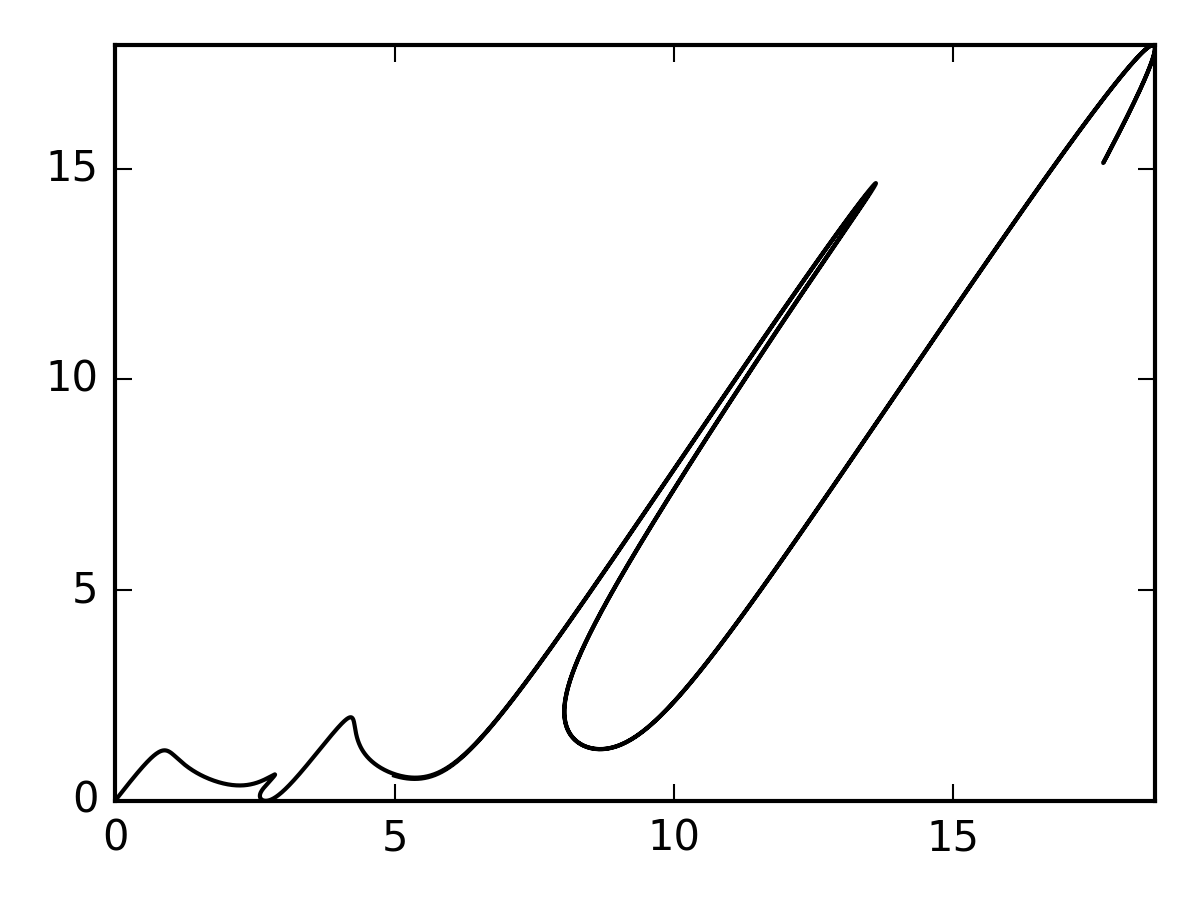}}
\put(135,353){(b)}
\put(137,343){$p_T$}
\put(265,262){$\theta_T$}
\put(290,250){\includegraphics[width=.32\textwidth,trim={10pt 10pt 0 0},clip]{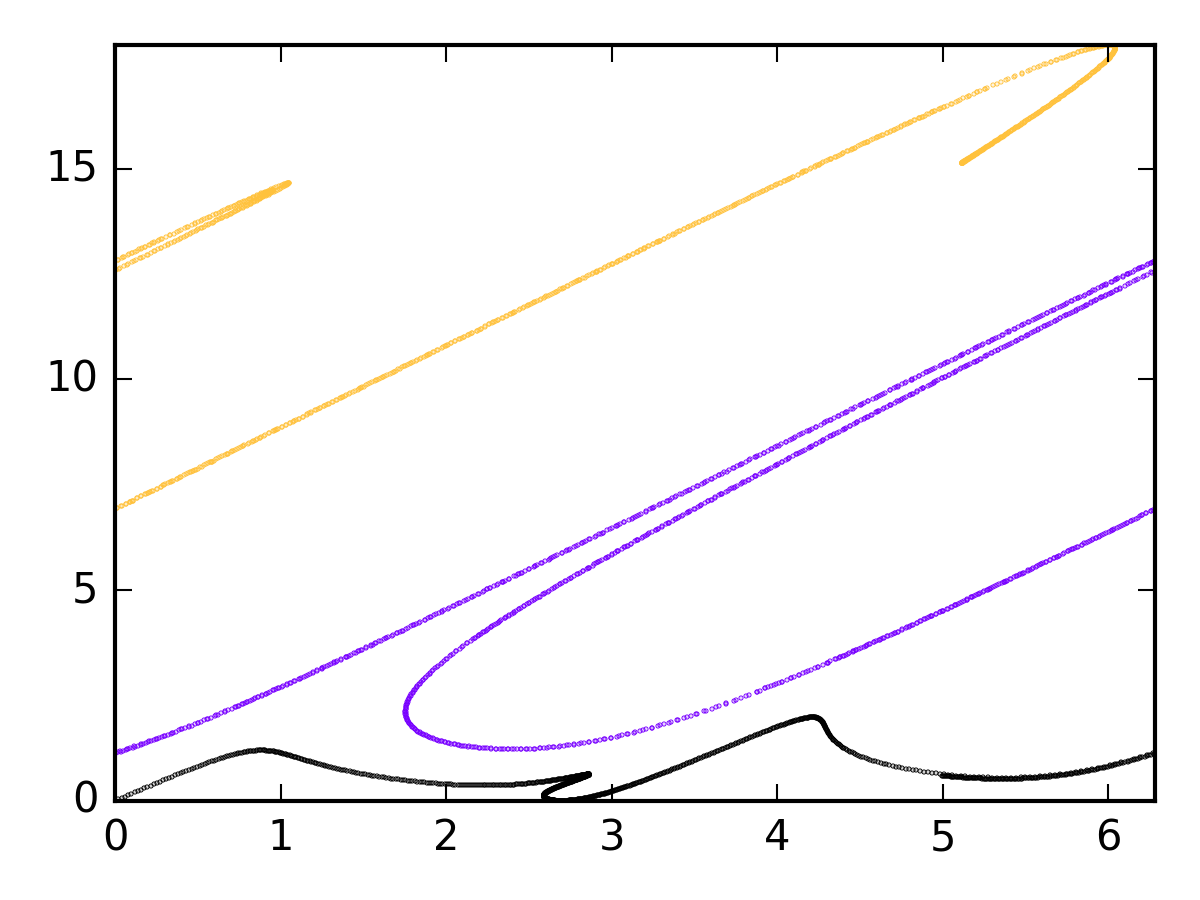}}
\put(305,353){(c)}
\put(307,343){$p_T$}
\put(424,249){$\theta_T$}
\put(-50,125){\includegraphics[width=.32\textwidth,trim={10pt 10pt 0 0},clip]{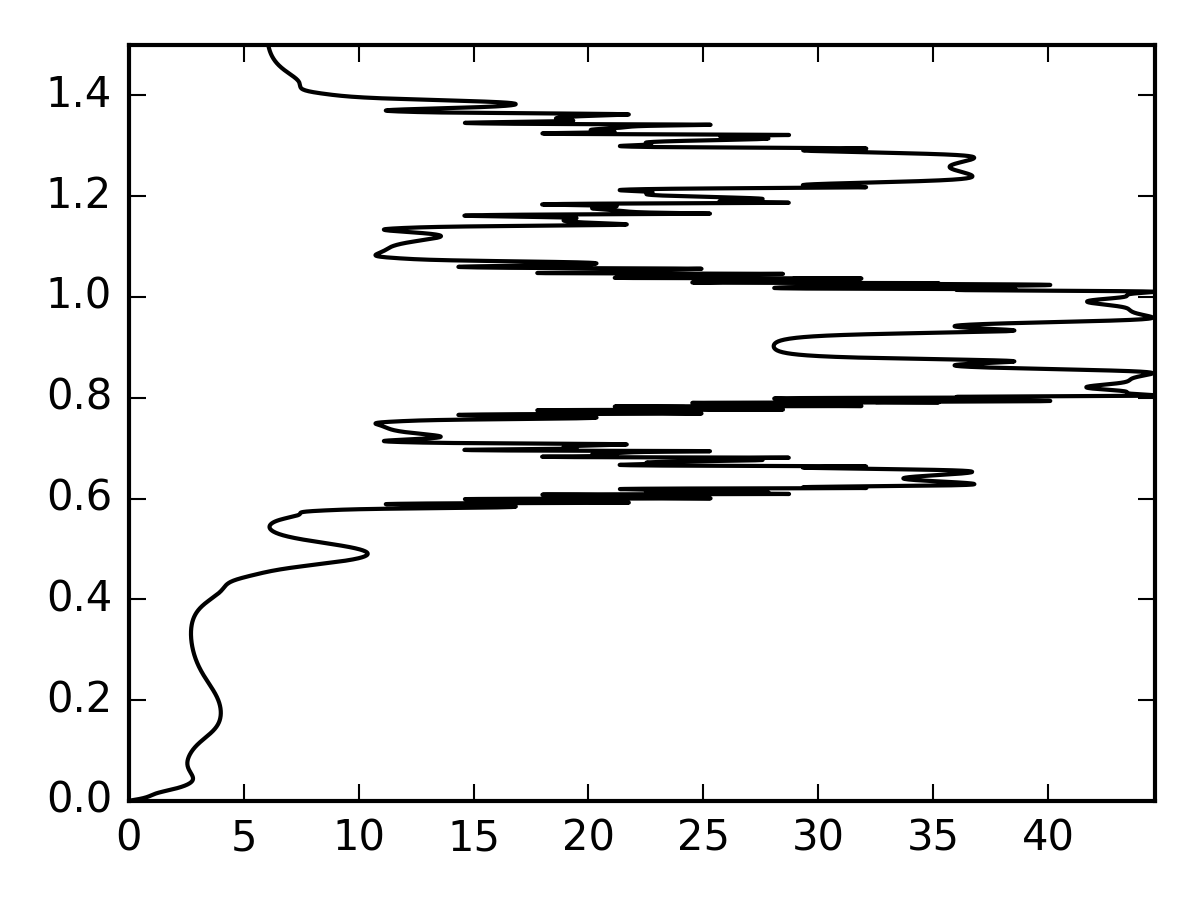}}
\put(-35,228){(d)}
\put(-33,218){$p_0$}
\put(95,137){$\theta_T$}
\put(70,228){\footnotesize $T = 4~\mu\mathrm{s}$}
\put(120,125){\includegraphics[width=.32\textwidth,trim={10pt 10pt 0 0},clip]{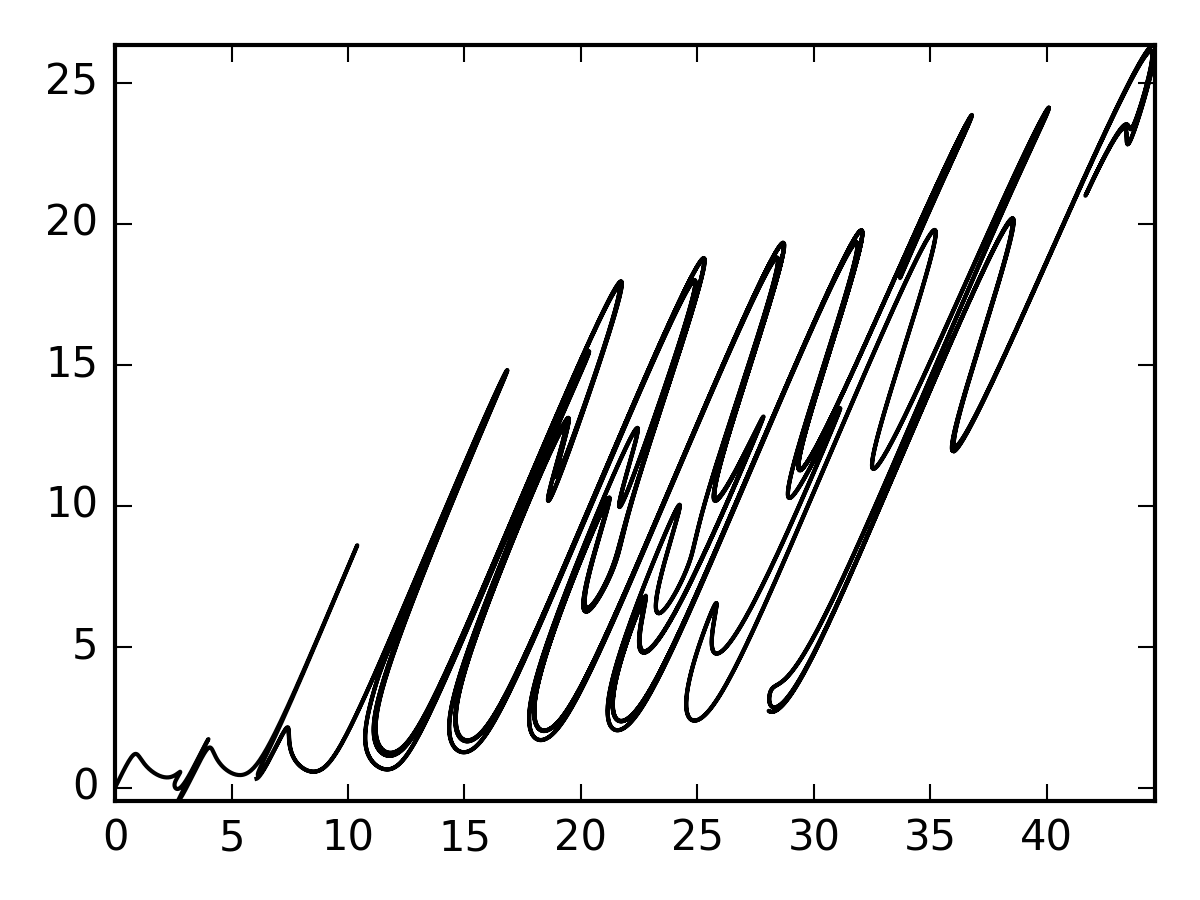}}
\put(135,228){(e)}
\put(137,218){$p_T$}
\put(265,137){$\theta_T$}
\put(290,125){\includegraphics[width=.32\textwidth,trim={10pt 10pt 0 0},clip]{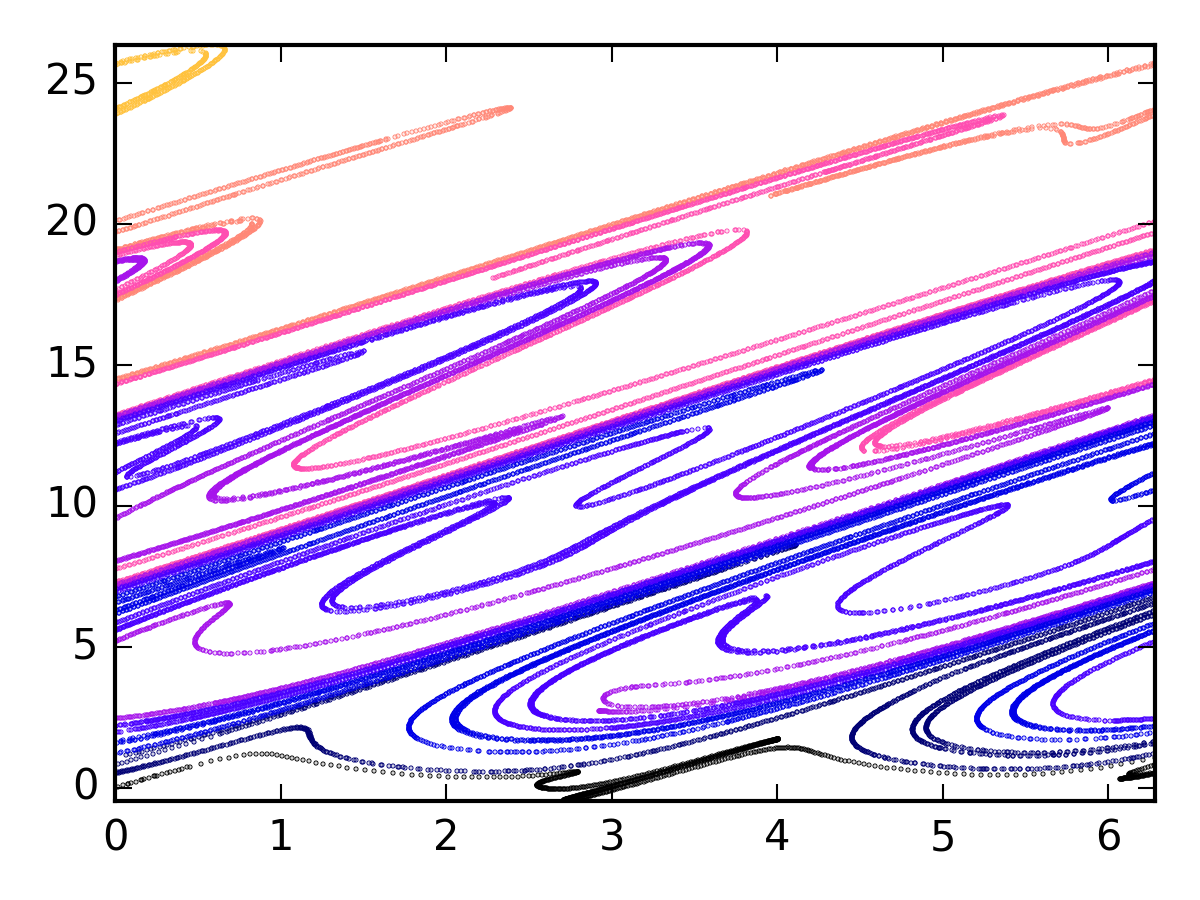}}
\put(305,228){(f)}
\put(307,218){$p_T$}
\put(424,124){$\theta_T$}
\put(-50,0){\includegraphics[width=.32\textwidth,trim={10pt 10pt 0 0},clip]{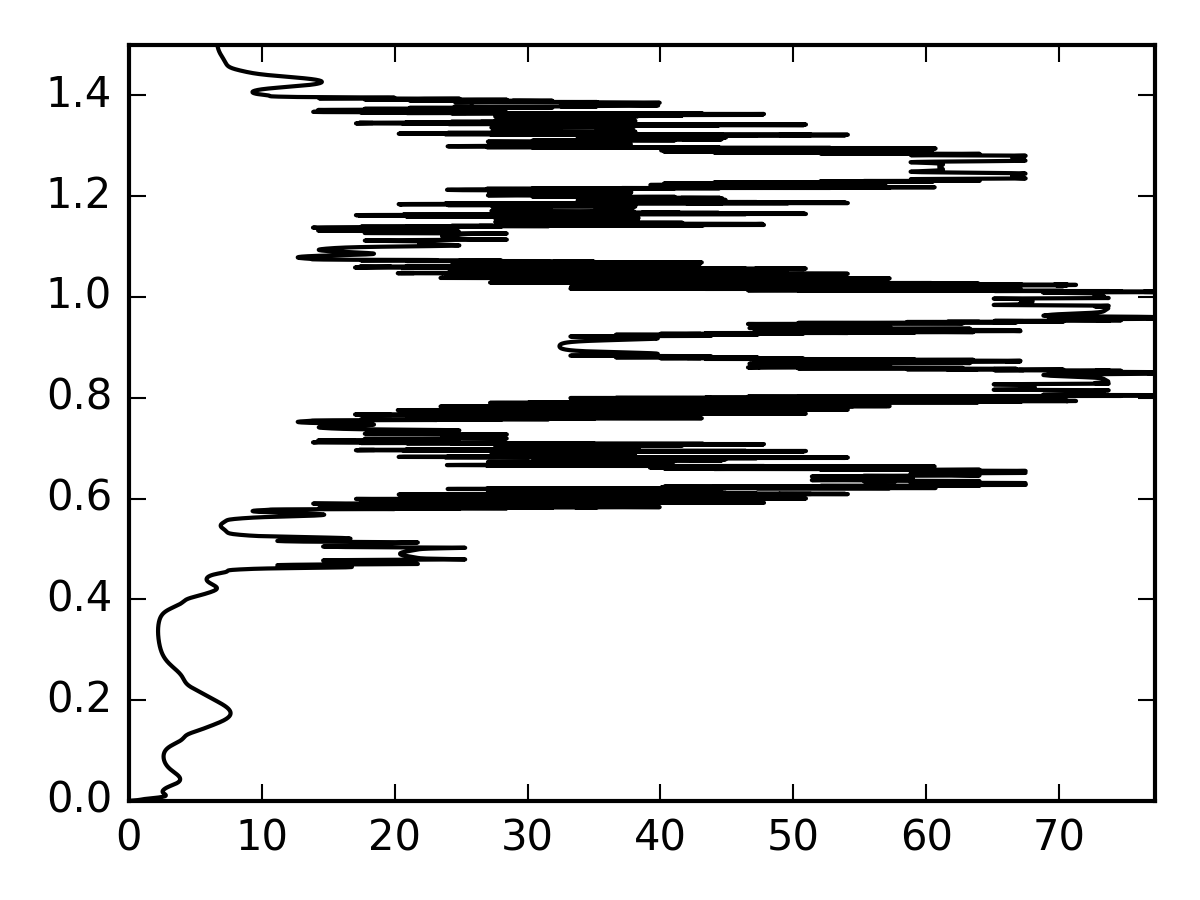}}
\put(-35,103){(g)}
\put(-33,93){$p_0$}
\put(95,12){$\theta_T$}
\put(70,103){\footnotesize $T = 5~\mu\mathrm{s}$}
\put(120,0){\includegraphics[width=.32\textwidth,trim={10pt 10pt 0 0},clip]{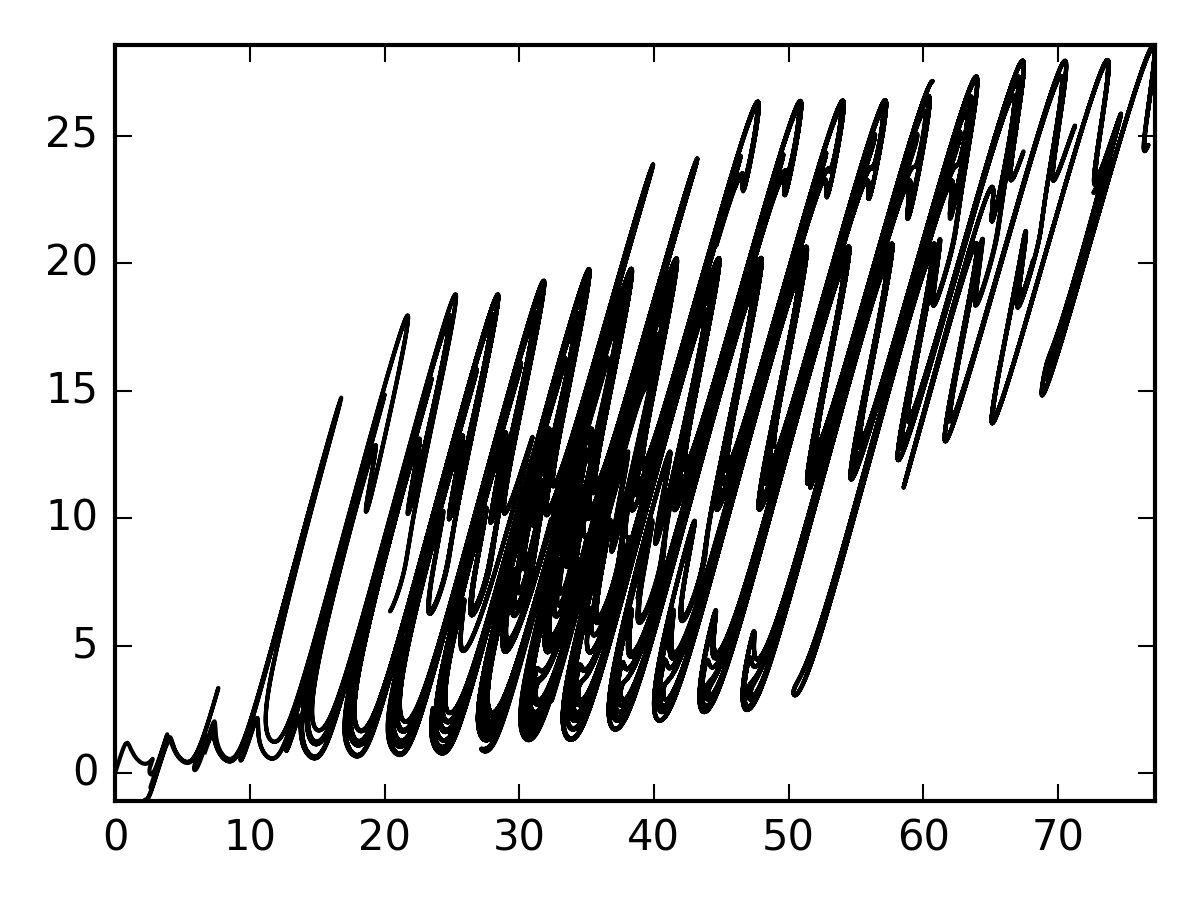}}
\put(135,103){(h)}
\put(137,93){$p_T$}
\put(265,12){$\theta_T$}
\put(290,0){\includegraphics[width=.32\textwidth,trim={10pt 10pt 0 0},clip]{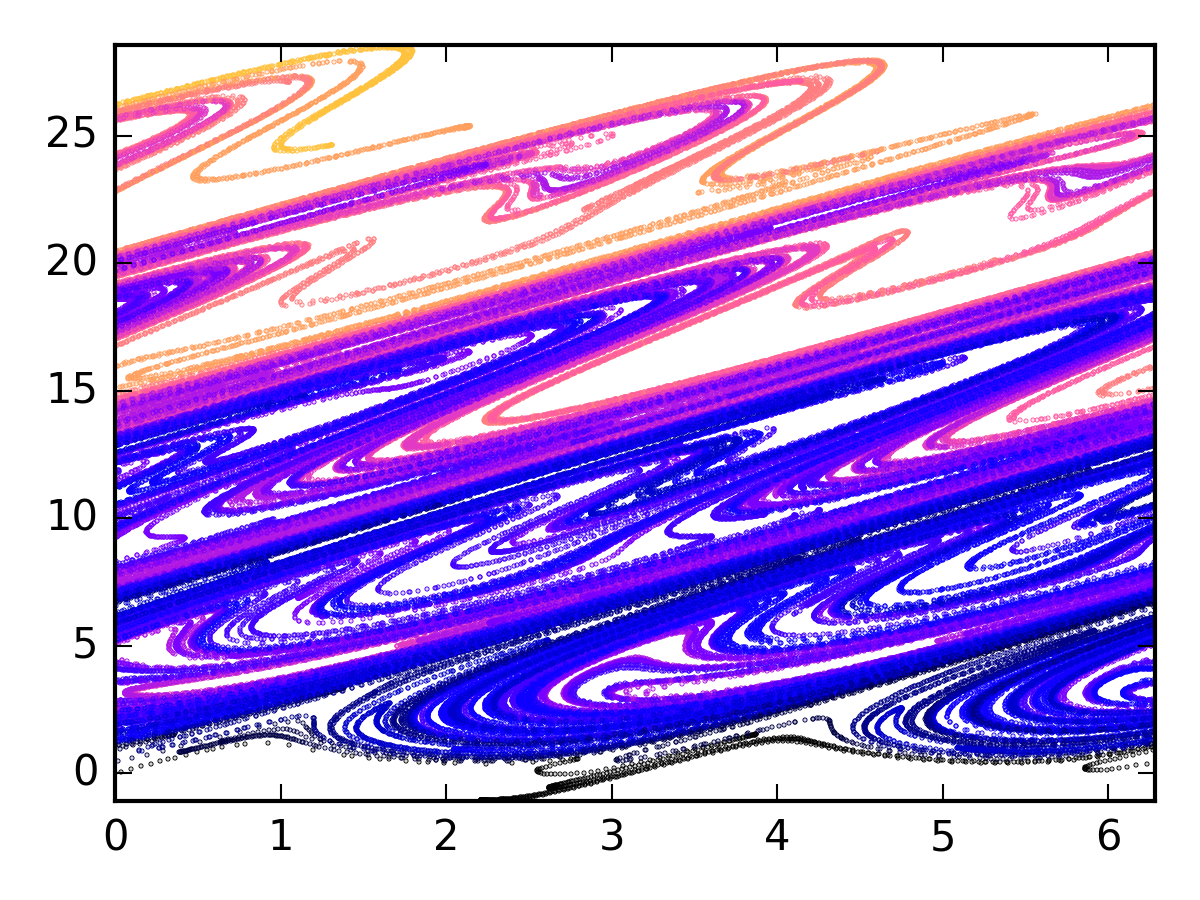}}
\put(305,103){(i)}
\put(307,93){$p_T$}
\put(424,-1){$\theta_T$}
\end{picture}
\caption{(Color online) The Lagrange Manifold describing multipaths originating at the excited state $\theta_0 = 0$ on $p_0 \in [0, 1.5]$, under evolution generated by \eqref{hxz} and \eqref{xz_kick}, with $\epsilon=0.99$ and $\tau_x = 1~\mu\mathrm{s} = \Lambda$, is shown at $T = 3~\mu\mathrm{s}$ (a-c), $T = 4~\mu\mathrm{s}$ (d-f), and $T = 5~\mu\mathrm{s}$ (g-i). The $y$--axis in the left column is $p_0$, and the $y$--axis in the center and right columns is $p_T$. The $x$--axis always denotes $\theta_T$, but in the right column it is shown mod-$2\pi$ (as is physically relevant for qubit states) rather than as the raw angle $\theta$. The pairs (b,c), (e,f), and (h,i), respectively, contain the exact same information represented differently. Winding counts in the right column are denoted by color, with lighter colors corresponding to more winds around the Bloch sphere. The manifold is only computed and plotted over positive $p_0$ because the phase--space has odd symmetry; the other half of the manifold for $\theta_0 = 0$ is identical except for being mirrored across the origin of the phase space. The number of catastrophes in this segment of the manifold grows from $9$ at $T=3~\mu\mathrm{s}$ (a-c), to roughly $140$ at $T=4~\mu\mathrm{s}$ (d-f), to approximately $2200$ at $T=5~\mu\mathrm{s}$ (g-i).}\label{fig-xzlm}
\end{figure*}

Let us consider the dynamics of $H^\star$ for $\epsilon \ll 1$, where both of the measurements $\tau_x \approx \tau_z$ remain weak, generating diffusive quantum trajectories. From animated surveys of the phase space in this regime (see appendix~\ref{sec-anim_strobo}) we know that disturbances to the integrable phase space are restricted to regions near $p = k\pi$ for integer $k$. This is clearly illustrated in Fig.~\ref{fig-earlyreson}, where we show a stroboscopic phase portrait for $\epsilon = 0.1$. The question which drives this subsection is: What is special about the values $p = k\pi$ when $\epsilon \ll 1$? 

\par Based on canonical perturbation theories \cite{BookPoincare, ClassQuillen, BookReichl,  BookArnoldClassical, BookGoldstein, BookF-M,BookJose}, including the theorem by Kolmogorov, Arnold, and Moser (KAM) \cite{KolmogorovAM, *KArnoldM, *KAMoser1, *KAMoser2}, we expect that the OP dynamics will not deviate substantially from the integrable dynamics of $H^{(0)}$ in this regime, \emph{except} near resonances. By ``resonance'', we refer to a rational number relationship between periodic motion from $H^{(0)}$ and the period $\Lambda$ of the first--order perturbation $\epsilon H^{(1)}$. 
We have $H^{(0)} = H^{(0)}(p) = E$, so that the frequency $\nu^{(0)}$ of integrable, periodic OPs is determined by computing $\nu^{(0)} = \partial_p H^{(0)} = p/\tau_x = \pm \sqrt{1 + 2\tau_x E}/\tau_x$. 
A resonance occurs, and so perturbation theory breaks down, wherever the condition
\be
\nu^{(0)}(E) \ell + 2\pi k /\Lambda = 0
\ee 
is satisfied, for $\ell = 0,\pm 2$ and any integer $k$ (see appendix~\ref{sec-resdetail} for details). For $\tau_x = \Lambda$, this relationship reduces to $ p \ell + 2 \pi k = 0$, which explains the appearance of resonances at integer multiples of $p = \pi$. 
Equivalently, those resonances appear where the period of the integrable motion
\be \label{period}
\tilde{T} = \int_0^{\tilde{T}} dt = \int_0^\pi \frac{d\theta}{\dot{\theta}} = \frac{\tau_x \pi}{p_0} \quad\rightarrow\quad p_0 = k \frac{\tau_x \pi}{\Lambda},
\ee
is an integer multiple of the kicking period itself.
Along the paths with resonant $p_0$ or $E$, the effects of the perturbation build over time, since the unperturbed motion is always ``in phase'' with the perturbing force, instead of averaging out and leaving the $H^{(0)}$ dynamics nearly unchanged, as would happen off--resonance. 
As these lines of $p_0$ are destroyed, e.g.~as shown in Fig.~\ref{fig-earlyreson}, chaos gets its first toehold in the phase space; the resonances generate the first major disruptions to the integrable tori as $\epsilon$ grows. 
\par The largest deviations from the LM generated by the flow of $H^{(0)}$ at low $\epsilon$ also occur in paths initialized near those same resonant $p_0$. 
$H^{(0)}$ is a quadratic function of $p$, so its LM cannot contain caustics \footnote{We are restricted to the form $H^\star = (p^2-1)a + p b$, where $a$ and $b$ must be real numbers if $H^\star$ only depends on $p$. Then the parabolic $H^\star$ only has one minimum with respect to $p$; that minimum is a fixed point, and the flow to either side is uniform in direction, with monotonically-increasing speed away from the fixed point. This phase space topology makes it impossible for the LM to fold back on itself into a caustic.}. 
The manifolds describing multipath behavior originating at the excited state $\theta_0 = 0$, both for $H^{(0)}$, and the full $H^\star$, are shown in Fig.~\ref{fig-LMreson}; the similarity between Figs.~\ref{fig-earlyreson} and \ref{fig-LMreson}(c) is immediately apparent. 
When $\epsilon$ is small, the formation of catastrophes in the LM corresponds with the $p_0$ forming resonance bands; these give the first interesting multipath behavior, \emph{and} will play a key role in turning the entire phase--space into a chaotic sea as $\epsilon$ grows larger. 
This is the first of many connections we will discuss between OP chaos and multipaths. 

\par The concept of resonance, in the context of OPs, should extend well beyond the specific example we discuss here. 
Starting with an integrable stochastic Hamiltonian, any additional periodic perturbations will generate dynamical disruptions along paths in the OP phase space which match the perturbation's period, much as we have seen above. 
This is true even when the perturbation is weak; from there the resonance phenomenon offers a relatively well--understood (from classical chaos theory) pathway to OP chaos as the perturbation is strengthened. 
A wide variety of other schemes which meet these criteria could be easily devised for qubit systems, using combinations of measurement(s) and Rabi drive.

\subsection{Chaos and Multipaths in the strong measurement regime \label{sec-strongreg}}

We now proceed to investigate the dynamics of $H^\star$ at values of $\epsilon$ near 1, corresponding to strong kicks (nearly projective $z$--measurements). 
Stroboscopic phase portraits for $\epsilon = 0.95,~0.98,$ and $0.99$ are shown in Fig.~\ref{fig-xzps}. 
These figures confirm that chaos overtakes larger portions of the phase space as $\epsilon$ grows. 
A few examples of strongly--chaotic paths at $\epsilon = 0.99$ are shown in Fig.~\ref{fig-xzexamples}, which serve to illustrate the qualitative effect of OP chaos. 
In these examples, we see groups of OPs which start at nearly identical quantum states; as time goes on, small deviations in their dynamics are magnified, until the states $\theta_t$ generated by these OPs are effectively uncorrelated. 
Qualitatively, we may understand that small deviations in diffusion between kicks get magnified by the kick itself; that is, a measurement kick probabilistically collapses the state to an eigenstate of $\sigma_z$ in this regime, and stochastic elements of the diffusion between kicks determine the probability for a path to go one direction or the other at the next kick. 
Thus, every kick effectively elevates the randomness inherent in the preceding diffusion step to the point that it manifests as chaotic unpredictability in the OPs (which are defined statistically, not on the basis of any one SQT). Thus we see that states prepared on the border of being easily distinguishable experimentally (we use $\delta\theta_0 = 0.01$ and $\delta p_0 = 0$ in Fig.~\ref{fig-xzexamples}) can lead to OPs which are ripped apart within 5--7 kicks (measurement cycles). 
In other words, the OPs are sufficiently sensitive to changes in initial state that small deviations in state preparation lead to OPs which diverge wildly from each other within experimentally--accessible time frames. 
We stress that although the OP formalism allows us to work with mathematics from classical Hamiltonian mechanics and chaos theory, the above intuition very much emphasizes the intrinsically quantum qualities of the qubit system, and effects of quantum measurement which give rise to all the dynamics we discuss.

\section{Implications of OP Chaos \label{sec-ch_mani}}

\par We will be able to elucidate the impact of OP chaos on the underlying SQTs by illustrating that there is a connection between manifold deformation, multipaths, and OP chaos. 
\subsection{Examples of Manifold Deformation}
\begin{figure}
\begin{picture}(150,515) 
\put(-3,495){\includegraphics[width=.7\columnwidth]{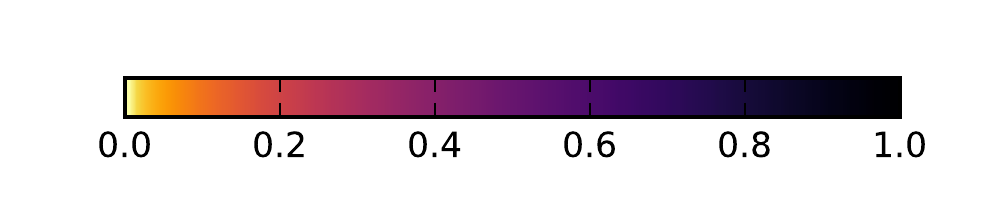}}
\put(-42,378){\includegraphics[width = .92\columnwidth, trim = {25pt 35pt 30pt 40pt},clip]{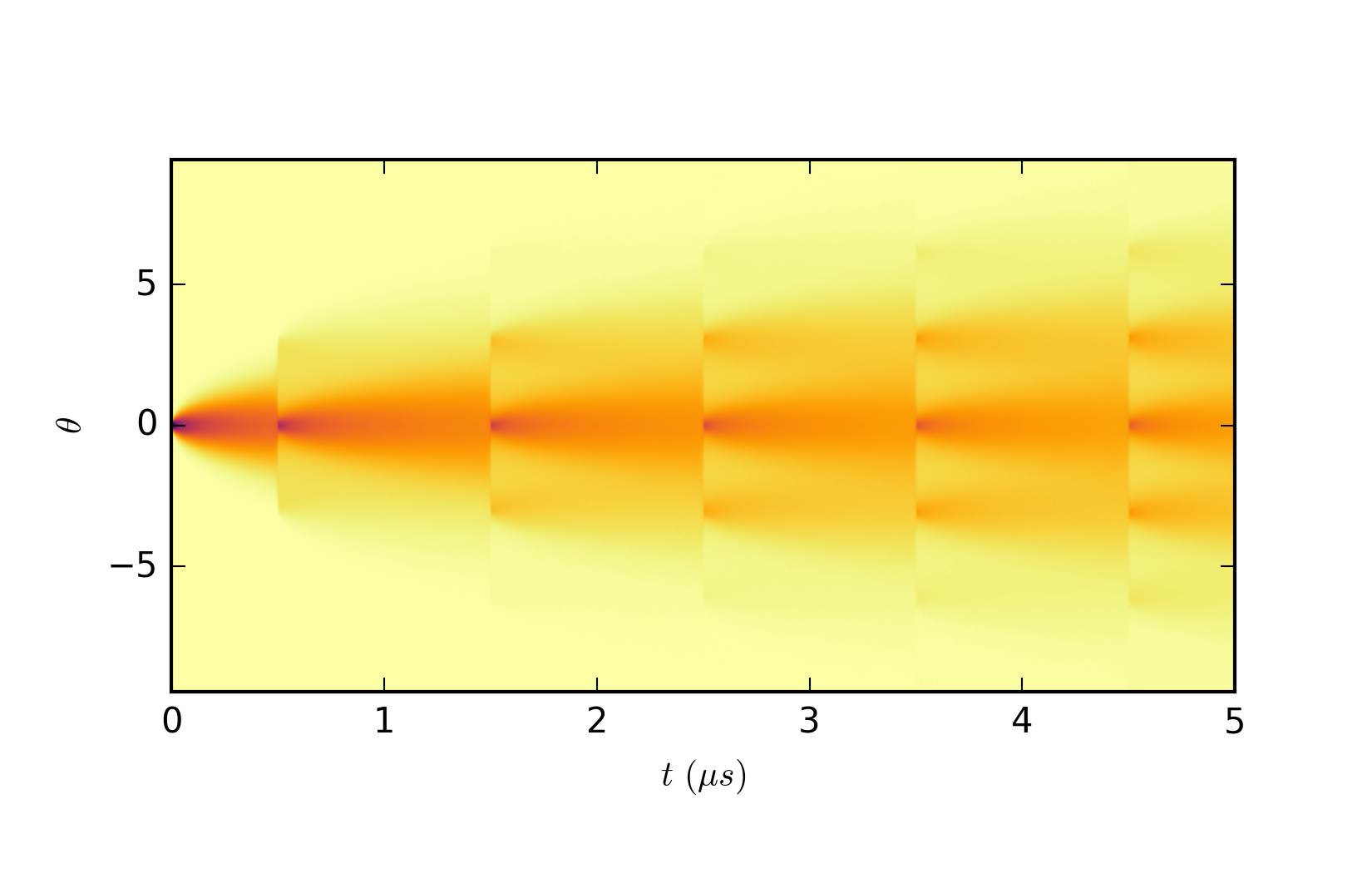}}
\put(-42,256){\includegraphics[width = .92\columnwidth, trim = {25pt 30pt 30pt 36pt},clip]{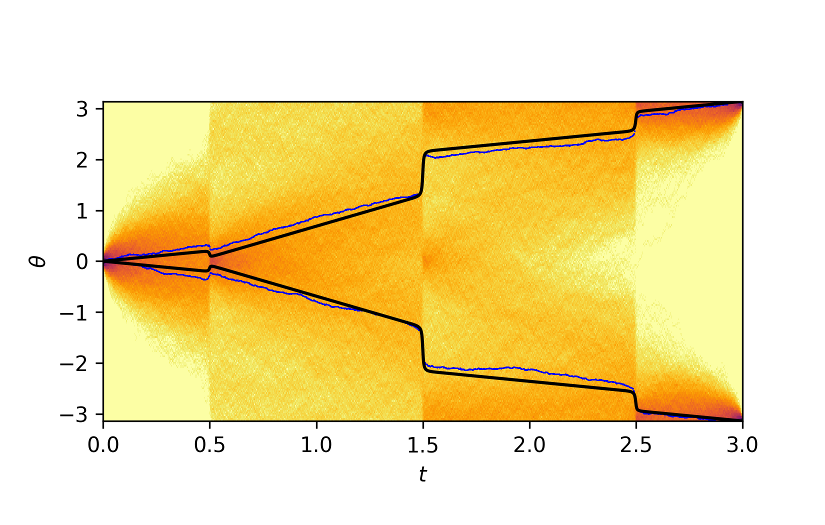}} 
\put(-42,134){\includegraphics[width = .92\columnwidth, trim = {25pt 30pt 30pt 36pt},clip]{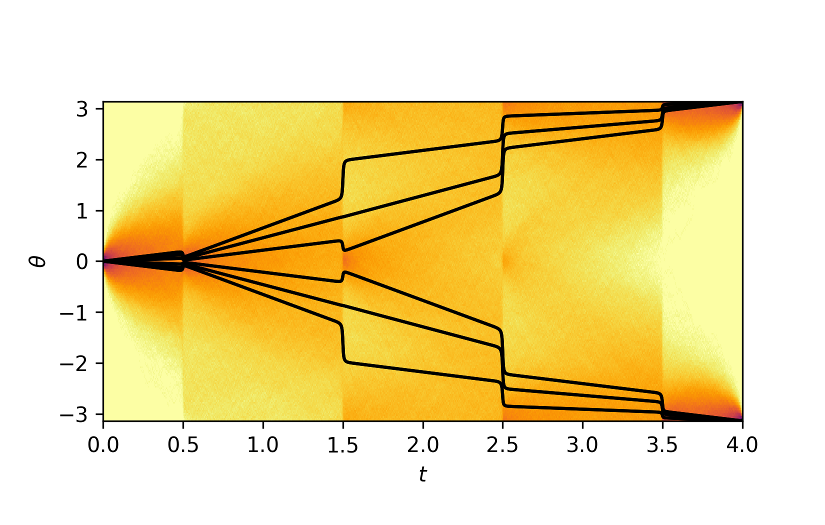}}
\put(-42,12){\includegraphics[width = .92\columnwidth, trim = {25pt 30pt 30pt 36pt},clip]{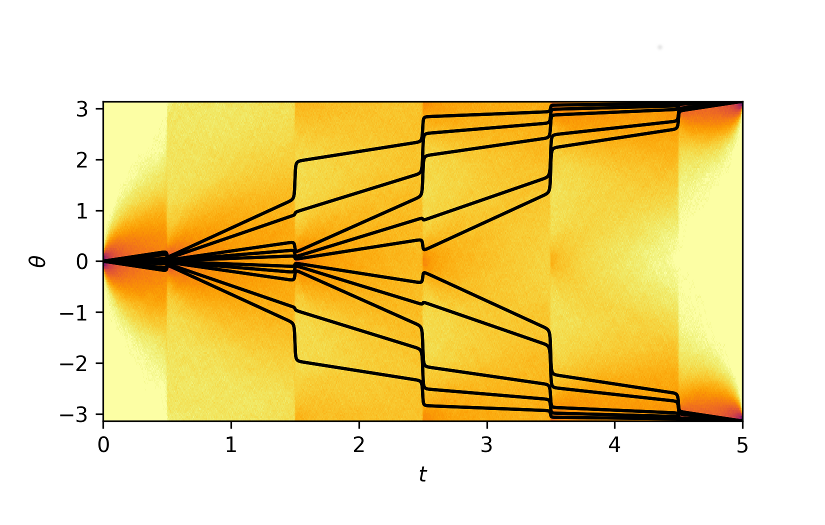}}
\put(-23,480){(a)}
\put(-23,360){(b)}
\put(-23,236){(c)}
\put(-23,115){(d)}
\put(65,0){$t ~(\mu\mathrm{s})$}
\put(-35,480){$\theta$}
\end{picture}
\caption{(Color online) We show plots for the density of simulated SQTs for \eqref{hxz}, initialized at $\theta_0 = 0$, with $\epsilon = 0.99$, $\tau_m = 0.025~\mu\mathrm{s}$, and $\tau_x = 1.0~\mu\mathrm{s} = \Lambda$. All $x$--axes are time in $\mu\mathrm{s}$, and all $y$--axes are $\theta$, either as the full angle (a), or mod-$2\pi$ (b-d). We show (a) the density without post--selection, and with post--selection on the ground state ($\theta_T = \pm\pi$) at (b) $T = 3~\mu\mathrm{s}$, (c) $T = 4~\mu\mathrm{s}$, and (d) $T = 5~\mu\mathrm{s}$. In (b) we show agreement between MLPs comptuted from simulated SQTs (blue) with the theoretical MLPs (black). The associated LMs may be found in Fig.~\ref{fig-xzlm}. Normalized density is shown on the colorbar, where 1 is the maximum trajectory density between boundary conditions, and 0 corresponds to no trajectories at all.
}\label{fig-xzdens}
\end{figure}
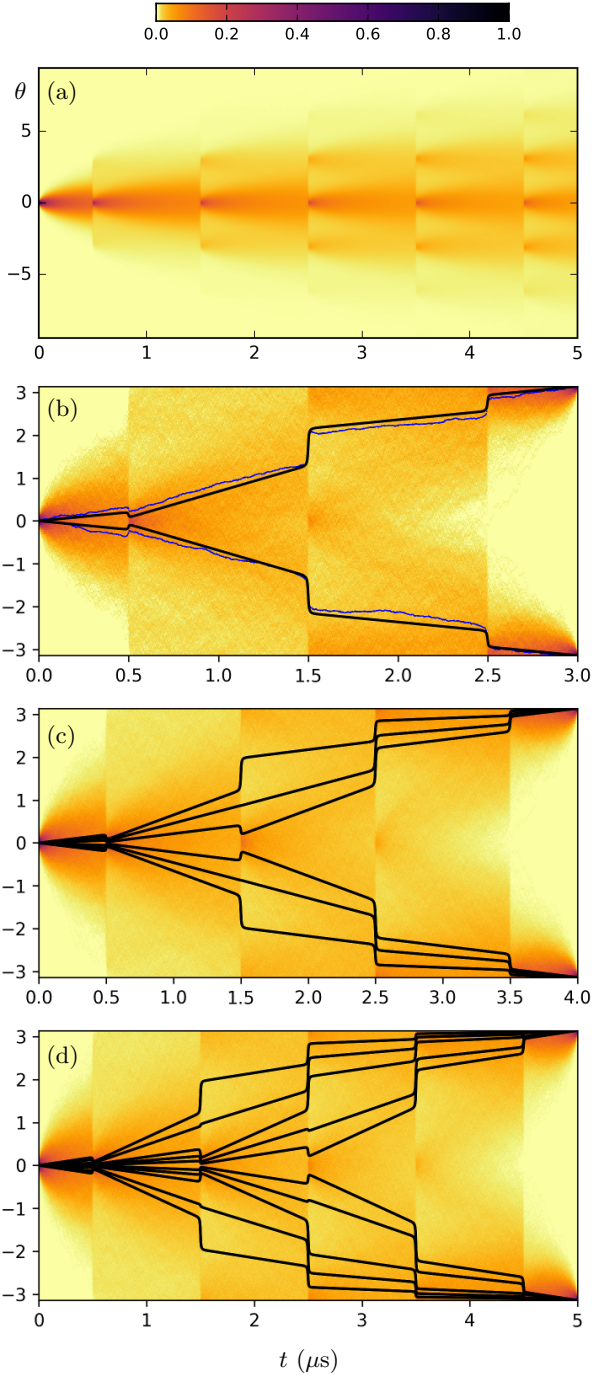

\begin{figure}
\includegraphics[width = \columnwidth]{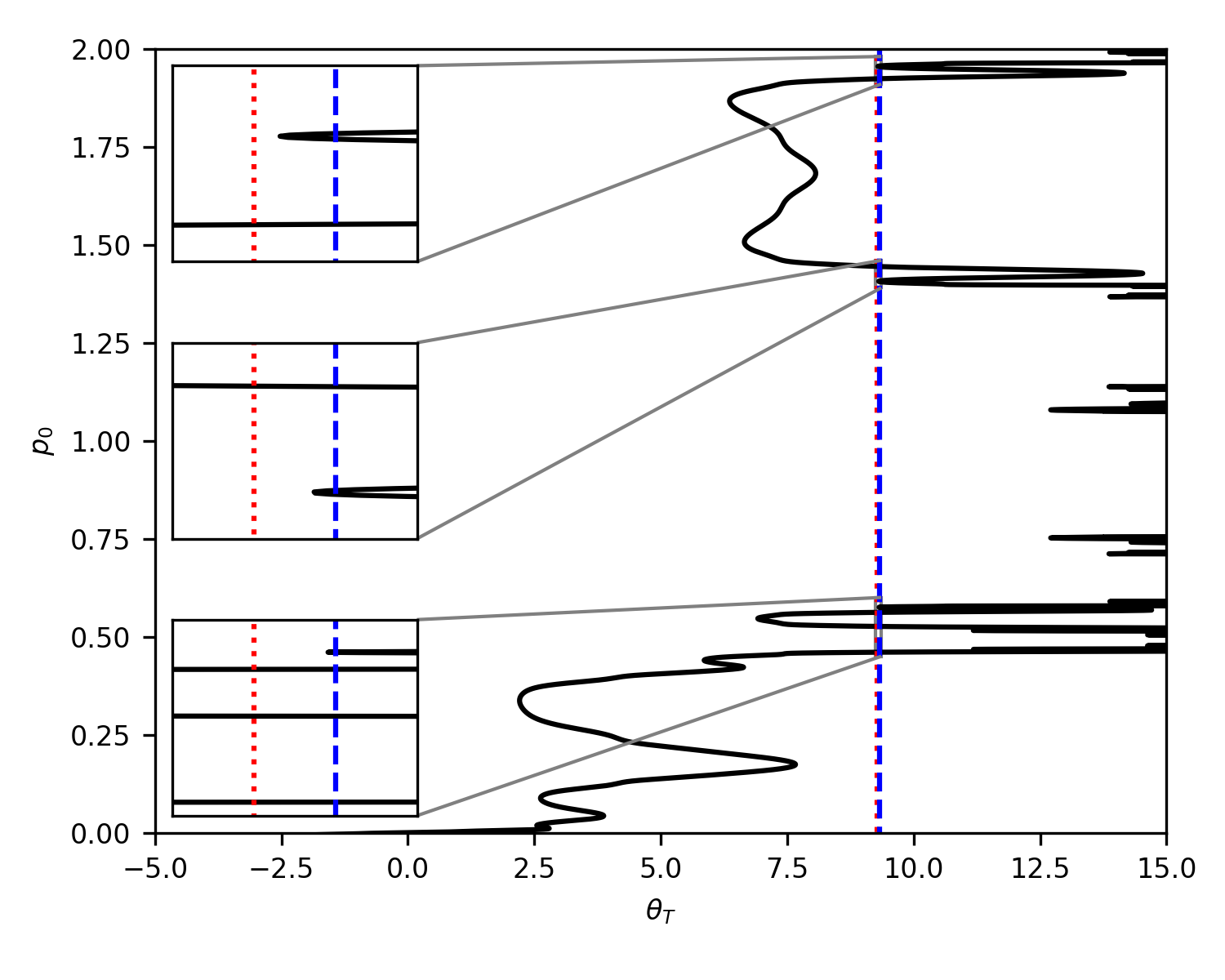} \\
\includegraphics[width = .45\columnwidth]{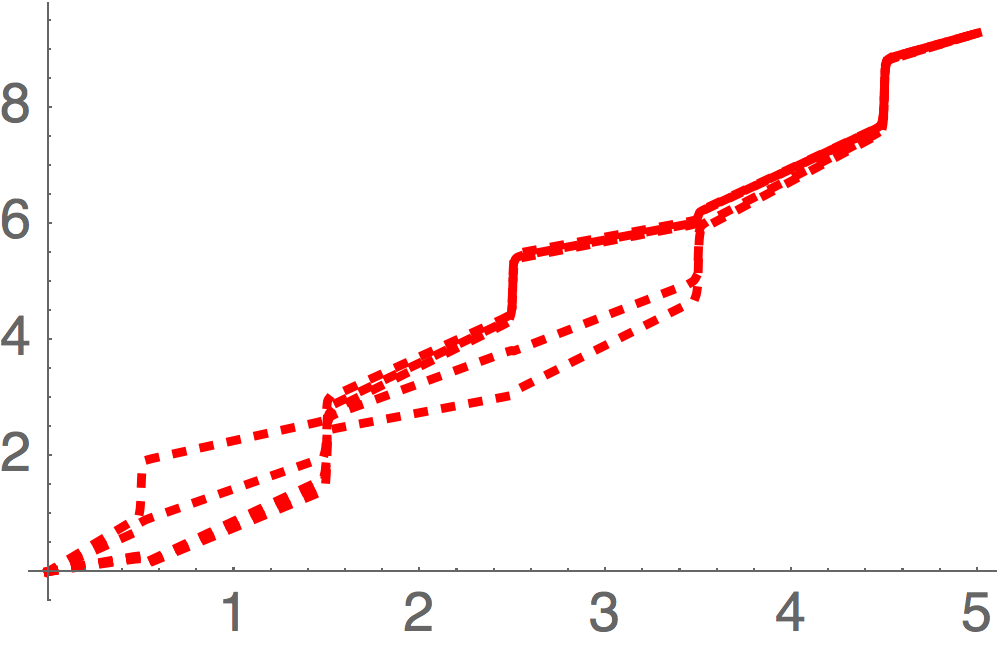}
\includegraphics[width = .45\columnwidth]{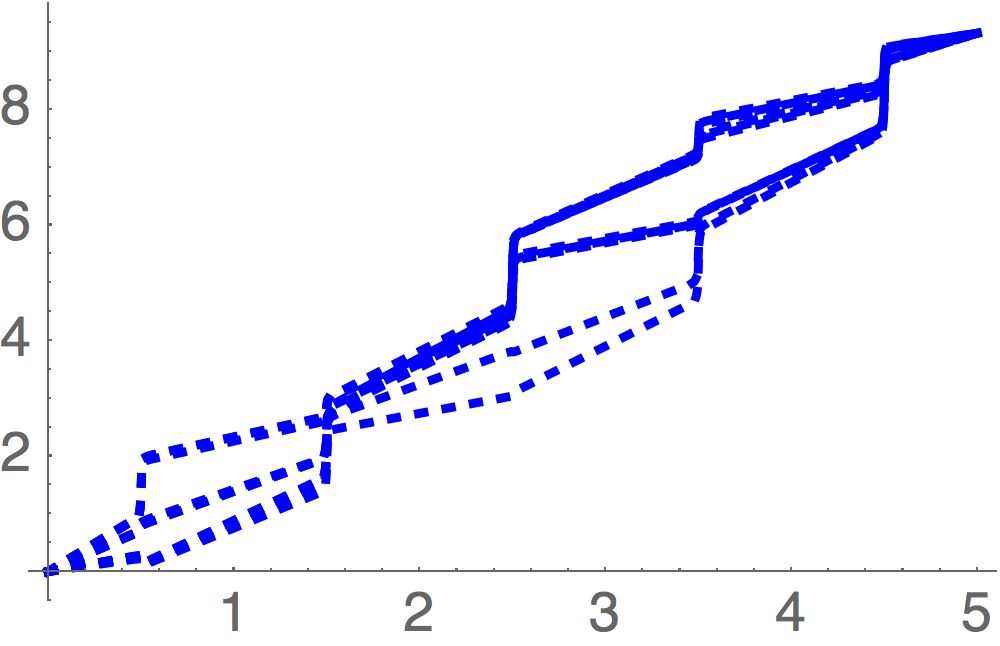} \\
\begin{picture}(0,5)
\put(-103,78){\small $\theta$}
\put(12,78){\small $\theta$}
\put(-68,0){\small $t$ ($\mu\mathrm{s}$)}
\put(46,0){\small $t$ ($\mu\mathrm{s}$)}
\put(95,118){(a)}
\put(-15,24){(b)}
\put(99,24){(c)}
\end{picture}
\caption{ (Color online) A segment of the LM from Fig.~\ref{fig-xzlm}(g), (initialized at $\theta_0 = 0$, after $T = 5~\mu\mathrm{s}$, with $\tau_m = 0.025~\mu\mathrm{s}$, $\tau_x = 1~\mu\mathrm{s} = \Lambda$ and $\epsilon = 0.99$), is shown in (a), in solid black. The vertical dotted red and dashed blue lines highlight two particular post--selections, at $\theta_T = 9.28$ and $\theta_T = 9.32$, respectively. We compare the number of multipaths linking the excited state to these two final states over the given time interval. The first of these final boundaries admits 5 OP solutions, shown in (b), whereas the second admits 11 OP solutions, shown in (c). This sharp change in the number of solutions existing between quite similar boundary conditions highlights a way that instabilities in the OP dynamics (and therefore the underlying distribution from which they are optimized) are exaggerated in conjunction with the OPs being chaotic. }\label{fig-manymulti}
\end{figure}
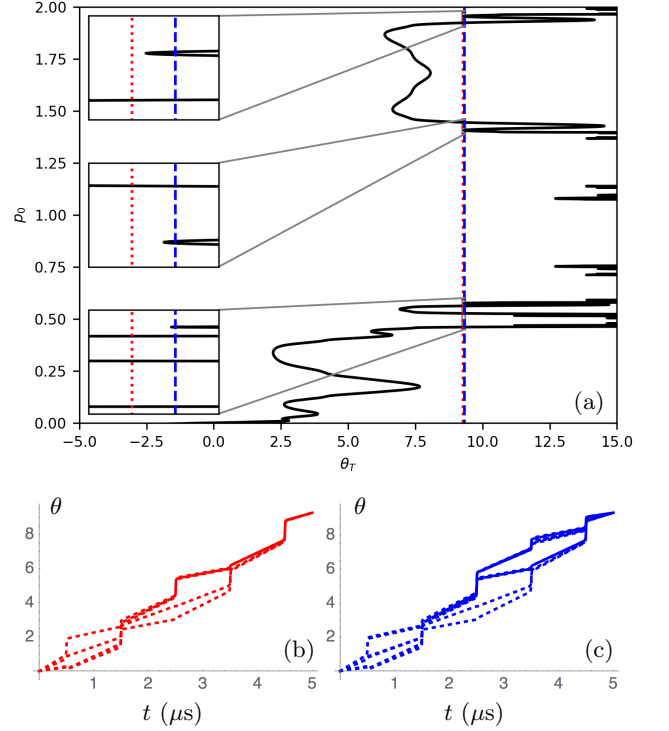

We show the LM in the strong--kick regime ($\epsilon = 0.99$), initialized at the excited state $\theta_0 = 0$ after 3, 4, and 5 kicks in Fig.~\ref{fig-xzlm}. 
It is immediately apparent that by the time even 5 kicks have taken place, the manifold contains a large number of catastrophes bounding wide caustic regions, especially at higher winding numbers \footnote{Winding counts are defined relative to $\theta_0$, and can be experimentally determined when we know the entire history of a SQT.}. 
Even among the relatively high--probability dynamics at lower $|p_0|$ and fewer winding counts (the calmest part of the LM), the number of multipaths grows quickly. 
Examples from this calmer region appear in Fig.~\ref{fig-xzdens}, where we show the OPs describing a bit flip. 
After five kicks, we see that there are already five OPs which make a bitflip  from the excited state to the ground state, in each direction (and that is only those which do it with half a winding count, never mind the rarer, but still observable, paths which orbit the Bloch sphere 1.5 times or more). 
All of the OPs shown in Fig.~\ref{fig-xzdens} are most--likely paths (MLPs) with similar probability weights, and are therefore all approximately equally physically significant \footnote{OPs are derived by extremizing the probability of trajectories moving between two states; as such, they can be MLPs, least-likely paths (LLPs), or even saddle-paths. 
Typically we are physically interested in locating MLPs, and existing methods for data analysis focus on extracting the MLP \cite{Weber2014, Mahdi2016}. 
The distinction between MLPs and LLPs is discussed more fully in our previous paper \cite{Lewalle2016}.}. 
The evolution of the quantum state under continuous measurement would experimentally be derived using a particular model to reconstruct the stochastic state evolution from the stream of readout results. 
OPs are effectively defined from data with respect to the pre-- and post-- selected density of ensembles of SQTs. Some examples are shown in Fig.~\ref{fig-xzdens}, including a two--path winding count group demonstrating good agreement between theory and simulation \footnote{We use a grouping algorithm to extract multipaths from SQT data, using the concept that was previously developed by Mahdi Naghiloo for this purpose. 
See \cite{Mahdi2016}.}. 
In showing that the OP equations of motion are chaotic, we show that evolutions with similar $\theta(t)$, $p(t)$, and $\mathbf{r}^\star(t)$ diverge from each other given a modest amount of time to evolve further. 
We have measured this divergence in the state $\theta$, but it also necessarily appears in the optimal readouts $\mathbf{r}^\star$, since $\mathbf{r}(t)$ is experimentally used to construct a trajectory $\theta(t)$. 
It is tempting to think that OP chaos ought to imply that the entire distribution of SQTs (without imposing a final boundary condition) should drastically change given small variations in the initial state; this is not necessarily true. 
Our OPs are derived under the assumption that a final boundary condition will be imposed, which makes them conceptually different from the global MLP (the OP which reaches the most--likely $\theta_T$ at the final time), or the average path. 
In fact, neither of the latter exhibit particularly striking behavior in the system defined by \eqref{kick} and \eqref{hxz}. 
We can see, however, that small changes in boundary conditions can drastically change the number of OP solutions. 
An example of this behavior appears in Fig.~\ref{fig-manymulti}. 
We confirm from Fig.~\ref{fig-xzdens} that the behavior of the OPs reflects the behavior of the underlying SQT distribution post--selected on the desired boundary conditions. 
We also see that we have a situation where the number of MLPs reaching most $\theta_T$ grows rapidly, such that it becomes harder to say which dynamics are actually the overall ``most-likely''. 
The MLP is simplest to interpret when only one or a few solutions exist, corresponding to well--defined routes visible in the underlying SQT density. 
Nonetheless, we may proceed knowing that multipaths with even large numbers of solutions reflect the features of the underlying post--selected trajectory density. 
Furthermore, from Fig.~\ref{fig-manymulti} we see that we may find examples of interesting OP behavior using the LM we use to find multipaths. 
We next devote some time to relate the LM's behavior directly to OP chaos.

\begin{figure}
\includegraphics[width=.95\columnwidth]{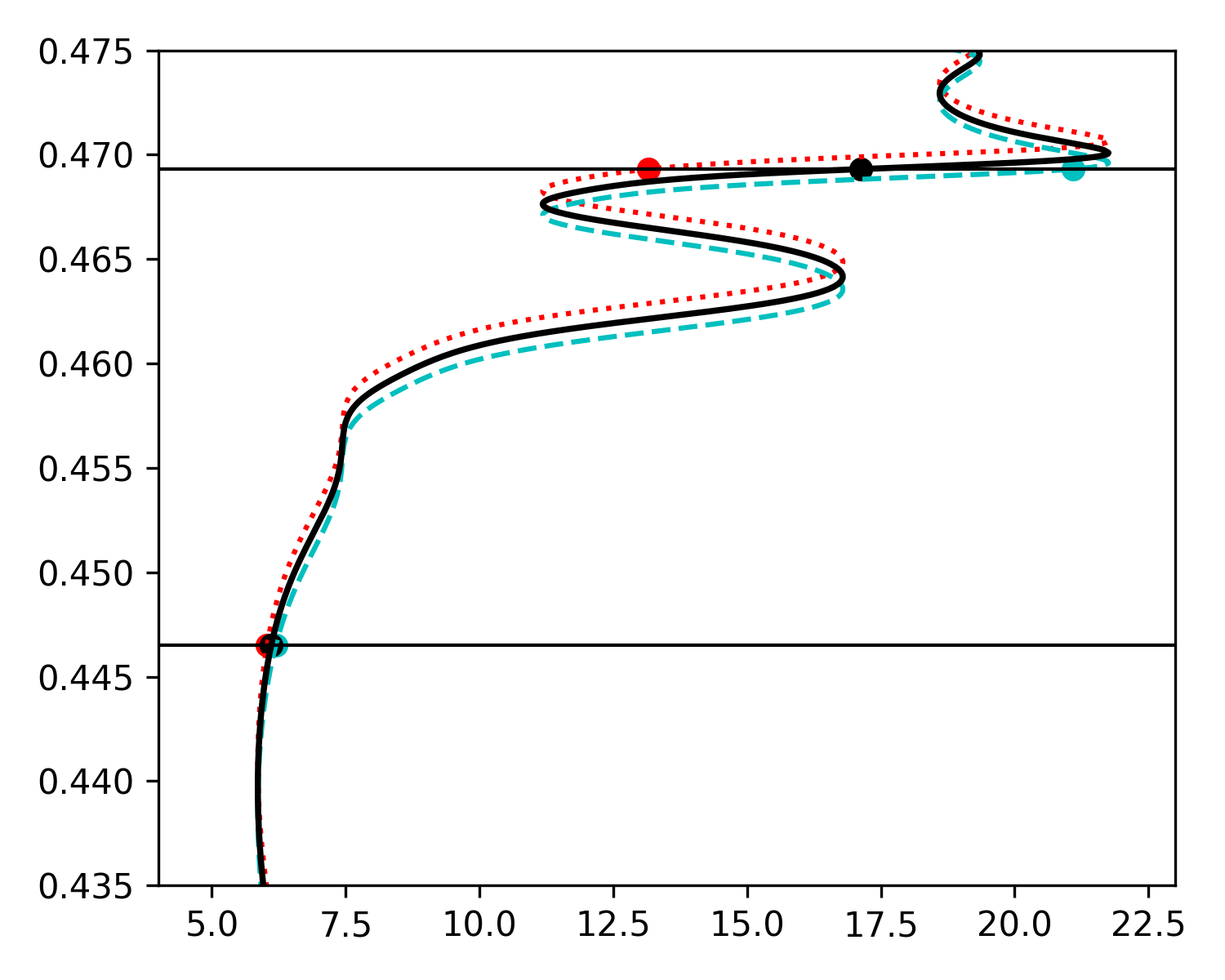}
\begin{picture}(0,0)
\put(-203,170){$p_0$}
\put(-30,20){$\theta_T$}
\end{picture}
\caption{(Color online) Three Lagrange manifolds, initialized at $\theta_0=0$ (solid black) and $\theta_0^\pm=\pm 0.01$ (dashed cyan and dotted red), are shown at $T=5~\mu\mathrm{s}$. A more comprehensive view of the central manifold is shown in Fig.~\ref{fig-xzlm} (a,d,g), as we are using the same parameters ($\tau_x = 1 ~\mu\mathrm{s} = \Lambda$, $\tau_m = 0.025~\mu\mathrm{s}$, and $\epsilon = 0.99$). In this region, the three manifolds have a similar shape, so the divergence of points with the same $p_0$ depends on the size of $|J_T|$. Two values of $p_0$ (shown as horizontal lines) and the corresponding points on each manifold are emphasized, highlighting that the points in the region with a larger $|J_T|$ have been torn much farther apart than those in the region with a smaller $|J_T|$, due to a small shift in the $p$ direction between the LMs.}\label{fig-case1}
\end{figure}
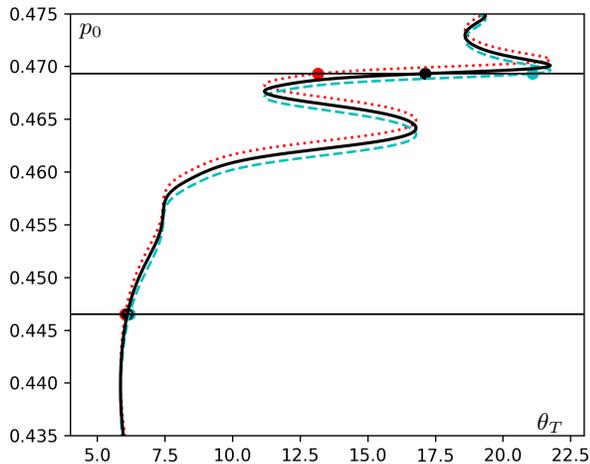

\begin{figure}
\includegraphics[width=.95\columnwidth]{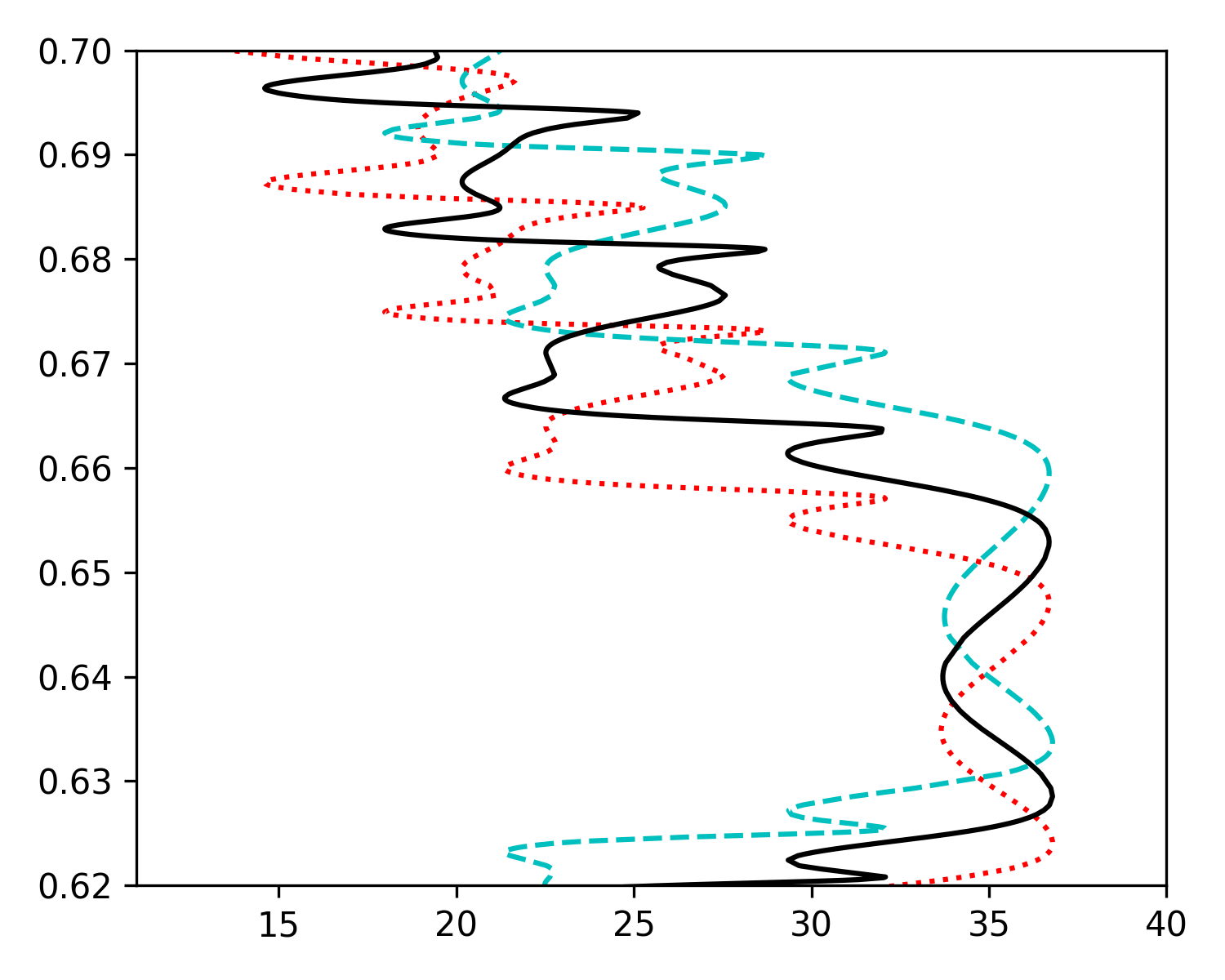}\\
\includegraphics[width=.95\columnwidth]{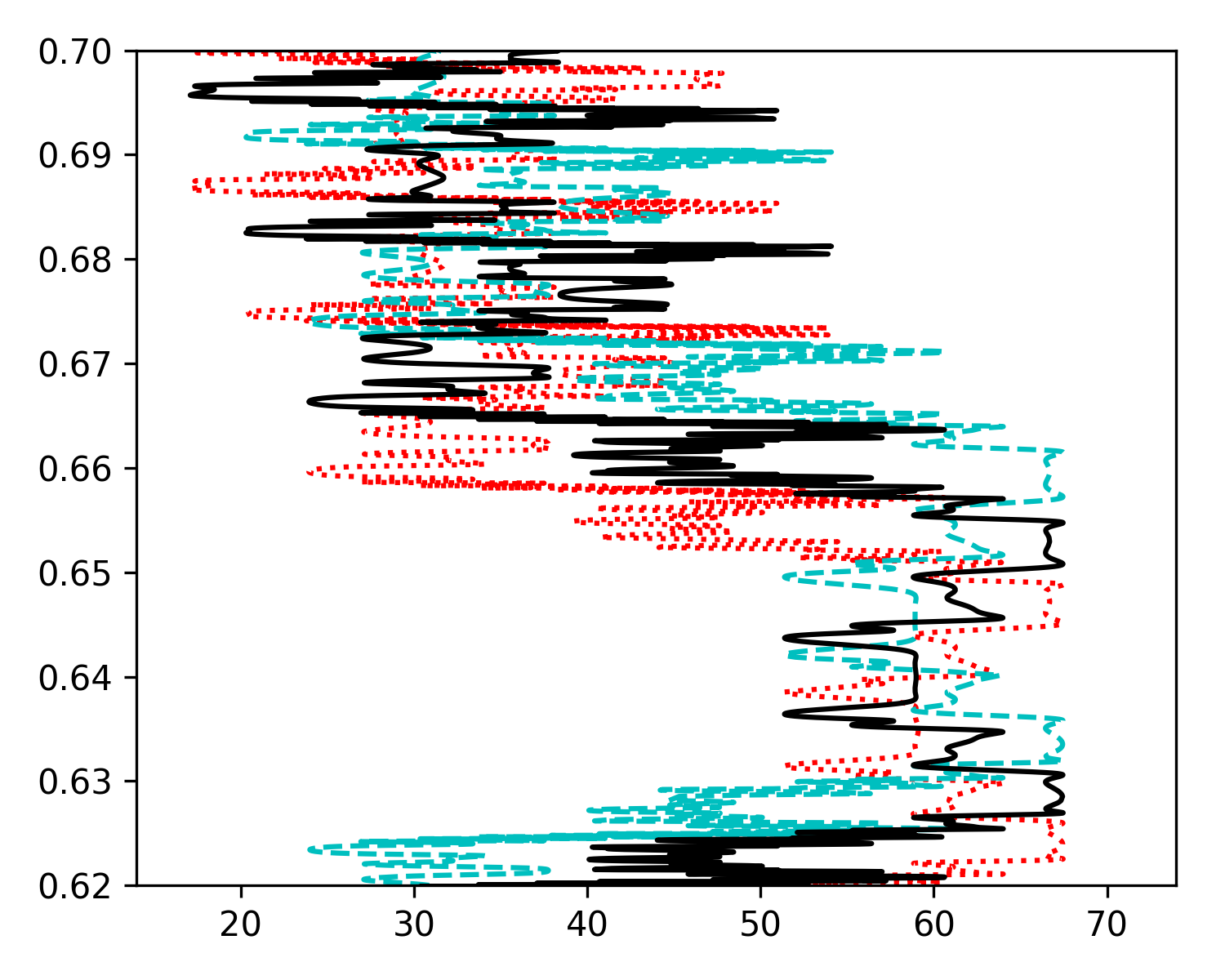}
\begin{picture}(0,0)
\put(-30,355){(a)}
\put(-30,168){(b)}
\put(-225,3){$p_0$}
\put(-225,190){$p_0$}
\put(-44,3){$\theta_T$}
\put(-33,190){$\theta_T$}
\end{picture}
\caption{(Color online) A particular segment of the LM from Fig.~\ref{fig-xzlm} is shown at $T=4~\mu\mathrm{s}$ (a), and $T=5~\mu\mathrm{s}$ (b), in solid black, with auxiliary LMs initialized $\pm 0.01$ radians away in dashed cyan and dotted red (as in Fig.~\ref{fig-case1}). We again use $\tau_x = 1 ~\mu\mathrm{s} = \Lambda$, $\tau_m = 25~\mathrm{ns}$, and $\epsilon = 0.99$. At $T=4~\mu\mathrm{s}$, the three manifolds still have a similar shape, but are shifted in the $p$ direction. In (b), one kick later, more catastrophes have formed, and are clustered on a scale smaller than the shift in $p$, such that paths with the same $p_0$ in different LMs are torn apart throughout the region shown, rather than only in a few segments of the LMs with large $|J_T|$.}\label{fig-case2}
\end{figure}

\subsection{Formal Connections between OP Chaos and Multipaths}

\begin{figure}
\includegraphics[width=.92\columnwidth, trim = {20pt 8pt 30pt 25pt}, clip]{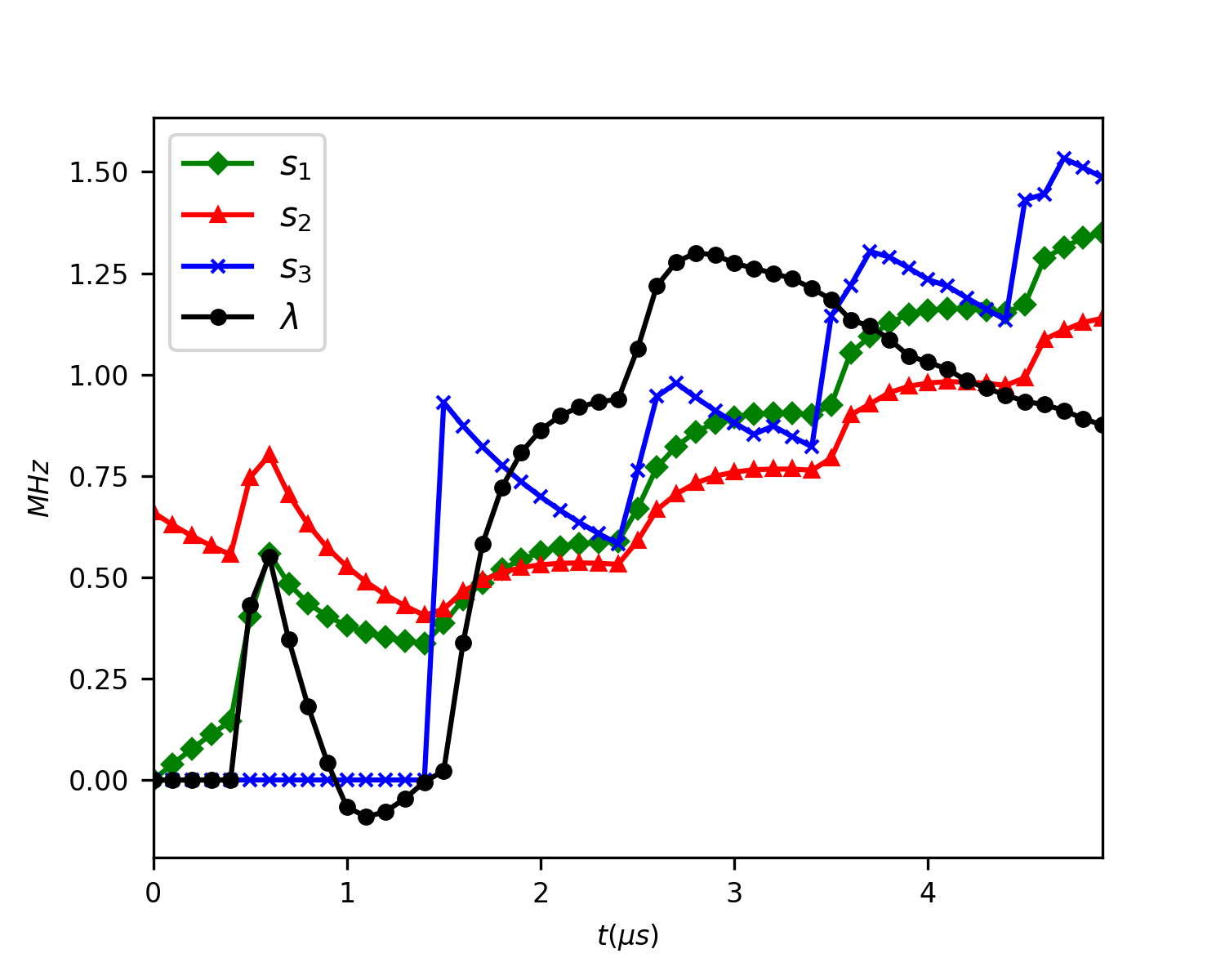} \\
\includegraphics[width=.92\columnwidth, trim = {20pt 8pt 30pt 25pt}, clip]{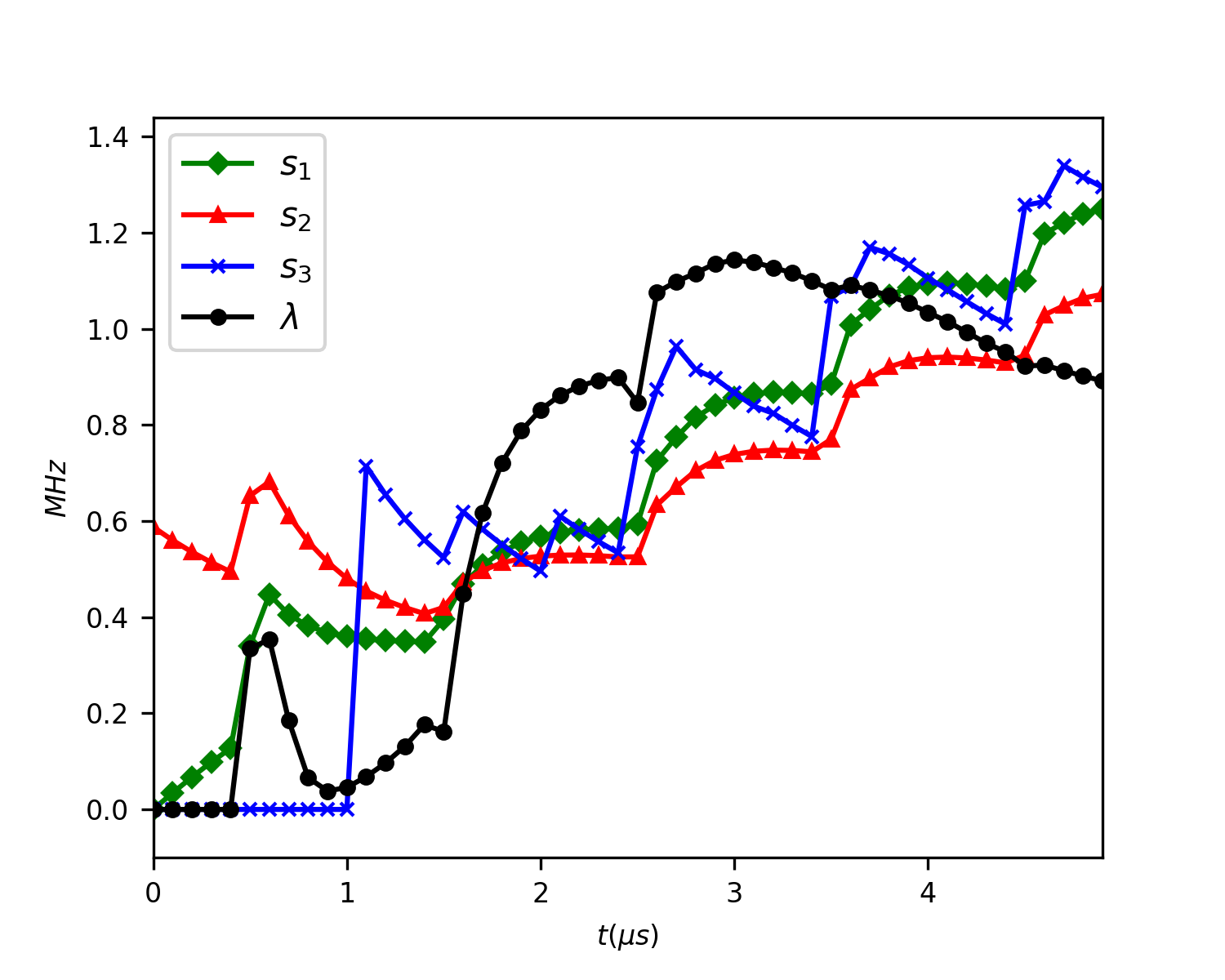}
\begin{picture}(0,0)
\put(-24,213){(a)}
\put(-24,26){(b)}
\put(-181,131.5){\tiny \emph{av}}
\put(-179.5,318){\tiny \emph{av}}
\put(-120,213){\footnotesize $\theta_0 = 0.000$, $p_0 \in [-2,2]$}
\put(-120,26){\footnotesize $\theta_0 = 0.286$, $p_0 \in [-2,2]$}
\put(-233,210){\footnotesize $\lambda_{av}$}
\put(-236,202){\tiny (MHz)}
\put(-233,18){\footnotesize $\lambda_{av}$}
\put(-236,10){\tiny (MHz)}
\end{picture}
\caption{(Color online) We plot three parameters $s_1$ \eqref{s1}, $s_2$ \eqref{s2}, and $s_3$ \eqref{s3}, defined in appendix~\ref{sec-lmdef}, which quantify the deformation of the manifold, along with the average LE $\lambda_{av}$ \eqref{lyap_av}. These ``stretching parameters'' are normalized relative to the LE so that the magnitude of the area under all the curves is the same over the interval shown. These parameters are plotted over the first $5~\mu\mathrm{s}$ for the system given by \eqref{hxz} and \eqref{xz_kick}, with $A = 0.99~\mu\mathrm{s}$, $\tau_x = 1.0~\mu\mathrm{s}$, and $\tau_m = 25~\mathrm{ns}$. The manifold was initialized at $\theta_0=0$ in (a), and $\theta_0 = 0.286$ in (b), the latter of which corresponds to Fig.~\ref{fig-xzexamples}(a). In both (a) and (b) we use auxiliary manifolds with $\theta_0 \pm 0.01$, and $p_0 \in [-2,2]$ to compute $\lambda_{av}$. We see that the length $L$ $(s_1)$ and Jacobian $J$ ($s_2$) for the manifold (green $\diamondsuit$ and red $\triangle$, respectively) are very closely related to each other, and that the number of catastrophes (see $s_3$, blue $\times$) grows sharply at each kick, following similar behavior. We see that by about the third or fourth kick, $\lambda_{av}$ \eqref{lyap_av} (black $\circ$) begins to saturate, and decreases (although it remains quite positive); this appears reasonable, since the distance \eqref{dist_av} is defined in a way that gives it a maximum value, whereas the other parameters plotted here may grow indefinitely. With the support of the arguments from section~\ref{sec-ch_mani}, the connection between the average LE and other parameters reinforces that chaos and complex multipath behaviors do not occur independently in this system; one implies the other across large sections of manifold (large enough to encompass virtually all reasonably--probable OP behaviors originating from a particular initial state). }\label{fig-str}
\end{figure}

\par The sharp growth in the complexity of the LM, due at least in part to sharp increases in the number of catastrophes with each kick, is born of the same resonance phenomenon which generates chaos in this system for low $\epsilon$ (compare Figs.~\ref{fig-earlyreson} and \ref{fig-LMreson}); it is natural to suppose that the subsequent LM complexity at larger $\epsilon$, and therefore the corresponding multipath behavior (Figs.~\ref{fig-xzlm} and \ref{fig-xzdens}), is connected with the chaotic properties of the OP dynamics we have described (e.g.~in Figs.~\ref{fig-xzps} and \ref{fig-xzexamples}).
\par The presence of large numbers of multipaths is a hallmark of OP chaos. 
When multipaths form, the LM overlaps itself, and in doing so gains areas of large $|J_T|$ (see Fig.~\ref{fig-xzlm} (a,d,g) for an example of this). 
Suppose three LMs are initialized at the states $\theta_0$ and $\theta_0^\pm=\theta_0\pm \delta\theta_0$. 
In order to avoid chaos, it is necessary that the LMs maintain similar shapes. 
We first consider the case where they do maintain similar shapes, except for some variations $\pm\delta\theta$ and $\pm\delta p$. After time $T$, the distance between points with the same $p_0$ is
\be \begin{split}
D_T &\sim \tfrac{1}{2}|\theta_T(p_0) - \theta_T^+(p_0+\delta p)| \\& \quad\quad + \tfrac{1}{2}|\theta_T(p_0) - \theta_T^-(p_0-\delta p)| + \delta \theta \\ &\approx \delta \theta_0 + \delta\theta + |J_T \: \delta p|.
\end{split} \ee
In regions where $|J_T|$ is large, the effect of small shifts in the $p$ direction between manifolds of similar shape is magnified, so nearby points in these regions will diverge. 
In other words, even a small difference $\delta p(t)$ between $p(t)$ and $p^\pm(t)$ generates very different values $\theta_T$ and $\theta_T^\pm$ where the manifold has large $|J_T|$. 
This situation is illustrated in Fig.~\ref{fig-case1}. A large $|J_T|$ can only appear when the manifold is either spreading over a very high range of winding counts, and/or forming many--layered caustics, with closely--spaced (relative to $p_0$) catastrophes.
It is also possible that adjacent LMs have unrelated shapes after some time, in which case the region in question is necessarily chaotic. 
This situation does not \emph{have to} occur in connection with the formation of many caustics; however, if the scale on which caustics form relative to $p_0$ is smaller than the shifts $\delta p$, LMs which maintained similar shapes become scrambled to the point that they no longer resemble each other as they evolve. 
This situation is shown in Fig.~\ref{fig-case2}. 
One particularly interesting feature which emerges from chaotic OP dynamics is that one may choose similar boundary conditions which sit on opposite sides of many catastrophes in the manifold; thus situations where small changes in final boundary conditions lead to dramatically different numbers of OP solutions, as shown in Fig.~\ref{fig-manymulti}, are relatively commonplace in such systems.

\par In appendix \ref{sec-lmdef} we define numerical ``stretching parameters'' which quantify aspects of a manifold's shape and deformation. 
The overall growth of the number of caustics in a representative segment of the manifold, and its overall level of deformation, are compared with the average LE \eqref{lyap_av} in Fig.~\ref{fig-str}. 
We see that the length of the manifold grows exponentially (see \eqref{s1}), and that this growth is closely matched by exponential growth in the average $J_t$ (see \eqref{s2}) and number of catastrophes (see \eqref{s3}).
This is in contrast with the rate of growth in catastrophes in the integrable case (for time--independent $\tau_x$ and $\tau_z$), which is only linear in time \cite{Lewalle2016}.
The average LE \eqref{lyap_av} over a large segment of the LM also grows exponentially for several kicks, and then levels off. 
This is expected, as the distance measure \eqref{dist_av} has a maximum value---our LEs can never actually grow indefinitely, because there is a limit to how far apart they can get on the Bloch sphere. 
Nonetheless, we see that there is enough chaos around the $\theta_0$ shown, across the range $p_0 \in [0,2]$ to obtain a function $\lambda_{av}(t)$ which is almost always positive and mostly increasing on $t \in (0,5] ~\mu\mathrm{s}$; its magnitude and shape are similar to those shown from particular examples in Fig.~\ref{fig-xzexamples}. 
Furthermore, we see in Fig.~\ref{fig-str} that the shapes of $\lambda_{av}(t)$ and the three parameters describing the LM are related, highlighting a general connection between multipaths and OP chaos. 

\section{Conclusions \label{sec-concl}}
We have have introduced a different kind of chaos in quantum systems, which may appear in open systems with dynamics due to continuous measurements. 
We do this by looking for chaos in the extremal--probability paths, rather than in ensembles of stochastic quantum trajectories directly. 
OPs are mathematically classical, and we therefore apply a classical definition of chaos, which implies a form of unpredictability even in deterministic systems; if paths with nearby initial conditions diverge exponentially, then knowledge of the long--term behavior of the system is severely limited by the precision with which initial states can be prepared. 
The OP formalism allows us to apply this classical definition of chaos to purely quantum systems, lacking any clear mechanical analog in the classical world. 
For example, we have demonstrated that such exponential divergence may occur in the OPs for a continuously--monitored qubit, where dynamics reminiscent of the kicked rotor are generated entirely by making measurements.
Lyapunov exponents computed among OPs show that OPs with initially--similar quantum states and optimal readouts can diverge to completely different states within relatively short times (i.e.~over intervals well within the coherence times of modern qubits, used in many experiments to implement the weak, continuous measurements we have considered here). 
We stress the instability of OPs is related to the instability of the underlying distribution of SQTs itself, when both initial and final boundary conditions are applied. 
With the boundaries properly taken into account, we have shown that chaotic OP behavior is connected with dramatic growth, over time, in the number of OP solutions meeting given sets of boundary conditions, as well as the possibility to see large differences in the number of OP solutions linking nearly--identical boundary conditions. 
These effects are themselves a form of unpredictability in the dynamics, which reflect instabilities in the dynamics of underlying SQTs and their statistics. 
We have been able to explain the onset of both the chaotic and multipath behaviors in terms of resonances between an integrable and perturbing part of the OP Hamiltonian. This concept should extend to any system with a time dependent and periodic perturbation.

\par Instabilities in the dynamics of stochastic quantum trajectories may have consequences for qubit control, error correction, and other general problems of interest in the larger quest for useful quantum control and information processing. 
Extensions of the work we have started here could generalize into a practical understanding of when and how OP chaos can occur, and how it impacts feedback control schemes or other useful tasks involving continuously--measured qubits. 
For instance, suppressing the kinds of dynamics we describe here underscores the need for effective feedback schemes to control complicated systems. 
Alternatively, a feedback control scheme itself (which is necessarily time--dependent) could lead to quite wild and unintended dynamics among rarer events which deviate from the intended behavior; an understanding of these dynamics could aid in designing robust schemes. 
We hope that further investigations of OP chaos in qubits, across a wider range of measurement schemes, can lead to a deeper understanding of these questions.

\begin{acknowledgments}
We acknowledge early contributions to this work made by Areeya Chantasri, along with numerous conversations at different stages of the project. 
We are also grateful to Miguel Alonso, Juan Atalaya, Justin Dressel, Shay Hacohen-Gourgy, Patrick Harrington, Sreenath K.~Manikandan, Alexander Korotkov, Leigh Martin, Joe Murphree, John Nichol, Sarada Rajeev, Irfan Siddiqi, Carlos Stroud Jr., and Jing Yang for helpful comments and discussions. 
Funding was provided through NSF grant no.~DMR-1506081 and US Army Research Office grant no.~W911NF-15-1-0496. 
PL acknowledges additional support from the US Department of Education grant No.~GR506598 as a GAANN fellow. 
Many of our numerical codes are written based on integration methods from \cite{NumRecC}, using Python 2.7.11.
\end{acknowledgments}


\appendix

\section{Review of State Update and Path Integral Formalisms for Continuous Weak Monitoring \label{sec-review}}
All of the information contained in this appendix can be found in other works, e.g.~Refs.~\cite{Korotkov1999, *Korotkov2001, *Korotkov2016} and \cite{Chantasri2013,Chantasri2015,Areeya_Thesis,Lewalle2016}, and is included here to make this paper more accessible to those not already familiar with the formalism. 

\subsection{Bayesian State Update}
As seen in \eqref{stateup} and the surrounding text, the state of our qubit $\rho$, given some readout, is updated through the application of some measurement operators according to
$
\rho(t+dt) = \mathcal{M} \rho(t) \mathcal{M}^\dag \left\lbrace \text{tr}\left(\mathcal{M} \rho(t) \mathcal{M}^\dag\right) \right\rbrace^{-1}.
$
Individual weak qubit measurements in the $xz$--plane of the Bloch sphere can be implemented with the measurement operator
\be \begin{split}
\mathcal{M}_\varphi = \: \exp & 
\left[ -\frac{(1+r_\varphi^2)\:dt}{4\tau_\varphi}\right] \bigg\lbrace \cosh\left[ \frac{r_\varphi\:dt}{2\tau_\varphi} \right] \mathtt{I} \\ & +
\sinh\left[ \frac{r_\varphi\:dt}{2\tau_\varphi} \right] \left( \sigma_z\cos\varphi + \sigma_x \sin\varphi \right)\bigg\rbrace,
\end{split} \ee
for $dt \ll \tau$. 
We have a $z$--measurement for $\varphi = 0$ and an $x$--measurement for $\varphi = \pi/2$, i.e.~we may define operators specifically for $\mathcal{X} = \mathcal{M}_{\varphi = \pi/2}$ and $\mathcal{Z} = \mathcal{M}_{\varphi=0}$ measurements, for use in the state update equation. These operator assignments can be shown to be equivalent to applying Bayes' rule, in the readout probability densities, to individual elements of the density matrix $\rho$ \cite{Korotkov1999, *Korotkov2001, *Korotkov2016, Chantasri2013, Lewalle2016}.

\par In the example developed in the main text, we are interested in monitoring both the observables $\sigma_x$ and $\sigma_z$ at the same time, i.e.~we are interested in the particular case where
\be 
\rho(t+dt) = \frac{\mathcal{Z}\mathcal{X} \rho(t) \mathcal{X}^\dag \mathcal{Z}^\dag}{\text{tr}\left(\mathcal{Z}\mathcal{X} \rho(t) \mathcal{X}^\dag \mathcal{Z}^\dag \right)}.
\ee
The probability density from which the readouts $\mathbf{r}$ are drawn is given by $\wp(\mathbf{r}|\rho) = \text{tr}\left(\mathcal{Z}\mathcal{X} \rho(t) \mathcal{X}^\dag \mathcal{Z}^\dag \right)$.
The two--measurement operator can be expanded to first order in $dt$, such that 
\be \begin{split}
\mathcal{Z}\mathcal{X} &\approx \left(\mathtt{I} + \hat{Z} dt \right) \left( \mathtt{I} + \hat{X} dt \right) \approx \mathtt{I} + \left( \hat{X} + \hat{Z} \right) dt,
\end{split} \ee
where $\hat{X} =  -(r_x - \sigma_x)^2/(4\tau_x)$ and $\hat{Z} =  -(r_z - \sigma_z)^2/(4\tau_z)$. Notice that to $O(dt)$, any dependence on the order of measurement operators disappears. 
Then an update equation for small $dt$
\be 
\dot{\rho} = \hat{\zeta}  - \rho(t) \text{tr}\left( \hat{\zeta} \right)
\ee
can be derived, where $\hat{\zeta} \equiv  [ \hat{X} + \hat{Z},\rho ]_+$, and the braces $[,]_+$ denote the anti--commutator. For the vector $\mathbf{q}$ of Bloch sphere coordinates, a dynamical system can be extracted by taking $\dot{\mathbf{q}} = \text{tr}(\dot{\rho} \sigma_{\mathbf{q}})$, which yields
\be\label{strato_xz} \begin{split}
 \dot{x} &= \frac{\left(1 - x^2 \right) r_x}{\tau _x} -\frac{x z r_z }{\tau _z},\\
\dot{y} &= -y \left(\frac{z r_z }{\tau _z}+\frac{x r_x}{\tau _x}\right), \\
\dot{z} &=  \frac{\left(1-z^2\right) r_z}{\tau _z}-\frac{x z r_x}{\tau _x}.
\end{split} \ee
It is easy to see that if $y = 0$, then $\dot{y} =0$ also, so that the $y$--evolution can be uncoupled from the system and neglected; we do this and work purely in the $xz$--plane of the Bloch sphere. 
The remaining $x$ and $z$ equations can be converted to polar coordinates $(R,\theta)$ in that plane; it is then simple to show that for perfect measurement efficiency (implicitly assumed above), and an initially pure state ($R = 1)$, that $\dot{R} = 0$, leaving only evolution in $\theta$, given by
\be \label{thdotF}
\dot{\theta} = \mathcal{F}[\theta,\mathbf{r},t] = \frac{r_x}{\tau_x} \cos\theta - \frac{r_z}{\tau_z} \sin \theta.
\ee

\subsection{The Stochastic Path Integral}
We now review the procedure developed in Refs.~\cite{Chantasri2013,Chantasri2015,Areeya_Thesis} to derive the OPs. 
We begin by writing down the joint probability associated with a path (a sequence of readouts $\lbrace \mathbf{r} \rbrace$ and their associated states $\lbrace \mathbf{q} \rbrace$), which may be expressed by
\begin{widetext}
\be \begin{split} \label{jointp}
\mathcal{P} 
(\lbrace \mathbf{q} \rbrace, \lbrace \mathbf{r} \rbrace|\mathbf{q}_i,\mathbf{q}_f) = \delta(\mathbf{q}_i - \mathbf{q}_0) \delta(\mathbf{q}_f - \mathbf{q}_n) 
\left[
\prod_{k=0}^{n-1} \wp(\mathbf{q}_{k+1}| \mathbf{q}_k,\mathbf{r}_k) \wp(\mathbf{r}_{k}|\mathbf{q}_k)\right] .
\end{split} \ee
The $\delta$--functions at the initial and final points apply the initial and final boundary conditions. The indices $k$ run over time, such that if $\rho_k = \rho(t)$, then $\rho(t+dt) = \rho_{k+1}$ and so on. 
We use $\wp(\mathbf{q}_{k+1}|\mathbf{q}_k,\mathbf{r}_k) = \delta(\mathbf{q}_{k+1} - \mathbf{q}_k - dt \mathcal{F}[\mathbf{q}_k,\mathbf{r}_k])$ is a deterministic update rule from e.g.~\eqref{strato_xz} or \eqref{thdotF}.
The readouts are stochastic, and drawn from the density $\wp(\mathbf{r}_k|\rho_k) = \text{tr}\left(\mathcal{Z}(r_z)\mathcal{X}(r_x) \rho_k \mathcal{X}^\dag(r_x) \mathcal{Z}^\dag(r_z) \right)$ discussed above. Recall that a $\delta$--function may be written
$
\delta(\mathbf{q}) = (2\pi i)^{-\text{dim}(\mathbf{q})} \int_{-i\infty}^{i\infty} d\mathbf{p} \: \exp\left[ - \mathbf{p}\cdot \mathbf{q} \right]
$, where $\text{dim}(\mathbf{q})$ is the dimension of $\mathbf{q}$, and $d\mathbf{p} = dp_1\; dp_2\; ... \; dp_{\text{dim}(\mathbf{q})}$.
We apply this identity to \emph{all} $\delta$--functions in \eqref{jointp}, such that
\be \begin{split}
\mathcal{P} &= \limit{n}{\infty} \limit{dt}{0} \mathcal{N} \idotsint\limits_{-i\infty}^{i\infty} \left(\prod_{k=0}^{n-1} d\mathbf{p}_k \right) \exp\bigg[ \mathcal{B}
+ \sum_{k=0}^{n-1} \left(-\mathbf{p}_k \cdot (\mathbf{q}_{k+1}  - \mathbf{q}_k - dt \: \mathcal{F}_k) + \ln \wp(\mathbf{r}_k|\mathbf{q}_k) \right) \bigg] \\
&= \int \mathcal{D}[\mathbf{p}] \exp \left[\mathcal{B} + \int_0^T dt \left( - \mathbf{p}\cdot \dot{\mathbf{q}} + \mathbf{p}\cdot \mathcal{F}[\mathbf{q},\mathbf{r}] + \mathcal{G}[\mathbf{q},\mathbf{r}] \right) \right] 
\\ &= \int \mathcal{D}[\mathbf{p}] \exp\left[\mathcal{B} + \int_0^T dt( H(\mathbf{q},\mathbf{p},\mathbf{r},t) - \mathbf{p}\cdot \dot{\mathbf{q}}) \right] 
= \int \mathcal{D}[\mathbf{p}] \exp\left(\mathcal{B}+ S[\mathbf{q},\mathbf{p},\mathbf{r}] \right)
\end{split} \ee \end{widetext}
for $\mathcal{N} = (2\pi i)^{-(n+2) \cdot\text{dim}(\mathbf{q})}$. We use the shorthand $\mathcal{B} = -\mathbf{p}_{-1}\cdot(\mathbf{q}_0 - \mathbf{q}_i) - \mathbf{p}_n\cdot(\mathbf{q}_n-\mathbf{q}_f)$ for the boundary terms, and the shorthand $\mathcal{G}$ for the expansion to $O(dt)$ of the log--probability for the readouts $\ln \wp(\mathbf{r}|\mathbf{q})$. The expansion of $\ln \wp(\mathbf{r}|\rho) = \ln \text{tr}\left(\mathcal{Z}\mathcal{X} \rho(t) \mathcal{X}^\dag \mathcal{Z}^\dag \right)$ to $O(dt)$, relevant to the example in the main text, yields
\be \label{gxz}
\mathcal{G}\:dt = -\left(\frac{r_x^2 - 2 r_x \sin\theta + 1}{2\tau_x}+\frac{r_z^2-2r_z \cos\theta + 1}{2\tau_z}\right) dt,
\ee
up to some constants which do not affect the dynamics, and which can be absorbed into $\mathcal{N}$. The expression \eqref{gxz} should be understood as the counterpart to \eqref{thdotF}. OPs extremize the path probability, i.e.~Hamilton's equations for the OPs emerge by demanding $\delta S = 0$, a constraint satisfied by solutions extremizing the path probability $\mathcal{P}$. See Refs.~\cite{Chantasri2013, Chantasri2015} for further details.

\section{Resonances disrupting integrability} 
Here we review resonances more formally, as they apply to the example from section~\ref{sec-xzkick}. Helpful external references which inform the following summary and analysis include \cite{KolmogorovAM, *KArnoldM, *KAMoser1, *KAMoser2, ClassQuillen, BookTabor, BookF-M, BookJose, BookArnoldClassical, BookGoldstein, Clineder}.

\subsection{Introduction to resonances in canonical perturbation theory \label{sec-resdetail}}
We decompose our stochastic Hamiltonian in powers of $\epsilon$, i.e.
$
H(\theta,p,t) = H^{(0)}(p) + \sum_{n=1}^{\infty} \epsilon^n H^{(n)}(\theta,p,t),
$ as we have done above.
Below we will assume that $\epsilon$ is small, and restrict our analysis to a first order perturbation, working only with $H \approx H^{(0)}+\epsilon H^{(1)} + O(\epsilon^2)$. 
When a Hamiltonian is integrable, it is generally possible to find a set of canonical coordinates called action--angle coordinates; the Hamiltonian, when transformed into these coordinates (the ``Kamiltonian'', $K$ \cite{BookGoldstein}), only depends on the new generalized momenta \cite{BookArnoldClassical}. 
We have assumed that $H^{(0)}$ is already in its action-angle coordinates. 
We will use our given coordinates $\lbrace \theta, p \rbrace = 1$, and hypothetical new coordinates $\lbrace \phi , J \rbrace = 1$ throughout our derivations. 
The braces denote the Poisson bracket. We will also use a generating function $G(\theta,J,t)$ of the second type \cite{BookGoldstein}, which transforms between the two sets of coordinates. Recall that 
\be \label{type2gen}
 p = \partl{G}{\theta}{},\quad \phi = \partl{G}{J}{}, \quad\text{and}\quad K = H + \partl{G}{t}{},
\ee
where $\dot{\phi} = \partial_J K$ and $\dot{J} = - \partial_\phi K$. The aim is to derive $K$ to first order, and the transformation from $\theta$ and $p$ to $\phi$ and $J$ which allows for $\dot{J}=0$ and $\dot{\phi} = \nu(J)$. 
In other words, we suppose $H^{(0)} + \epsilon H^{(1)}$ is also integrable, and search for the $G$ which transforms to a new Kamiltonian $K(J)$ in action--angle coordinates. 
However, this procedure will fail at resonances, indicating that even to first order, the Hamiltonian is no longer integrable at certain points (the transformation to new action--angle coordinates cannot be found). 
Our primary interest is not in resolving this issue analytically, but merely in seeing where and how this transformation becomes impossible. 
Derivations similar to the one below can be found in e.g.~\cite{ClassQuillen,BookJose,BookF-M}.
\par We suppose that $G$ can be expanded in powers of $\epsilon$, such that $G \approx \theta J + \epsilon \:G^{(1)}(\theta,J,t)$, where $G^{(0)} = \theta J$ gives the identity transformation. Then from \eqref{type2gen} we have
\be 
p \approx J + \epsilon \partl{G^{(1)}}{\theta}{}, \quad\text{and}\quad \theta \approx \phi - \epsilon \partl{G^{(1)}}{J}{}.
\ee
Putting these coordinates into $H(\theta,p,t)$ and expanding gives
\be
H(\theta,p,t) \approx H(\phi,J,t) + \epsilon \lbrace G^{(1)} , H^\star \rbrace + O(\epsilon^2),
\ee
and then inserting the expansion of $H$ itself, and throwing out terms to second order in $\epsilon$, gives
\be 
H(\theta,p,t) \approx H^{(0)}(J) + \epsilon H^{(1)}(\phi,J,t) + \epsilon \lbrace G^{(1)},H^{(0)}\rbrace.
\ee
The Kamiltonian $K \approx K^{(0)} + \epsilon K^{(1)}$ is then given, to first order in $\epsilon$, by 
\be \begin{split}
K^{(0)} & = H^{(0)}(p=J) \\
K^{(1)} & - H^{(1)}(\theta=\phi,p=J,t) =\lbrace G^{(1)}, H^{(0)} \rbrace + \partl{G^{(1)}}{t}{}.
\end{split} \ee
We will now assume that our phase space is $2\pi$-periodic in $\theta$ and/or $\phi$, and that the time--dependent perturbation is also periodic, with some period $\Lambda$. Then we can write $H$ and $G$ as Fourier series
\be \begin{split}
H^{(1)} &= \sum_{\ell, k} \zeta_{\ell,k}(J) e^{i \ell \theta} e^{2i\pi k t/\Lambda}, \: \text{ and} \\
G^{(1)} &= \sum_{\ell,k} \xi_{\ell,k}(J) e^{i \ell \theta} e^{2i\pi k t/\Lambda}.
\end{split} \ee
We put these into the expression for $K^{(1)}$ above. We assume that $K^{(1)}$ is now only a function of $J$ (meaning that we have found the action-angle coordinates), such that $K^{(1)}(J)$ can be absorbed into $\zeta_{0,0}(J)$. Then we have 
\be \label{resonance}
\frac{i\zeta_{\ell,k}(J)}{\ell \nu^{(0)}(J) + \frac{2\pi k}{\Lambda} } =  \xi_{\ell,k}(J).
\ee
This equation contains the resonance condition we are interested in. 
Resonances occur where $\ell \nu^{(0)}(J) + \frac{2\pi k}{\Lambda} = 0$, for any integer $\ell$ and $k$. 
The Fourier coefficients $\xi_{\ell,k}$ of even the first order generating function $G^{(1)}$, which attempts to recast the perturbed Hamiltonian into a clearly--integrable form, diverge. 
This effectively means that the canonical transformation to action--angle coordinates cannot be completed where a resonance condition appears. 
The KAM theorem \cite{KolmogorovAM, *KArnoldM, *KAMoser1, *KAMoser2} is largely concerned with 1) understanding how close to a resonance a path must be to become chaotic, 2) proving that integrable tori of $H^{(0)}$ are in fact approximately preserved so long as they aren't too close to the resonances, and 3) formally showing how to actually construct a convergent perturbation theory away from the resonances.
We will not concern ourselves with the details of their results overmuch below, except to note that resonances with low $\ell$ and $k$ (near 0) typically affect or destroy a larger neighborhood of nearby integrable orbits than those with larger $\ell$ or $k$. 
These qualitative features of the dynamics are visible throughout our numerical studies, detailed in a series of animations, included in the supplementary materials and described in section~\ref{sec-anim_strobo}. 

\subsection{More formal analysis of the two-measurement example \label{sec-xzperturb}}
We here consider the system from section \ref{sec-xzkick} in the perturbative regime (both measurements are still relatively weak) more formally. 
Recall the notation and equations in \eqref{xz_kick}, \eqref{hexpansion}, and in their surrounding text. 
Below, we will take $\tau_x = 1$, thereby handling all times in units of $\tau_x$; the kicking period will also taken to be unity ($\Lambda = 1$). We reiterate that $\tau_m \ll 1$, such that the weak kicks are narrow compared with their repetition period.
The OP Hamiltonian \eqref{hxz} is $H^\star= H^{(0)}(p) + \epsilon H^{(1)}(\theta,p,t) + \epsilon^2 H^{(2)}(\theta,p,t) + ... $, or
\be \begin{split}
H^\star(\theta,p,t)  = \frac{p^2-1}{2} +\left( \sum_{n=1}^\infty \epsilon^n g^n(t) \right) \tilde{H}(\theta,p) \end{split} \ee 
for \be 
 \tilde{H} \equiv \frac{p^2-1}{2} \sin^2\theta - p\sin\theta\cos\theta.
\ee
We may truncate the series to a desired order in $\epsilon$, and when $\epsilon \ll 1$ (weak kicks, corresponding to $A \ll \tau_x$), the first or second order approximation of the Hamiltonian will reflect the dynamics quite well.
Below we undertake the actual Fourier expansion implied by \eqref{resonance}, to see how much more we can learn analytically about its range of applicability in this system.

\par We will need to decompose both $\tilde{H}$ (which is periodic in $\theta$) and $\sum_n \epsilon^n g^n(t)$ (which is periodic in time). We start with $\tilde{H}$. Recall that the Fourier form and coefficients may be defined as
\be
\left(\sum_n \epsilon^n g^n \right)\tilde{H}(\theta,p) = \sum_{\ell = -\infty}^{\infty} C_\ell(p,t) e^{i\ell \theta} 
\ee  \be 
C_\ell = \frac{1}{2\pi} \int_0^{2\pi} \left( \sum_n \epsilon^n g^n \right) \tilde{H}(\theta,p) e^{-i \ell \theta} d\theta.
\ee
We find the coefficients
\be 
C_0 = \left( \sum_n \epsilon^n g^n \right)\frac{p^2-1}{4}, \ee \be 
C_{\pm2} = \left( \sum_n \epsilon^n g^n \right) \left( \frac{1-p^2}{8} \mp \frac{p}{4i}\right),
 \ee
with those for all other $\ell$ vanishing. We now perform a similar computation, expanding the above coefficients into Fourier form in $t$, such that
\be 
C_\ell(p,t) = \sum_k \zeta_{\ell,k}(p) e^{2i \pi k t} = \sum_{k,n} \zeta_{\ell,k}^{(n)}(p) e^{2i \pi k t}.
\ee
This implies that we may write new coefficients
\be 
\zeta_{0,k}^{(n)}(p) = \epsilon^n \frac{p^2-1}{4} \int_0^1 g^n(t) e^{-2i\pi k t} dt, \:\text{ and}
\ee \be 
\zeta_{\pm2,k}^{(n)}(p) = \epsilon^n \left( \frac{1-p^2}{8} \mp \frac{p}{4i}\right) \int_0^1 g^n(t) e^{-2i\pi k t} dt.
\ee
Evaluating these expressions requires the result 
\be \label{cnk} \begin{split}
\mathcal{C}_{n,k} &\equiv \int_0^1 g^n(t) e^{-2i\pi k t} dt \\ &
= 2 (-1)^k \int_0^\frac{1}{2} \exp \left[ - \frac{n x^2}{2 \tau_m^2} \right] \cos(2 \pi k x) dx
\\ &= (-1)^k \sqrt{\frac{\pi}{2 n}}\tau_m \exp \left[  - \frac{2 \pi^2 k^2 \tau_m^2}{n} \right] \\ &\quad \times \left( \text{erf} \left[ \frac{\beta_{n,k}^+}{\tau_m \sqrt{8 n}} \right] + \text{erf} \left[ \frac{\beta_{n,k}^-}{\tau_m \sqrt{8 n}} \right] \right)
\end{split} \ee
where we have defined
$ \beta_{n,k}^\pm \equiv n \pm 4 i k \pi \tau_m^2,$ the error functions according to 
\be 
\text{erf}(z) \equiv \frac{2}{\sqrt{\pi}} \int_0^z e^{-u^2} du,
\ee
and used the form \eqref{kick}.
With $\tau_m$ much narrower than $\tau_x=1$ we may approximate $\mathcal{C}_{n,k}$ by extending the integration bounds, i.e.
\begin{widetext}
\be \begin{split}
\mathcal{C}_{n,k} &= (-1)^k \int_{-\frac{1}{2}}^\frac{1}{2} \exp \left[ - \frac{n x^2}{2 \tau_m^2} \right] \cos(2 \pi k x) dx \\ &\approx  (-1)^k \int_{-\infty}^\infty \exp \left[ - \frac{n x^2}{2 \tau_m^2} \right] \cos(2 \pi k x) dx = (-1)^k \tau_m \sqrt{\frac{2\pi}{n}} \exp\left[- \frac{2 k^2 \pi^2 \tau_m^2}{n} \right].
\end{split} \ee
The entire Hamiltonian can be written, to order $O(\epsilon^N)$, as
\be \begin{split} \label{Hexpand}
H^\star &= H^{(0)}(p) + \sum_\ell \sum_k \sum_n \zeta_{\ell,k}^{(n)}(p) e^{i(\ell \theta + 2\pi k t)} \\
& = H^{(0)}(p)+\sum_k \sum_n \left(\zeta_{0,k}^{(n)}(p) + \zeta_{-2,k}^{(n)}(p) e^{-2i\theta} + \zeta_{2,k}^{(n)}(p) e^{2i\theta} \right) e^{2i\pi k t} \\
&= H^{(0)}(p) + \left( \frac{p^2-1}{4} (1- \cos(2\theta))- \frac{p}{2} \sin(2\theta)\right)\sum_{k = -\infty}^\infty \sum_{n=1}^N \epsilon^n\:\mathcal{C}_{n,k} e^{2i\pi k t}.		
\end{split}\ee
\end{widetext}
This is still an \emph{exact} result if $N \rightarrow \infty$ and the exact form of $\mathcal{C}_{n,k}$ is used. It is however easily set up to truncate any of the expressions down to a specific order $N$ in $\epsilon$. The range of $k$ which makes substantial contributions to its sum at a given $n$ should run roughly proportional to $\sqrt{n}/\tau_m$, such that the $k$-sum could also be truncated, although the number of relevant terms will remain large (this is clearly valid based on the approximate form of $\mathcal{C}_{n,k}$). An approximate version of the Hamiltonian, based on a truncated Fourier series, is useful in that it can be more easily studied analytically to a desired order in $\epsilon$.
\par Resonance phenomena are not relevant for $\ell$ and $k$ where $\zeta_{\ell,k} = 0$, but the analysis above shows that we have more than enough non--zero coefficients at play in this system to generate all the resonances of interest using \eqref{resonance}. Specifically, $\nu^{(0)} = p$ and $\Lambda = 1$ gives resonances at $p = 2\pi k/\ell$ for $\ell = 0,\pm 2$ and any integer $k$, with emphasis on $|k|$ near zero. This predicts the onset of chaos at integer multiples of $\pi$, including the fixed point of $H^{(0)}$ along $p = 0$, and is entirely consistent with our observations in Figs.~\ref{fig-earlyreson} and \ref{fig-LMreson}; we elaborate further in appendix~\ref{sec-anim_strobo}.

\par This analysis does not preclude the formation of resonances beyond those we just described using \eqref{resonance}. The expression \eqref{resonance} only includes matches in period between the zeroth and first order (in $\epsilon$) parts of $H^\star$. For larger $\epsilon$, the first--order approximation of the full Hamiltonian is no longer a good representation of the dynamics. It is known that resonant tori will give way to alternating elliptic and hyperbolic fixed points (see e.g.~section 7.2 of the text by Ott \cite{BookOtt}, and the Poincar\'{e}--Birkhoff theorem). Paths about the elliptic fixed points are closed curves in the phase space (representing periodic orbits), which themselves develop resonances with the perturbation (which may become more pronounced in numerical studies at larger values of $\epsilon$). 

\section{Description of supplementary animations}
We give a complete list of the animations included in the supplementary materials here, along with captions to clarify the details and context of each.
All animations from the system of section~\ref{sec-xzkick} can be found at the link \href{https://drive.google.com/file/d/1cx__Aggt40s3r8ueTe8LlqZyAWLb5NV1/view?usp=sharing}{\tt here} (films collected in single {\tt .pdf}) or \href{https://drive.google.com/drive/folders/12LgI0dCiSjRYoHO9oiWDaXm7S0PzM7Y1?usp=sharing}{\tt here} (individual {\tt .mp4} files). See article's main arXiv page for links if viewing in print.

\subsection{Path pairs in phase space}
The two videos below superpose the evolution of particular path--pairs over the dynamic phase space context in which they evolve. We see the paths, which start next to each other in the phase space get pulled apart over time, in an illustration of the basic definition of chaos. 
\par {\texttt{VT\_psani1.mp4}} -- This video shows the phase space ($\theta$ on the $x$--axis, $p$ on the $y$--axis) of the system described in section~\ref{sec-xzkick}, animated in time. Different colors represent different stochastic energies at any given moment, and the separatrix at any given time is shown in light green. It sits along the $p=0$ line when the system is like a rotor, and briefly flares out with each kick. The red and blue dots track the evolution of the paths of the same colors in Fig.~\ref{fig-xzexamples}(a). All operating parameters are the same as in that figure ($\tau_x = 1~\mu\mathrm{s}$, $A = 0.99~\mu\mathrm{s}$, $\tau_m = 25~\mathrm{ns}$).
\par {\texttt{VT\_psani3.mp4}} -- This video tracks the paths in Fig.~\ref{fig-xzexamples}(b). All other details are identical to those in the video above.

\subsection{Stroboscopic phase portraits \label{sec-anim_strobo}}
We here discuss animated stroboscopic phase portraits for the system described in section~\ref{sec-xzkick}. Each frame is a stroboscopic portrait at a different value of $\epsilon$, with the animation running over increasing values of $\epsilon$. In all of the films below, $\tau_x = 1.0~\mu\mathrm{s}$, such that $A$ (in $\mu\mathrm{s}$) is numerically equivalent to $\epsilon$ (they are used interchangeably here). We continue using $\tau_m = 25~\mathrm{ns}$ throughout. All initial conditions, arranged on a mesh, run to $T = 100~\mu\mathrm{s}$, and are plotted at the strobe times unless they have diverged (the numerical integration of the path's value is no longer certain to be correct to within some tolerance). Strobe times occur at every integer time between kicks; at these times, paths are plotted as a point, with the point's color denoting its LE at that time. Black dots correspond to $\lambda = 0$. Cool earth colors range over $\lambda \in [0,-0.25]~\mathrm{MHz}$, growing lighter across that range, with every point $\lambda < -0.25~\mathrm{MHz}$ plotted at the extreme end (lightest gray-brown hue) of that cool color bar. Likewise, warm colors range over $\lambda \in [0,0.25]~\mathrm{MHz}$, growing lighter across that range, with every point $\lambda > 0.25~\mathrm{MHz}$ plotted at the extreme end (lightest yellow hue) of that warm color bar. 

\par {\texttt{VT\_chonset\_survey.mp4}} -- The evolution of a large swath of the phase space over low values of $\epsilon$ is shown. Resonances at $p = \pm2\pi,\pm3\pi,\pm4\pi$ are immediately visible, followed by those at $p = \pm\pi$, and then those at $p = 0$. In general, we see similar dynamics at all of these resonances, where the initially ($\epsilon = 0$) flat line in phase space at a resonant $p_0$ opens into a series of stable islands around an elliptic fixed point (separated by hyperbolic fixed points). As $\epsilon$ increases, resonances form within the periodic islands themselves; these resonances gradually destroy the periodic orbits forming the island, until the entire region of phase space taken up by the island is effectively a completely--chaotic sea. This process is generic to chaotic Hamiltonians with resonances \cite{BookOtt}, the kicked rotor or standard map being one of the classic examples \cite{BookJose}. Although we only see its early stages in this particular film, the process will run to conclusion in subsequent ones. There is considerable variability in how fast the different lines generate elliptic islands and hyperbolic fixed points, and in how large those island get, thereby disrupting the rotor behavior (relative to the unperturbed case) nearby. Generally, we can already see that islands forming at higher $|p|$ in the phase space (corresponding to rarer events \cite{Lewalle2016}) undergo this process faster than those at lower $|p|$, relative to changes in $\epsilon$.

\par {\texttt{VT\_chonset\_reson.mp4}} -- Once again we animate the phase space over low values of $\epsilon$. This time, our emphasis is specifically on the simplest resonances. Initial conditions generating each frame are chosen specifically for $p_0 = 0,\pi/3,\pi/2,2\pi/3,\pi,3\pi/2,2\pi,3\pi$. Auxiliary lines shifted by $\pm 0.2$ from the main list of $p_0$ are also included for context. This gives a clear look at the relative ``strengths'' of different resonances, emphasizing the different rates, relative to changes in $\epsilon$, at which islands from resonances grow, generate internal resonances, and are broken apart. Although only the $p =0, \pi, 2\pi,$ and $3\pi$ resonances appear in the first-order expression \eqref{resonance}, we can begin to see qualitatively similar effects happening, with periodic islands of half the size and spacing, at the $p=3\pi/2$ line by the end of this video, suggesting that a wider variety of resonances come into play when higher orders of $H^{(n)}$ become relevant to the dynamics.

\par {\texttt{VT\_chevolv\_narrow.mp4}} -- We zoom in on the formation and evolution of the longest-lived islands in the phase-space, at $p = 0,~ \pm\pi$, over moderate to high values of $\epsilon$. An increasing number of resonances within the stable islands formed from the simpler ones are visible. These gradually eat away at the main islands as $\epsilon$ grows. Those which formed around $p = \pm \pi$ are almost completely destroyed by the end of the video.

\par {\texttt{VT\_chfinal\_pi.mp4}} -- We slow down the animation over the destruction of the islands around $p = \pi$ at high $\epsilon$, so that they can be viewed in detail. Note the numbers of sub-islands forming within the main one; there is a ``countdown'' which occurs in the number of alternating elliptic and hyperbolic fixed points which emerge from the larger island as it is destroyed. (See e.g.~Ott \cite{BookOtt}, section 7.2 for details.) That is, the main island is gradually destroyed through the emergence first of a period--6 island chain, followed by a period--5 island chain, and so on, down to a period--2 sub--island pair just visible as the last periodic remnants in that part of the phase portrait disintegrate. 

\par {\texttt{VT\_chfinal\_center.mp4}} -- Finally, we conclude with a detailed look at the destruction of the final remaining periodic islands in the phase space, around $p = 0$, at values of $\epsilon$ extremely close to one (the stronger measurement ``kicks'' are nearly perfectly projective). 

\section{Details on numerical methods}
In Fig.~\ref{fig-str} we plot a number of ``stretching parameters'' which quantify various properties of the LM's shape. We define those parameters explicitly in section~\ref{sec-lmdef}, and then make some general comments about the numerical computation of the LM in section~\ref{sec-manrefine}. The three stretching parameters we describe are designed for numerical use, but are very much an outgrowth of the ideas laid out in section~\ref{sec-ch_mani} of the main text.

\subsection{Numerically Quantifying Manifold Deformation \label{sec-lmdef}}
\par The first parameter quantifying the deformation of the manifold examines the degree to which the manifold ``stretches out'' relative to its initial configuration, and is given by
\be \label{s1}
s_1(t) = \frac{1}{t}\ln\left(\frac{L(t)}{L(0)}\right),
\ee
where the length of the LM is given by
\be \label{length}
L(t)=\sum_{i=0}^{N-1}\sqrt{(\theta_{i+1}(t)-\theta_{i}(t))^2+(p_0^{i+1}-p_0^{i})^2}.
\ee
Numerically, the manifold is defined in terms of a discrete string of points, indexed by $i$.
When the parameter $s_1(t)$ is positive, it indicates an exponential rate of growth of the length of the manifold; by definition this means that states which start near to each other on the manifold spread out dramatically over intervals where $s_1$ sustains a positive value over time. Since the LM remains continuous, this should happen in conjunction with growth in the second parameter we define, which is given by
\be \label{s2}
s_2(t) = \frac{1}{t}\ln \left(J_{av}(t)+1\right).
\ee
We define $J_{av}(t)$ as the average $J_t$ of the manifold, calculated as the weighted average
\be \label{J_av}
J_{av}(t)=\frac{1}{2}\sum_{i=1}^{N-1}w_i(t)\left(\left|J_i^+(t)\right|+\left|J_i^-(t)\right|\right)
\ee
where
\be \begin{split} \label{Jpm} 
J^+_i(t)=\frac{\theta_{i+1}(t)-\theta_{i}(t)}{{p_0^{i+1}-p_0^{i}}}, \quad
J^-_i(t)=\frac{\theta_{i}(t)-\theta_{i-1}(t)}{{p_0^{i}-p_0^{i-1}}},
\end{split} \ee
and the weights $w^{i}(t)$ are the fraction of the range of initial momenta taken up by each segment, i.e.
\be \label{weights}
w_i(t) = \tfrac{1}{2}\left(p_0^{i+1}-p_0^{i-1} \right).
\ee
These weights will all be equal as long as the chosen initial momenta are evenly spaced, but are required to compensate for the fact that this is in general not the case (see appendix~\ref{sec-manrefine}). We showed above how chaos is related to $\left|J_t\right| \gg 1$; sustained growth in $s_2$ implies exponential growth in $\left|J_t\right|$ across the relevant segment of the manifold. 
Following the arguments of section~\ref{sec-ch_mani}, we expect this to be connected to the formation of chaotic regions and higher numbers of catastrophes. We capture this last feature with a third parameter
\be \label{s3}
s_3(t) = \frac{1}{t}\ln\left(1+N_c\right(t)),
\ee
where $N_c(t)$ is the number of catastrophes in the manifold at time $t$, i.e. the number of places where $J_t=0$. Sustained positive values and growth in $s_3(t)$ imply that the number of catastrophes is increasing exponentially; this necessarily implies a corresponding amount of growth in multipath solutions connecting the affected boundary conditions.
\par Finally, it is useful to define an average LE of the manifold, for the purposes of having an explicit measure of chaos in a manifold segment to compare against the above parameters describing aspects of the LM's shape and deformation over time. We initialize three manifolds with initial coordinates $\theta_0$ and $\theta^\pm_0=\theta_0\pm\delta\theta_0$. The average distance between points is obtained using the same weighting as above, combined with \eqref{dist}, to give
\be\begin{split} \label{dist_av}
D_{av}(t) =& \tfrac{1}{2}\sum_{i=1}^{N-1}w_i(t)\sqrt{\left(\delta x_i^+(t)\right)^2 + \left(\delta z_i^+(t)\right)^2} \\ &+ \tfrac{1}{2}\sum_{i=1}^{N-1}w_i(t)\sqrt{\left(\delta x_i^-(t)\right)^2 + \left(\delta z_i^-(t)\right)^2},
\end{split}\ee
where $\left(\delta x^\pm_i (t)\right)^2 = (\sin \theta_i(t) - \sin\theta^\pm_i(t))^2$ and $\left(\delta z^\pm (t)\right)^2 = (\cos \theta_i(t) - \cos\theta_i^\pm(t))^2$ and the weights $w_i(t)$ are given in \eqref{weights}. The average LE can then be calculated using
\be \label{lyap_av}
\lambda_{av}(t) = \frac{1}{t}\ln\left(\frac{D_{av}(t)}{D_{av}(0)}\right),
\ee
in analogy with \eqref{lyap}.

\subsection{Manifold refinement methods \label{sec-manrefine}}
We have shown, e.g.~in Fig.~\ref{fig-str}, that length of a LM in OP phase space may increase dramatically in time. While this may happen to some degree in integrable systems, simply due to the LM stretching across many windings about the Bloch sphere or spiraling around an elliptic fixed point, the effect is far less predictable and far more pronounced in the chaotic systems which are our topic now. In order to perform good plotting and analysis of the LM after a given time interval, the resolution of paths in the LM must be adequate at the final time. The point resolution of the LM has to be especially good near final boundary conditions of interest for a multipath, if we are to catch all of the solutions in a multipath group and find the $p_0$ which lead to the desired $\theta_T$ with high precision. It should be apparent that this raises considerable numerical difficulties, since it is not obvious, a priori, where in the range of $p_0$ a high density of paths should be initialized in order to obtain a good LM after integration. Furthermore, the number of paths required may quickly become prohibitively large for timely computation as the interval over which the LM needs to be integrated grows. To add to the complications, there are certain paths where $p(t)$ diverges to $\pm \infty$ (or close enough to stop a numerical integration), which must be handled carefully to avoid wasting time or crashing certain types of integrators.
\par We resolve these issues by developing a process to refine the manifold; that is, we have written algorithms which run a preset number of initial conditions forward, determine where there are gaps between points in the final manifold which are unacceptably large, and then runs more points in the neighborhood of the relevant initial conditions so as to fill in the final manifold up to the desired resolution. We have used the Python programming language, and a mix of fourth--order Runge--Kutta and Bulirsch--Stoer integration (see \cite{NumRecC}) to do this. The algorithm may iterate many times until the final manifold passes some resolution tests over its entire final range. Some version of this process is required to obtain the graphics shown in Figs.~\ref{fig-LMreson},~\ref{fig-xzlm},~\ref{fig-xzdens},~\ref{fig-manymulti},~\ref{fig-case1},~\ref{fig-case2}, and \ref{fig-str}. Such an algorithm necessarily results in a manifold sampled over points that are unevenly-spaced in $p_0$, which motivates the use of weighting factors \eqref{weights} in evaluating shape properties of the LM.
\par We highlight an aspect of the stretching we have shown particularly using \eqref{s1} however, which is the sheer number of paths required to get a usable manifold after even moderate $T$ for the strongly chaotic regime of larger $\epsilon$. For $\epsilon = 0.99$, a single manifold for $\theta_0 = 0$ and $p_0 \in [0,2]$, used to find the paths at $T = 4.0~\mu\mathrm{s}$ as shown in Fig.~\ref{fig-xzdens}(c), or make a plot like those in Figs.~\ref{fig-xzlm} or \ref{fig-str}(a), ends up requiring integration of $20,236$ initial conditions over the time interval. This can be done to quite high precision on a personal computer within a few hours. By adding one more kick at these same parameters, i.e. going $T = 5.0~\mu\mathrm{s}$ as shown in Fig.~\ref{fig-xzdens}(d), that number jumps to $311,710$ OPs required to construct the LM; this may be integrated precisely on a personal computer in 1--2 days. It should quickly be apparent how this growth becomes a problem for numerical computation; getting a good manifold after even $7$ or $8$ $\mu\mathrm{s}$ in the strongly--chaotic regime could take weeks or months without high--powered computational facilities.

\section{SQT diffusion with two measurements \label{sec-diffuse}}
The mathematical context of many of the objects we use here is discussed in the literature on stochastic mathematical methods; see, e.g., the book by Gardiner \cite{BookGardiner2} for further details. Suppose we are given a stochastic differential equation (SDE, a Langevin equation)
\be \label{lang_strato}
\dot{x} = A(x,t) + B(x,t)\xi(t),
\ee
in Stratonovich form. As discussed elsewhere \cite{Gambetta2008, Chantasri2013, Lewalle2016}, the SDEs we obtain from a Bayesian approach become equivalent to the Stratonovich form of the SDEs we would obtain from a Stochastic Master Equation (SME) approach \cite{BookWiseman, Jacobs2006} if we make a simplification by assuming the noise is white, specifically by substituting in $r_q(t) = q(t) + \sqrt{\tau_q} \xi(t)$ where $\xi(t) = dW(t)/dt$ and $dW(t)$ is a Weiner process. Given a one--dimensional SDE in Stratonovich form \eqref{lang_strato}, the corresponding FPE is given by \cite{BookGardiner2}
\be \label{fpe_strato} \begin{split}
\partl{\wp}{t}{} &=  - \partl{}{x}{} (A \wp) + \frac{1}{2} \partl{}{x}{} \left( B \partl{}{x}{}(B\wp) \right) \\
& = \left( \frac{B}{2} \partl{B}{x}{2} + \frac{1}{2} \left( \partl{B}{x}{} \right)^2 - \partl{A}{x}{} \right) \wp\\
&\quad +  \left( \frac{3 B}{2} \partl{B}{x}{} - A \right) \partl{\wp}{x}{} + \frac{B^2}{2} \partl{\wp}{x}{2},
\end{split}\ee
for $\wp = \wp(\theta_t,t|\wp(\theta_0,0))$.

\par Let us apply this formula to study the diffusion under two measurements with ``kicking'', as treated in section~\ref{sec-xzkick}. The equation of motion is
\be \begin{split}
\mathcal{F} & = \frac{r_x}{\tau_x} \cos\theta - \frac{r_z}{\tau_z} \sin \theta \\ &= \sin\theta \cos\theta \left( \frac{1}{\tau_x} - \frac{1}{\tau_z} \right) + \frac{\xi_x}{\sqrt{\tau_x}} \cos\theta - \frac{\xi_z}{\sqrt{\tau_z}} \sin\theta,
\end{split} \ee
where we have simplified the noise by taking $r_x = \sin \theta + \sqrt{\tau_x}\xi_x$ and $r_z = \cos\theta + \sqrt{\tau_z}\xi_z$. We let $\mathbf{B} = (B_x,B_z)$, and take appropriate dot products in \eqref{fpe_strato} to obtain the FPE
\be\begin{split} \label{full_fpe_xzkick}
\partl{\wp}{t}{} = & \frac{3}{2} \left(\frac{1}{\tau_x}-\frac{1}{\tau_z} \right)\left( \sin^2\theta - \cos^2\theta \right) \wp \\ &+\frac{5}{2} \cos\theta \sin\theta \left(\frac{1}{\tau_x}-\frac{1}{\tau_z} \right) \partl{\wp}{\theta}{} \\ &+\frac{1}{2}\left( \frac{\cos^2\theta}{\tau_x} + \frac{\sin^2\theta}{\tau_z} \right) \partl{\wp}{\theta}{2},
\end{split}\ee
where $\wp(\theta,t)$ is the probability distribution at a given time, which is always contingent on having evolved forward from some given initial distribution. The term attached to $\partial_\theta^2\wp$ is effectively a diffusion constant; notice that for small $\tau_z$ (e.g. at a kick) the diffusion constant grows very large, meaning that for a short time trajectories may jump across large distances in the state-space.
Note also that \eqref{full_fpe_xzkick} reduces to 
\be \label{fpe_xzsimple}
\partl{\wp}{t}{} = \frac{1}{2\tau} \partl{\wp}{\theta}{2},
\ee 
when $\tau_x = \tau = \tau_z$. Thus we see that when the two-measurement system reduces to a simple rotor, the underlying SQTs undergo isotropic diffusion. This is consistent with theoretical results from elsewhere \cite{Lewalle2016,Chantasri2017}, as well as observations in the original experimental implementation of this system with fixed measurement strengths \cite{Leigh2016}. It sits in contrast with the more complex case \eqref{full_fpe_xzkick} where the measurement strengths are unequal, and coefficients in the FPE are state dependent (thereby privileging collapse to one set of eigenstates over the other). Periodic strengthening of the measurement ostensibly results in an overall faster rate of diffusion to higher winding numbers. The diffusion is no longer isotropic when $\tau_x \neq \tau_z$, and shorter $\tau$ corresponds directly to a faster diffusion rate (bigger diffusion constant) in a particular direction.

\section{A Simple Model of OPs in the projective kicking limit\label{sec-projlim}}
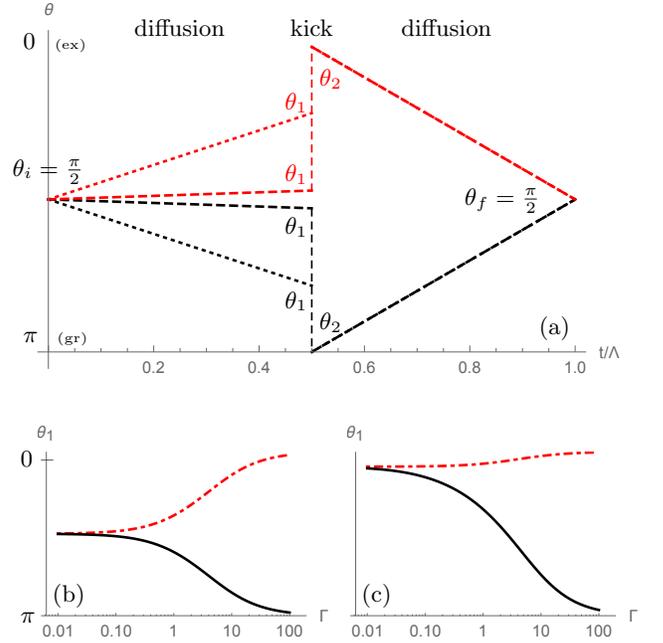
\begin{figure}
\begin{picture}(240,250)
\put(10,100){\includegraphics[width = .9\columnwidth]{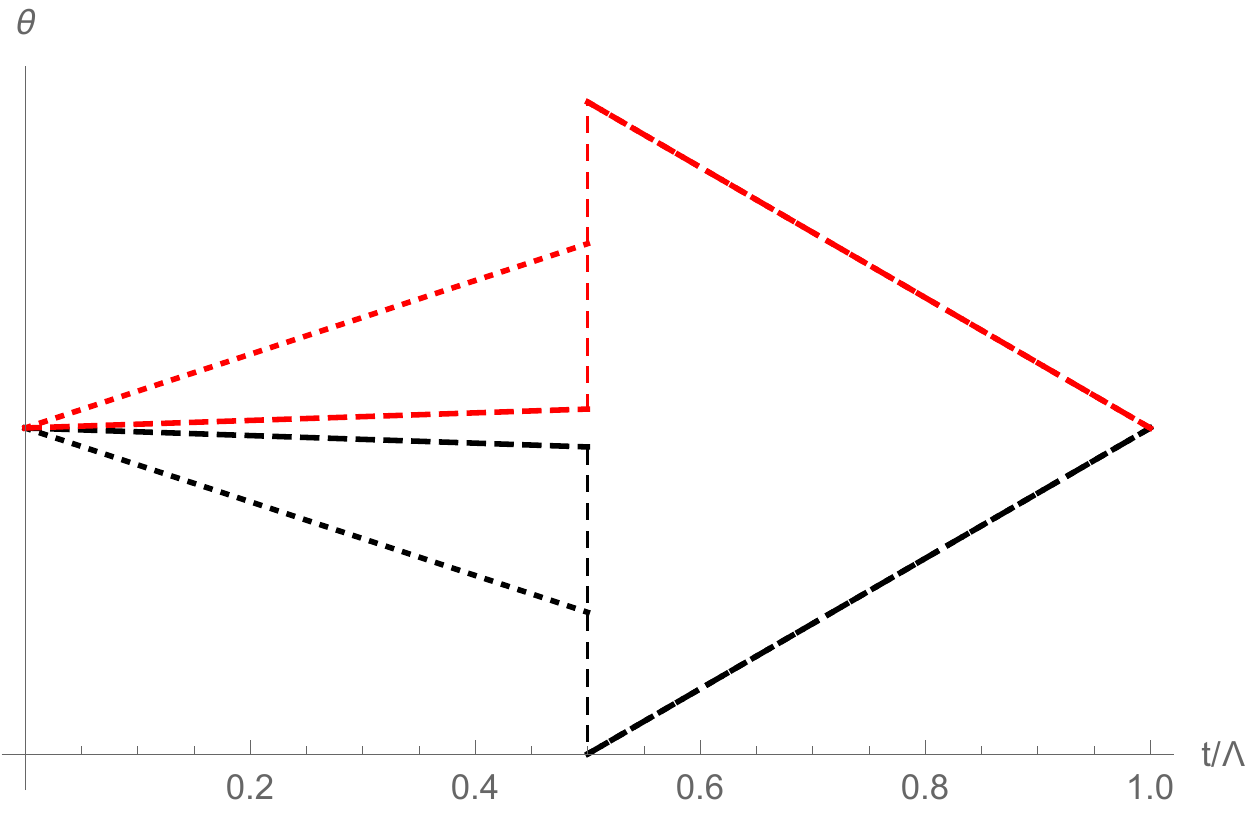}}
\put(10,0){\includegraphics[width = .45\columnwidth]{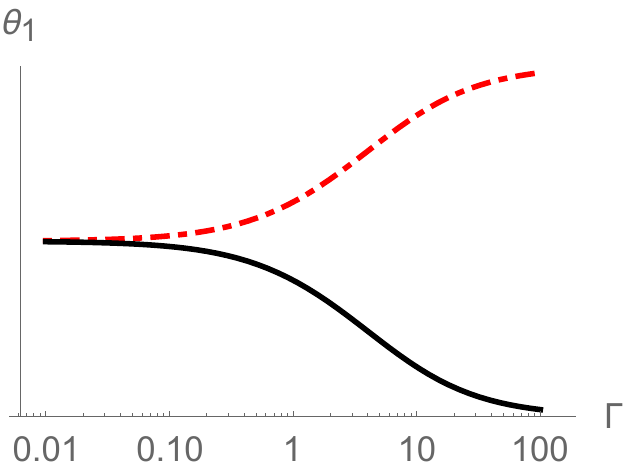}}
\put(127,0){\includegraphics[width = .45\columnwidth]{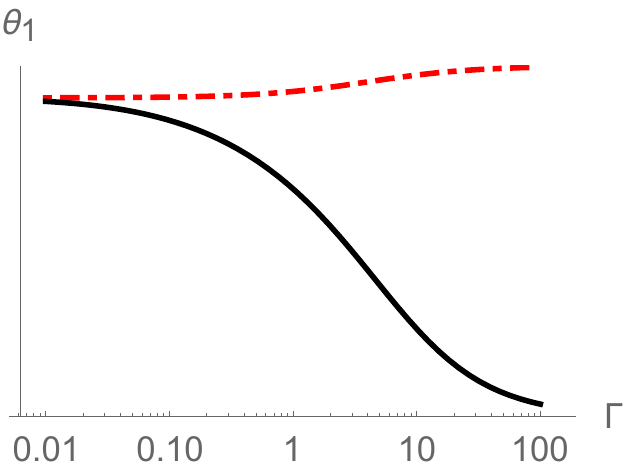}}
\put(5,225){\small $0 \:\:$ {\tiny (ex)}}
\put(5,114){\small $\pi \:\:$ {\tiny (gr)}}
\put(1,177){\small $\theta_i = \tfrac{\pi}{2}$}
\put(47,230){\small diffusion}
\put(148,230){\small diffusion}
\put(106,230){\small kick}
\put(172,166){\small $\theta_f = \tfrac{\pi}{2}$}
\put(104,175){\small \color{red} $\theta_1$}
\put(104,202){\small \color{red} $\theta_1$}
\put(104,155){\small $\theta_1$}
\put(104,127){\small $\theta_1$}
\put(117,119){\small $\theta_2$}
\put(117,211){\small \color{red} $\theta_2$}
\put(200,117){(a)}
\put(16,16){(b)}
\put(133,16){(c)}
\put(4,67){\small $0$ --}
\put(4,8){\small $\pi$}
\end{picture}
\caption{In (a) we show OPs for the situation discussed in appendix~\ref{sec-projlim}. Paths are allowed to diffuse isotropically for a time interval $\Lambda/2$, are measured projectively along $\sigma_z$, and are then allowed to diffuse isotropically again for another $\Lambda/2$. We show the symmetric case, where OPs diffuse from $\theta_i = \tfrac{\pi}{2}$ to $\theta_1$, collapse to either $\theta_2 = 0$ (red) or $\pi$ (black), and then diffuse again to the post-selected state $\theta_f = \tfrac{\pi}{2}$. The amount of diffusion allowed before each kick is characterized by $\Gamma = \Lambda / \tau$. The dashed paths are allowed to diffuse relatively little before the kick ($\tau \gg \Lambda$, $\Gamma = 0.2$ is shown), whereas the dotted paths are allowed more diffusion before the kick ($\tau \ll \Lambda$, $\Gamma = 5$ is shown). In the lower plots we show solutions optimizing the value of $\theta_1$ according to the transcendental equations \eqref{transcendental}, with $\theta_i = \tfrac{\pi}{2}$ in (b), and $\theta_i = 0.286$ in (c). The path through the excited state is shown in dash-dotted red, and the path through the ground state is shown in solid black. As $\Gamma$ grows, more diffusion is allowed between each kick, and the jump made by the OP at the kick is reduced (the optimal $\theta_1$ is close to $\theta_i$ for $\Gamma \ll 1$, and close to $\theta_2$ for $\Gamma \gg 1$). 
}\label{fig-proj}
\end{figure}

We here consider the optimal dynamics in the limit where $\tau_m\rightarrow 0$, and $\epsilon\rightarrow 1$ (see eq.~\eqref{kick}); these parameters correspond to diffusion under equal measurement strengths, periodically punctuated by instantaneous, perfectly--projective $z$--measurements. We will use a simplified model which ignores winding counts, considering two MLPs over a single kicking period $\Lambda$. That is, we prepare an initial state $\theta_i$, allow for diffusion over a time interval $\Lambda/2$ to some $\theta_1$, perform a projective $z$--measurement resulting in a state $\theta_2$, and then again allow for isotropic diffusion over a duration $\Lambda/2$, post--selected on a neighborhood around some final state $\theta_f \in [0,\pi]$. We will assume that $\theta_i \in [0,\pi]$, and that the projective measurement kick may either collapse the state to $\theta_2 = 0$ or $\theta_2 = \pi$. This situation is represented in Fig.~\ref{fig-proj}(a). Assuming we initialize the state at $\theta_i$ (i.e. the probability density is a delta function), the probability density to reach some $\theta_1$ right before the kick (from solving \eqref{fpe_xzsimple}) is given by
\be \label{purediff1}
\wp(\theta_1|\theta_i) = \sqrt{\frac{\tau}{\pi \Lambda}} \exp \left[ - \frac{\tau}{\Lambda} (\theta_1 - \theta_i)^2 \right].
\ee
The discrete probability to collapse to $\theta_2 = 0$ or $\pi$ based on the previous diffusion step is given by
\be \label{projkick}
P(\theta_2|\theta_1) = \bigg\lbrace \begin{array}{cl}
 \cos^2(\theta_1/2) & \text{ to move to } \theta_2 = 0 \\
 \sin^2(\theta_1/2) & \text{ to move to } \theta_2 = \pi.
\end{array}
\ee
Diffusion after the kick looks like \eqref{purediff1}, i.e.
\be 
\wp(\theta_f|\theta_2) = \sqrt{\frac{\tau}{\pi \Lambda}} \exp \left[ - \frac{\tau}{\Lambda} (\theta_f - \theta_2)^2 \right].
\ee
Combining these two diffusion steps and intermediate jump, we may construct two probability densities $\wp_{ex}$ and $\wp_{gr}$, the first of which is associated with a path that goes through the excited state ($\theta_2 = 0$), and the second of which goes through the ground state ($\theta_2 = \pi$). These read
\be \begin{split}
\wp_{ex} &= \cos^2\left( \frac{\theta_1}{2}\right) \exp \left[ - \frac{\tau}{\Lambda} \left\lbrace \theta_f^2 + (\theta_1 - \theta_i)^2 \right\rbrace\right], \text{ and} \\
\wp_{gr} &= \sin^2\left( \frac{\theta_1}{2}\right) \exp \left[ - \frac{\tau}{\Lambda} \left\lbrace (\theta_f - \pi)^2 + (\theta_1 - \theta_i)^2 \right\rbrace\right].
\end{split} \ee
Note that to account for winding numbers around the Bloch sphere, we would additionally have to let $\theta_1 \rightarrow \theta_1 + 2\pi \ell$ and sum over all integers $\ell$. This implies that our system reaches equilibrium much faster than if it were on the real line.
\par We have shown through the connection of \eqref{hrotor} and \eqref{fpe_xzsimple} that the OP dynamics over isotropic diffusion are straight lines, i.e.~when $\tau_x = \tau = \tau_z$, the OP goes as $\theta(t) = \theta_0 + p_0\: t/\tau$. Therefore, we understand that the probabilities above describe OPs which go from $\theta_i \rightarrow \theta_f$ via a straight line from $\theta_i \rightarrow \theta_1$, a jump from $\theta_1\rightarrow\theta_2$, then another straight line from $\theta_2 \rightarrow \theta_f$. The remaining question is: what is the value of $\theta_1$ which optimizes the probability density? (What $\theta_1$ does the OP go through?) This can be computed by taking $\partial_{\theta_1} \ln \wp = 0$, and solving for the optimal value of $\theta_1$. The solutions are given according to the transcendental equations
\be \begin{split} \label{transcendental}
\tan \left( \frac{\theta_1}{2} \right) &+ \frac{2}{\Gamma} (\theta_1 - \theta_i) = 0 \text{ for } \wp_{ex}, \text{ or } \\
\cot \left( \frac{\theta_1}{2} \right) &- \frac{2}{\Gamma} (\theta_1 - \theta_i) = 0 \text{ for } \wp_{gr},
\end{split} \ee
where we have defined $\Gamma \equiv \Lambda / \tau$. The parameter $\Gamma$ is dimensionless, and since $\tau^{-1}$ sets the rate of diffusion between kicks, we understand that $\Gamma \ll 1$ represents a situation in which very little diffusion is allowed between kicks, whereas when $\Gamma \gg 1$ SQTs diffuse widely between kicks. In the main text we have emphasized examples in the intermediate regime where $\Gamma = 1$. Note also that $\Gamma$ scales the range of $p$ at which resonances appear (see eq.~\eqref{period}). The solutions to \eqref{transcendental} for two different $\theta_i$ are plotted in Fig.~\ref{fig-proj}(b,c). There we see that the value of $\theta_1$ which the OP takes is very close to $\theta_i$ when $\Gamma \ll 1$, and is very close to $\theta_2$ when $\Gamma \gg 1$. This is an intuitive result; if wide diffusion has occurred prior to a kick ($\Gamma \gg 1$), it is probable to find paths which have already diffused to the eigenstates they will collapse to when kicked, and these paths which make only a small jump under the projective measurement are optimal. However, if very little diffusion is allowed to occur before a measurement kick ($\Gamma \ll 1$), trajectories will not have been able to diffuse to the eigenstates of the kick, and the OP is forced to jump much further.


\bibliography{refs}

\end{document}